\documentclass[a4paper,fleqn,usenatbib,useAMS]{mnras}

\usepackage[T1]{fontenc}
\usepackage{ae,aecompl}

\usepackage{graphicx}	% Including figure files
\usepackage{amsmath}	% Advanced maths commands
\usepackage{amssymb}	% Extra maths symbols
\usepackage{multirow}
\usepackage{longtable}
\usepackage{threeparttable}

\usepackage{graphicx}	% Including figure files
\usepackage{multicol}        % Multi-column entries in tables
\usepackage{bm}		% Bold maths symbols, including upright Greek
\usepackage{pdflscape}	% Landscape pages

\title[Origins of the shocks in SCCs]{Origins of the shocks in high-mass starless clump candidates}

\author[F.-Y. Zhu et al.]{
Feng-Yao Zhu,$^{1,2}$\thanks{E-mail: zhufy@zhejianglab.com}
Junzhi Wang,$^{3}$\thanks{E-mail: junzhiwang@gxu.edu.cn}
Yaoting Yan$^{4}$
Qing-Feng Zhu$^{5}$
Juan Li$^{2}$
\\
$^{1}$Research Center of Intelligent Computing Platforms, Zhejiang Laboratory, Hangzhou, 311100, PR China\\
$^{2}$Shanghai Astronomical Observatory, Chinese Academy of Sciences, Shanghai, 200030, PR China\\
$^{3}$Guangxi Key Laboratory for Relativistic Astrophysics, School of Physical Science and Technology, \\
Guangxi University, Nanning 530004, PR China\\
$^{4}$Max-Planck-Institut f\"{u}r Radioastronomie, Auf dem H\"{u}gel 69, 53121 Bonn, Germany\\
$^{5}$Astronomy department, University of Science and Technology of China, Hefei, 230026, PR China\\
}

\date{Accepted XXX. Received YYY; in original form ZZZ}

\pubyear{2020}

\begin{document}
\label{firstpage}
\pagerange{\pageref{firstpage}--\pageref{lastpage}}
\maketitle

\begin{abstract}
Shocks are abundant in star-forming regions, and are often related with star formation. In our previous observations toward 100 starless clump candidates (SCCs) in the Galaxy, a sample of 34 SCCs associated with shocks is identified. In this work, we perform mapping observations of the SiO 2-1, 3-2, HC$_3$N 10-9, HCO$^+$ 1-0, H$^{13}$CO$^+$ 1-0, and H41$\alpha$ lines toward 9 out of the detected sources by using IRAM 30-m radio telescope to study the origins of the shocks in the SCCs. We find shocks in three sources (BGPS 3110, 3114, and 3118) are produced by collisions between the expanding ionized gas and ambient molecular gas, instead of by the star formation activity inside SCCs. On the other hand, shocks in the other six sources are related to star formation activity of SCCs. The signatures of protostellar outflows are clearly shown in the molecular lines toward BGPS 4029, 4472, 5064. Comparing our results with the previous ALMA observations performed in the same region, the shocks in BGPS 3686 and 5114 are also likely to be due to protostellar activity. The origin of shock in BGPS 5243 is still unclear although some features in the SiO spectra imply the presence of protostellar activity.
\end{abstract}


%%%%%%%%%%%%%%%%%%%%%%%%%%%%%%%%%%%%%%%%%%%%%%%%%%

%%%%%%%%%%%%%%%%% BODY OF PAPER %%%%%%%%%%%%%%%%%%

\section{Introduction} \label{sec:intro}

Massive stars (M$>8~M_\odot$) are important in the evolution of galaxies, but the origin of massive stars are under debate \citep{mot18}. Massive starless clumps in molecular clouds could be precursors of protostellar cores \citep{kir21}. A sample of massive clumps have been identified as Starless Clump Candidates (SCCs) in the 1.1 mm continuum Bolocam Galactic Plane Survey (BGPS) \citep{svo16}. Since high-mass protostars whose strong feedbacks can disturb the ambient environments are unlikely to be embedded in the SCCs of this sample \citep{svo16,cal18}, these SCCs represent the environments at the early stage of star formation \citep{svo16,mat17,cal18}.

Shocks are abundant in star-forming regions, and the processes of star formation are generally associated with shocks. The jets and molecular outflows from protostellar activities with supersonic speed can lead to shocks, and the expansion of H II region which is ionized by massive main-sequence stars may also cause shocks \citep{deh10,zhu15b,taf21}. Even in infrared dark clouds (IRDCs) regarded as the early stage of star formation, shocks were also detected \citep{mot18,li19}. The formation of shocks in early star-forming regions is commonly attributed to outflows from the protostars deeply embedded in molecular clouds. \citet{svo19} presented the ALMA Band 6 1.3 mm continuum and spectral line survey toward 12 high-mass SCCs. The shocks detected in these clumps are all caused by the outflows from low- and intermediate-mass protostars. However, some previous works suggested that colliding flows and cloud-cloud collision could also lead to low-velocity and widespread shocks \citep{jim10,ngu13}. \citet{cse16} performed a spectral line survey of a sample of 430 sources from the APEX telescope large area survey of the Galaxy (ATLASGAL). The SiO lines attributed to low-velocity shocks from cloud-cloud collisions are detected in some infrared-quiet clumps. In addition, a part of shocks in W43-MM1 ridge are also probably associated with colliding flows \citep{lou16}.

In order to detect the shocks in star-forming regions, SiO rotational lines are often used \citep{jim10,dua14,cos18,svo19}. The SiO abundance can be increased by fast shocks up to several orders of magnitude higher than the abundance in dormant regions \citep{gus08}. This makes the SiO molecule be a good shock tracer. In previous works, SiO lines were able to probe high-velocity ($v_s\geq20$ km s$^{-1}$) and low-velocity shocks ($v_s<10$ km s$^{-1}$) in star-forming regions \citep{ngu13,dua14,cse16}. The high-velocity shocks are thought to be caused by protostellar outflows, and the low-velocity ones could originate from colliding gas flows or weak outflows from low- and intermediate-mass protostars \citep{lou16}. The SiO lines attributed to high-velocity shocks always have broad profiles while the profiles of the lines attributed to low-velocity shocks are narrow \citep{lou16,cse16}. Some previous studies showed that the SiO emission can contain both broad and narrow components \citep{ngu11,ngu13,san13}.

%Observations of SiO lines are useful to identify \textbf{protostellar} sources since current infrared surveys are incomplete to detected embedded protostars with low luminosities \citep{svo19,zhu20}.

We recently performed single-dish observations of SiO lines (J=1-0, 2-1 and 3-2) toward 100 SCCs \citep{zhu20}. The detection rate of the SiO lines is about 30$\%$. About 30$\%$ detected SiO lines have a broad line width with full width at zero power (FWZP) greater than 20 km s$^{-1}$, while about 40$\%$ of the SiO detections have narrow line widths with FWZP less than  10 km s$^{-1}$. However, the origins of the shocks indicated by narrow SiO line profiles are unclear and have several possible candidates, such as outflows from low-mass protostars and cloud-cloud collisions \citep{cos18}. In the sources with detections of broad SiO lines, whether widespread low-velocity shocks exist also needs to be checked \citep{ngu13}. Then whether the shocks in these sources mainly result from star formation activities or from large-scale movement of clouds is unknown.

The identification of the main origin or the percentages of different origins of shocks in SCCs is crucial. This knowledge can aid in the study of the evolutionary processes in the early stage of massive star formation. In some cases, early phases of massive star formation have been found to exhibit bipolar outflows leading to shocks \citep{tan16,fen16}. This suggests that outflows and accretion processes could have begun in the early stage of massive star formation \citep{tan16}. If the majority of shocks in SCCs are caused by outflows, then accretion and outflow processes should be common in the SCCs with SiO detections as the early stage of star formation. Alternatively, determining the percentage of shocks resulting from collisions between large-scale gas components in SCCs is also helpful to evaluate the significance of these events in the early-stage star formation environment.

%\textbf{The identification of the main origin or the percentages of different origins of shocks in SCCs is crucial. This knowledge can aid in obtaining a more accurate sample of starless clumps by distinguishing protostellar sources from other SCCs. If the majority of shocks in SCCs are caused by protostellar outflows, then only those SCCs lacking shocks should be considered as starless sources. Otherwise, inaccuracies may arise in the estimation of certain properties of starless clumps, as they are based on a flawed sample. For example, the lifetime of starless clumps may be overestimated, as it is calculated from the numbers of starless and protostellar clumps \citep{svo16}. Thus, the determination of the origin of shocks in SCCs is crucial for understanding the properties and formation of starless clumps.}

The molecular lines of HCO$^+$ 1-0, HC$_3$N 10-9, and the isotope H$^{13}$CO$^+$ 1-0 are often used to trace  dense gas in star-forming regions \citep{beu07,cos18,liu20}, which are also used for the studies of  extragalactic star-forming activities \citep{mic18,nay20}. HCO$^+$ and H$^{13}$CO$^+$ 1-0 lines can trace dense gas within clumps and  extended gas  surrounding  clumps as well, while (sub-)millimeter HC$_3$N lines  better trace dense cores in clumps \citep{nis17,liu20} than HCO$^+$ and H$^{13}$CO$^+$ 1-0 lines. With mapping observations for these lines, the velocity integrated flux distribution and velocity field of high-density gas can be revealed. The potential outflows could also be recognized by the high-velocity emission of the HCO$^+$ 1-0 line \citep{lop13,liu20}. Comparing the spatial distributions of these dense-gas and outflow tracers with that of the SiO lines, the origins of the shocks in the SCCs can be studied and revealed. Moreover, hydrogen recombination lines which are useful to detect ionized gas \citep{woo89} can be used to distinguish shocks caused by expansion of ionized gas from those due to the star formation activities in SCCs, especially since the influence of classical H II regions might be missed in classifying massive starless and star-forming clumps \citep{svo16}.
%And the interactions between shocks and the environments could be investigated from the comparison.

In this work, we present the mapped observations of the SiO 2-1, 3-2, HC$_3$N 10-9, HCO$^+$ 1-0, H$^{13}$CO$^+$ 1-0, and H41$\alpha$ lines toward 9 SCCs using IRAM 30m radio telescope. These SCCs have been observed in our previous work \citep{zhu20} and are associated with the SiO emissions. The positions and the morphologies of the dense clumps are displayed. And the origins of the shocks in these SCCs are studied and discussed. The organization is as follows. In Section \ref{sec:method}, the details of the observations and the selected sample are described. The results of the observations are shown in Section \ref{sec:result}. The discussions and analyses of the results are presented in Section \ref{sec:discussion}. In Section \ref{sec:conclusion}, the summary is written.

\section{Source sample and Observations} \label{sec:method}

\subsection{Sample}

The sample in the current work consists of 9 sources selected from the 100 SCCs in our previous single-dish observations of the SiO 1-0, 2-1 and 3-2 lines using KVN 21m telescopes \citep{zhu20}.  These SCCs were identified by using indicators of star formation including compact 70 $\mu m$ sources, mid-IR color-selected YSOs, H$_2$O and CH$_3$OH masers, and ultra-compact H II regions in \citet{svo16}. The selected sources were detected with SiO 2-1 line peak stronger than 60 mK in \cite{zhu20}, which guarantee SiO 2-1 detection in mapping observations. The SiO 2-1 line peak is only 33 mK in BGPS 3686, but it is one of the brightest sources in the SiO 3-2 line. So BGPS 3686 is included in the selected sample.  In addition, the SiO line widths in these sources are also considered. We prefer to choose sources with various SiO line widths in order that different kinds of shock origins may be discovered.

The properties of the sources are listed in Table \ref{table_source}. The distances of these sources range from 1.85 to 5.21 kpc. The V$_{lsr}$ measured from single-dish H$_2$CO 2$_{12}$-1$_{11}$ observations is considered as the systematic velocity of the sources \citep{zhu20}. The mass surface densities and the gas masses of the selected sources are provided by \citet{gin13}. The velocity-integrated SiO 2-1 line intensities obtained from previous observations are also listed.

\begin{table*} \tiny %\footnotesize
\centering
\caption{The starless clump candidates observed in the current work. The $\int T_{mb}dv$ represents the velocity-integrated main-beam temperature of the SiO 2-1 line spectra in the previous single-dish observations.}\label{table_source}
\begin{threeparttable}
\begin{tabular}{|cccccccc|}
\hline
\multirow{2}*{Source} & Ra & Dec & Distance & v$_{\textrm{lsr}}$\tnote{\textit{b}} & Mass & H$_2$ Column density & $\int T_{\textrm{mb}}dv$ \\
 & hh:mm:ss\tnote{\textit{a}} & dd:mm:ss\tnote{\textit{a}}  & kpc & km s$^{-1}$ & M$_\odot$ & cm$^{-2}$ & K km s$^{-1}$ \\
\hline
BGPS 3110 & 18:20:16.27 & -16:08:51.13 & 2.005 & $17.32\pm0.03$ & 370.8 & $1.72\times10^{22}$ & $0.935\pm0.070$ \\
BGPS 3114 & 18:20:31.50 & -16:08:37.80 & 1.848 & $22.87\pm0.02$ &  5371.3 & $5.89\times10^{22}$ & $0.548\pm0.053$ \\
BGPS 3118 & 18:20:16.17 & -16:05:50.72 & 2.001 & $16.88\pm0.02$ &  365.4 & $9.36\times10^{21}$ & $0.659\pm0.078$ \\
BGPS 3686\tnote{\textit{c}} & 18:34:14.58 & -09:18:35.84 & 2.939 & $78.14\pm0.29$ &  404.7 & $6.01\times10^{21}$ & $0.454\pm0.079$ \\
BGPS 4029\tnote{\textit{c}} & 18:35:54.40 & -07:59:44.60 & 3.539 & $81.43\pm0.30$ &  473.4 & $9.65\times10^{21}$ & $0.936\pm0.081$ \\
BGPS 4472 & 18:41:17.32 & -05:09:56.83 & 3.216 & $47.71\pm0.08$ & 130.3 & $5.76\times10^{21}$ & $1.071\pm0.108$ \\
BGPS 5064 & 18:45:48.44 & -02:44:31.65 & 5.210 & $100.80\pm0.10$ & 1958.2 & $4.77\times10^{21}$ & $0.541\pm0.049$ \\
BGPS 5114\tnote{\textit{c}} & 18:50:23.54 & -03:01:31.58 & 3.681 & $64.51\pm0.35$ & 821.9 & $7.02\times10^{21}$ & $0.598\pm0.053$ \\
BGPS 5243 & 18:47:54.70 & -02:11:10.72 & 5.210 & $96.06\pm0.13$ &  421.9 & $4.75\times10^{21}$ & $0.270\pm0.040$ \\
\hline
\end{tabular}
\begin{tablenotes}
\item \textbf{Notes.} The locations and distances of the sources are given in \citet{cal18}. The masses, H$_2$ column densities, and $\int T_{\textrm{mb}}dv$ are provided in \citet{zhu20}.
\item{\tnote{\textit{a}}} The coordinates are epoch J2000.0.
\item{\tnote{\textit{b}}} v$_{\textrm{lsr}}$ are H$_2$CO line central velocities for corresponding sources given in \citet{zhu20}.
\item{\tnote{\textit{c}}} BGPS 3686, 4029, and 5114 were also observed in \citet{svo19}. Their names in that previous work are G22695, G24051, and G30120, repectively.
\end{tablenotes}
\end{threeparttable}
\end{table*}

%\begin{tabular}{||||}
%  \hline
%  % after \\: \hline or \cline{col1-col2} \cline{col3-col4} ...
%   &  &  \\
%   &  &  \\
%   &  &  \\
%  \hline
%\end{tabular}

%\begin{figure}
%  \centering
%  % Requires \usepackage{graphicx}
%  \includegraphics[scale=0.5]{numbersources.eps}
%  \caption{ The distance distribution of the sample. The vertical dashed line represents the average distance.}\label{fig:distance}
%\end{figure}

\subsection{Observations and Data reduction}

We performed mapping observations toward 9 SCCs using IRAM 30-m telescope in April 2020. We used 3 mm (E0) and 2 mm (E1) bands of the Eight Mixer Receivers (EMIR) simultaneously and the FTS backend with an 195 kHz channel width and an 8 GHz band width for the observations. The frequency ranges are 84.5-92.3 GHz and 128.0-135.8 GHz for the 3 mm and 2 mm bands, respectively. The spectrometer resolution is typically about 0.67 km s$^{-1}$ in the 3 mm band, and is about 0.45 km s$^{-1}$ in the 2 mm band. The OTF PSW observing mode was used to cover an area of $2'\times2'$ for each source. This angular size corresponds to an area of $1.16-9.19$ pc$^2$ for different sources. The beam sizes were $\sim28''$ and $\sim20''$ for the 3 mm and 2 mm bands, respectively. The locations of SCCs given in \citet{svo16} correspond to the centers of observational fields. The reference positions are 1 degree offset from each source in RA. The dump time is 0.5 second, while the sampling interval is $6''$. The system temperatures are typically 100 K and 120 K for the 3 mm and the 2 mm bands. The precipitable water vapour (pwv) with the average conditions is about 4.0 mm. The  accuracy  of absolute flux calibration is better than 20$\%$. The line frequencies, the beam sizes and the main-beam efficiencies corresponding to the observed molecular and atomic lines are listed in Table \ref{table_line}.

The observations were reduced with CLASS reduction package (https://www.iram.fr/IRAMFR/GILDAS). Linear baselines were removed from the spectra. The rms noise levels in T$_{mb}$ are typically 35 mK and 50 mK per pixel in the 3 mm and 2 mm bands, respectively. Because the integration times and weather condition for the observations toward the sources are not significantly different, the rms noise levels of the main-beam brightness temperature corresponding to the molecular lines are consistent for different sources.

\begin{table} \tiny %\footnotesize
\centering
\caption{The frequencies, the beam sizes and the beam efficiencies corresponding to the observed molecular transitions for IRAM 30m telescope.}\label{table_line}
\begin{threeparttable}
\begin{tabular}{|ccccc|}
\hline
Transition & Frequency [MHz] & Beam size [$''$] & B$_{\textrm{eff}}$ & S$_\nu$/T$^*_A$ [Jy/K] \\
\hline
SiO 2-1 & 86847.00 & 28 & 0.81 & 5.8 \\
SiO 3-2 & 130268.71 & 19 & 0.75 & 6.1 \\
HCO$^+$ 1-0 & 89188.53 & 28 & 0.81 & 5.8 \\
H$^{13}$CO$^+$ 1-0 & 86754.29 & 28 & 0.81 & 5.8 \\
HC$_3$N 10-9 & 90979.02 & 27 & 0.81 & 5.8 \\
H41$\alpha$ & 92034.43 & 27 & 0.81 & 5.8 \\
%He41$\alpha$ & 92071.94 & 27$''$ & 0.81 & 5.8 \\
\hline
\end{tabular}
\begin{tablenotes}
\item{\textbf{Notes.}} The beam size, B$_{\textrm{eff}}$, and S$_\nu$/T$^*_A$ are derived from https://publicwiki.iram.es/Iram30mEfficiencies.
\end{tablenotes}
\end{threeparttable}
\end{table}

%\begin{table} \tiny %\footnotesize
%\centering
%\caption{The frequencies, the beam sizes and the beam efficiencies corresponding to the observed molecular transitions for IRAM 30-m telescope.}\label{table_line}
%\begin{tabular}{|ccccc|}
%\hline
%Transition & Frequency [MHz] & Beam size & B$_{eff}$ & S$_\nu$/T$^*_A$ [Jy/K] \\
%\hline
%SiO 2-1 & 86847.00 & 28$''$ & 0.82 & 6.0 \\
%SiO 3-2 & 130268.71 & 18$''$ & 0.64 & 7.5 \\
%HCO$^+$ 1-0 & 89188.53 & 27$''$ & 0.82 & 6.0 \\
%H$^{13}$CO$^+$ 1-0 & 86754.29 & 28$''$ & 0.82 & 6.0 \\
%HC$_3$N 10-9 & 90979.02 & 27$''$ & 0.82 & 6.0 \\
%H41$\alpha$ & 92034.43 & 26$''$ & 0.82 & 6.0 \\
%He41$\alpha$ & 92071.94 & 26$''$ & 0.82 & 6.0 \\
%\hline
%\end{tabular}
%\end{table}

\section{results} \label{sec:result}

In this work, we present  SiO 2-1, 3-2, HCO$^+$ 1-0, H$^{13}$CO$^+$ 1-0, HC$_3$N 10-9,  and H41$\alpha$ lines toward 9 SCCs. The SiO lines are used to study the shocked gas, while  HCO$^+$ 1-0, H$^{13}$CO$^+$ 1-0 and HC$_3$N 10-9  lines are used as dense gas tracers. The H41$\alpha$ line is used to analyze the ionized gas in H II regions.

The properties of the observed molecular and hydrogen recombination lines toward the SCCs are written in Table \ref{table_property1}. They are measured within the regions that cover the compact parts of clumps and shocked gas, which are indicated by the black dashed circles shown in Figure \ref{fig:BGPS3686_HCO+_HC3N_blrd}-\ref{fig:BGPS3118_HCO+_H41a}. A single Gaussian distribution is used to estimate the properties from line profiles. The derived values in Table \ref{table_property1} correspond to the main velocity component in cases where multiple velocity components were identified. Moreover, the uncertainties of absolute flux calibration are not included in the uncertainties listed in Table \ref{table_property1}. %It is necessary to note that the regions mentioned above are not the regions indicated by the dashed circles in Figure \ref{fig:BGPS3686_HCO+_HC3N_blrd}-\ref{fig:BGPS3118_HCO+_H41a}.

\begin{figure}
  \centering
  \includegraphics[scale=0.45]{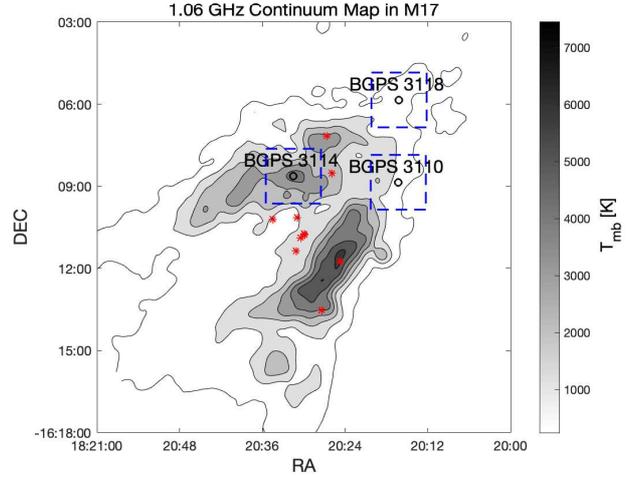}
  \caption{The 1.06 GHz continuum map toward M17 H II regions \citep{beu16}. The coordinates of RA and DEC are epoch J2000.0. The coverage areas of our observations are indicated by the blue dashed rectangles. The positions of the observed SCCs are shown by the black circles. The high-mass stars earlier than O9V are indicated by red asterisks \citep{hof08}.}\label{fig:ctnmap}
\end{figure}

The molecular lines are detected in all of the sources. The H41$\alpha$ line is only detected in three SCCs (BGPS 3110, 3114 and 3118). According to the 1.06 GHz continuum map toward M17 H II region provided by the THOR project \citep{beu16} and presented in Figure \ref{fig:ctnmap}, the SCCs, BGPS 3110, 3114, and 3118, are located in or near the H II region.
%In addition, the He41$\alpha$ line is clearly detected in BGPS 3110 and 3114.

%The gas kinetic temperatures calculated from the NH$_3$ lines provided by \citet{svo16} are also given in Tables \ref{table_property1}. The kinetic temperatures for BGPS 3110 and 3118 with the H41$\alpha$ detections are higher than the temperatures for the other sources. The simulations of \citet{hos06} and \citet{zhu15b} show that the FUV radiation from ionizing stars can heat the neutral gas near H II regions. It is also displayed in \citet{zha20} that the dust temperatures of the starless clumps associated with H II regions are higher.

\begin{table*} \tiny %\footnotesize
\centering
\caption{The results of Gaussian fitting of the molecular lines toward the SCCs.}\label{table_property1}
\begin{threeparttable}
\begin{tabular}{|c|c|ccc|ccc|ccc|}
\hline
   &  & \multicolumn{3}{c}{SiO 2-1} & \multicolumn{3}{c}{SiO 3-2} & \multicolumn{3}{c}{H$^{13}$CO$^+$ 1-0} \\
\hline
\multirow{2}*{Source} & T$_k$\tnote{\textit{a}} & V$_{lsr}$ & FWHM & $\int T_{\textrm{mb}}dv$ & V$_{lsr}$ & FWHM & $\int T_{\textrm{mb}}dv$ & V$_{lsr}$ & FWHM & $\int T_{\textrm{mb}}dv$ \\
 & [K] & [km s$^{-1}$] & [km s$^{-1}$] & [K km s$^{-1}$] & [km s$^{-1}$] & [km s$^{-1}$] & [K km s$^{-1}$] & [km s$^{-1}$] & [km s$^{-1}$] & [K km s$^{-1}$] \\
\hline
BGPS 3110 & 25.17 & $17.61\pm0.09$ & $3.50\pm0.21$ & $0.44\pm0.02$ & 18.03$\pm$0.19 & $4.29\pm0.37$ & $0.39\pm0.04$ & $17.46\pm0.05$ & $3.10\pm0.11$ & $1.25\pm0.04$ \\
BGPS 3114 & - & $22.47\pm0.16$ & $3.91\pm0.39$ & $0.42\pm0.04$ & $22.98\pm0.17$ & $4.42\pm0.41$ & $0.39\pm0.03$ & $22.43\pm0.16$ & $3.77\pm0.39$ & $0.33\pm0.03$ \\
BGPS 3118 & 23.68 & $17.13\pm0.37$ & $3.42\pm0.82$ & $0.14\pm0.03$ & ... & ... & ...\tnote{\textit{b}} & $17.28\pm0.04$ & $2.55\pm0.10$ & $0.92\pm0.03$ \\
BGPS 3686 & 14.70 & $77.18\pm1.83$ & $9.20\pm4.32$ & $0.14\pm0.06$ & $78.43\pm1.71$ & $13.76\pm4.04$ & $0.31\pm0.08$ & $77.30\pm0.05$ & $1.78\pm0.11$ & $0.51\pm0.03$ \\
BGPS 4029 & 11.87 & $83.48\pm0.88$ & $15.90\pm2.10$ & $0.56\pm0.06$ & $83.58\pm1.65$ & $14.70\pm3.88$ & $0.29\pm0.07$ & $81.36\pm0.03$ & $2.02\pm0.06$ & $0.88\pm0.02$ \\
BGPS 4472 & 14.29 & $47.15\pm0.73$ & $17.19\pm1.73$ & $0.75\pm0.07$ & $47.29\pm0.95$ & $15.63\pm2.26$ & $0.56\pm0.07$ & $46.91\pm0.04$ & $1.43\pm0.09$ & $0.39\pm0.02$ \\
BGPS 5064 & 15.80 & $100.79\pm0.40$ & $7.65\pm0.95$ & $0.33\pm0.04$ & $100.00\pm0.69$ & $5.62\pm1.64$ & $0.22\pm0.06$ & $100.76\pm0.05$ & $1.56\pm0.11$ & $0.44\pm0.03$ \\
BGPS 5114 & 14.12 & $65.67\pm0.13$ & $2.69\pm0.30$ & $0.20\pm0.02$ & $64.95\pm0.37$ & $4.81\pm0.87$ & $0.19\pm0.03$ & $65.62\pm0.06$ & $2.96\pm0.15$ & $0.63\pm0.03$ \\
BGPS 5243 & 12.69 & $94.49\pm0.45$ & $7.05\pm1.05$ & $0.23\pm0.03$ & $96.15\pm0.61$ & $5.35\pm1.43$ & $0.15\pm0.03$ & $95.84\pm0.03$ & $1.61\pm0.07$ & $0.46\pm0.02$ \\
\hline
  &  & \multicolumn{3}{c}{HCO$^+$ 1-0} & \multicolumn{3}{c}{HC$_3$N 10-9} & \multicolumn{3}{c}{H41$\alpha$} \\
\hline
\multirow{2}*{Source} &   & V$_{lsr}$ & FWHM & $\int T_{\textrm{mb}}dv$ & V$_{lsr}$ & FWHM & $\int T_{\textrm{mb}}dv$ & V$_{lsr}$ & FWHM & $\int T_{\textrm{mb}}dv$  \\
 &   & [km s$^{-1}$] & [km s$^{-1}$] & [K km s$^{-1}$] & [km s$^{-1}$] & [km s$^{-1}$] & [K km s$^{-1}$] & [km s$^{-1}$] & [km s$^{-1}$] & [K km s$^{-1}$]  \\
 \hline
 BGPS 3110 &   & $17.28\pm0.05$ & $3.97\pm0.11$ & $22.22\pm0.53$ & $17.33\pm0.04$ & $2.90\pm0.09$ & $2.92\pm0.08$  & $18..69\pm0.34$ & $26..26\pm0.81$ & $2.47\pm0.07$ \\
 BGPS 3114 &   & $22.64\pm0.06$ & $4.66\pm0.15$ & $12.66\pm0.36$ & ... & ... & ...\tnote{\textit{c}}  & $17.40\pm0.19$ & $30.04\pm0.44$ & $10.30\pm0.13$ \\
 BGPS 3118 &   & $16.84\pm0.04$ & $2.90\pm0.09$ & $11.95\pm0.33$ & $17.27\pm0.03$ & $2.16\pm0.06$ & $1.69\pm0.04$  & $31.99\pm0.69$ & $32.54\pm1.62$ & $1.77\pm0.08$ \\
 BGPS 3686 &   & $78.33\pm0.11$ & $7.53\pm0.26$ & $3.97\pm0.12$ & $78.36\pm0.68$ & $6.41\pm1.61$ & $0.27\pm0.06$ \\
 BGPS 4029 &   & $81.54\pm0.10$ & $4.59\pm0.23$ & $3.12\pm0.14$ & $81.21\pm0.04$ & $1.69\pm0.11$ & $0.35\pm0.02$ \\
 BGPS 4472 &   & $46.68\pm0.13$ & $6.01\pm0.30$ & $2.50\pm0.11$ & $47.06\pm0.09$ & $2.80\pm0.22$ & $0.47\pm0.03$  \\
 BGPS 5064 &  & $100.65\pm0.12$ & $3.64\pm0.27$ & $5.94\pm0.39$ & $100.93\pm0.14$ & $2.69\pm0.32$ & $0.20\pm0.02$ \\
 BGPS 5114 &   & $66.38\pm0.31$ & $6.59\pm0.72$ & $4.84\pm0.46$ & $65.86\pm0.10$ & $3.07\pm0.24$ & $0.28\pm0.02$ \\
 BGPS 5243 &   & $96.40\pm0.15$ & $3.80\pm0.34$ & $2.31\pm0.18$ & $96.18\pm0.08$ & $1.54\pm0.19$ & $0.10\pm0.01$ \\
\hline
\end{tabular}
\begin{tablenotes}
\item{\tnote{\textit{a}}} The kinetic temperatures in the SCCs are given in \citet{svo16}.
\item{\tnote{\textit{b}}} The signial-to-noise ratio of SiO 3-2 line is lower than 3 in the measured region, but this line is clearly detected in the southwest of BGPS 3118 as shown in Figure \ref{fig:SiO_map1}.
\item{\tnote{\textit{c}}} The HC$_3$N 10-9 line can be detected in BGPS 3114 if the measured region is much shrunk.
\end{tablenotes}
\end{threeparttable}
\end{table*}

\subsection{The sources without H41$\alpha$ detections}

The optically thin  HCO$^+$ 1-0 emission at the line wings, comparing with simultaneously obtained H$^{13}$CO$^+$ 1-0,  is used to determine protostellar outflows. Most of  HCO$^+$ 1-0 line emission is optically thick in star-forming regions while the H$^{13}$CO$^+$ 1-0 line is usually optically thin \citep{cal18,liu20}. When the HCO$^+$ 1-0 emission is optically thin, the intensity ratio of HCO$^+$ to H$^{13}$CO$^+$ lines tends to be high due to the typically large $^{12}$C$/^{13}$C elemental ratio, which is $\sim68$ in interstellar medium (ISM) \citep{col20}. In such mapping observations, a gas component associated with optically thin HCO$^+$ 1-0 emission is usually devoid of H$^{13}$CO$^+$ 1-0 detection. Compared with the H$^{13}$CO$^+$ line, the optically thick components and the contamination from nearby clumps with different velocities can be distinguished from the real high-velocity components of the HCO$^+$ line.   On the other hand, if H$^{13}$CO$^+$ 1-0 is detected as well, even though the velocity of  HCO$^+$ 1-0 emission can be far away from the main line center, such emission should be from another clump, which is the case for BGPS 3686, 5064, and 5243 with two main clumps  closely located in each source.

%The optically thin  HCO$^+$ 1-0 emission at the line wings, comparing with simultaneously obtained H$^{13}$CO$^+$ 1-0,  is used to determine protostellar outflows. Most of  HCO$^+$ 1-0 line emission is optically thick in star-forming regions while the H$^{13}$CO$^+$ 1-0 line is usually optically thin \citep{cal18,liu20}. Optically thin  HCO$^+$ 1-0 component should have large HCO$^+$/H$^{13}$CO$^+$ 1-0 line ratio \citep{col20}, normally without H$^{13}$CO$^+$ 1-0 detection in such mapping observation. Compared with the H$^{13}$CO$^+$ line, the optically thick components and the contamination from nearby clumps with different velocities can be distinguished from the real high-velocity components of the HCO$^+$ line.   On the other hand, if H$^{13}$CO$^+$ 1-0 is detected as well, even though the velocity of  HCO$^+$ 1-0 emission can be far away from the main line center, such emission should be from another clump, which is the case for BGPS 3686, 5064, and 5243 with two main clumps  closely located in each source.

%The optically thin  HCO$^+$ 1-0 emission at the line wings, comparing with simultaneously obtained H$^{13}$CO$^+$ 1-0,  is used to determine protostellar outflows.

Although  high-velocity SiO emission is also useful to search for outflows \citep{svo19}, it is difficult to show spatial distributions of high-velocity gas clearly  in the current observations, due to its relatively weaker emission. Thus, the SiO emission is only used to present the distribution of shocked gas.  Since there are several sources with more than one clump identified by optically thin tracer H$^{13}$CO$^+$ 1-0 line, HC$_3$N 10-9 line, with higher critical density and higher upper level energy than that of H$^{13}$CO$^+$ 1-0,  is used to trace the compact regions in clumps where the potential dense cores associated with outflows may be embedded \citep{nis17,liu20}. The upper energy levels ($E_u$) are 4.2 K and 24.0 K for H$^{13}$CO$^+$ 1-0 and HC$_3$N 10 -9 transitions, respectively. At a kinetic temperature of 10 K, the corresponding critical densities of the two transitions are $6.2\times10^4$ cm$^{-3}$ and $1.6\times10^5$ cm$^{-3}$, separately \citep{shi15}. We describe the detailed information for individual sources.

%\textbf{In this section, the HC$_3$N 10-9 line is used to trace the compact regions in clumps where the potential dense cores may be embedded \citep{nis17,liu20}. For the sources where multiple dense clumps are found, the distribution of high-density gas is shown by the flux density of H$^{13}$CO$^+$ 1-0 line to present the locations of these clumps. The HCO$^+$ and H$^{13}$CO$^+$ 1-0 lines are both dense gas tracers. However, the HCO$^+$ 1-0 line may be optically thick while the H$^{13}$CO$^+$ 1-0 line is usually optically thin \citep{cal18,liu20}. Then the H$^{13}$CO$^+$ 1-0 line is better to show the distribution of high-density gas by using the flux density. For indicating the evidence of protostellar outflows, the HCO$^+$ 1-0 line is used since this line is also useful to detect outflows \citep{liu20}. The HCO$^+$ emission from the shocked gas should be optically thin and high-velocity. By comparing the HCO$^+$ and H$^{13}$CO$^+$ spectra, the optically thin and high-velocity components of HCO$^+$ emission are recognized as the potential indicator of outflows. Additionally, although the high-velocity SiO emission is also useful to search for outflows, its relatively weaker intensity in the current observations is difficult to show spatial distributions of high-velocity gas clearly. So in this work, the SiO emission is only used to present the distribution of shocked gas.}% Moreover, there is no hydrogen recombination lines detected in BGPS 3686, 4029, 4472, 5064, 5114, and 5243.}

\subsubsection{BGPS 3686}

\begin{figure}
  \centering
  \includegraphics[scale=0.40]{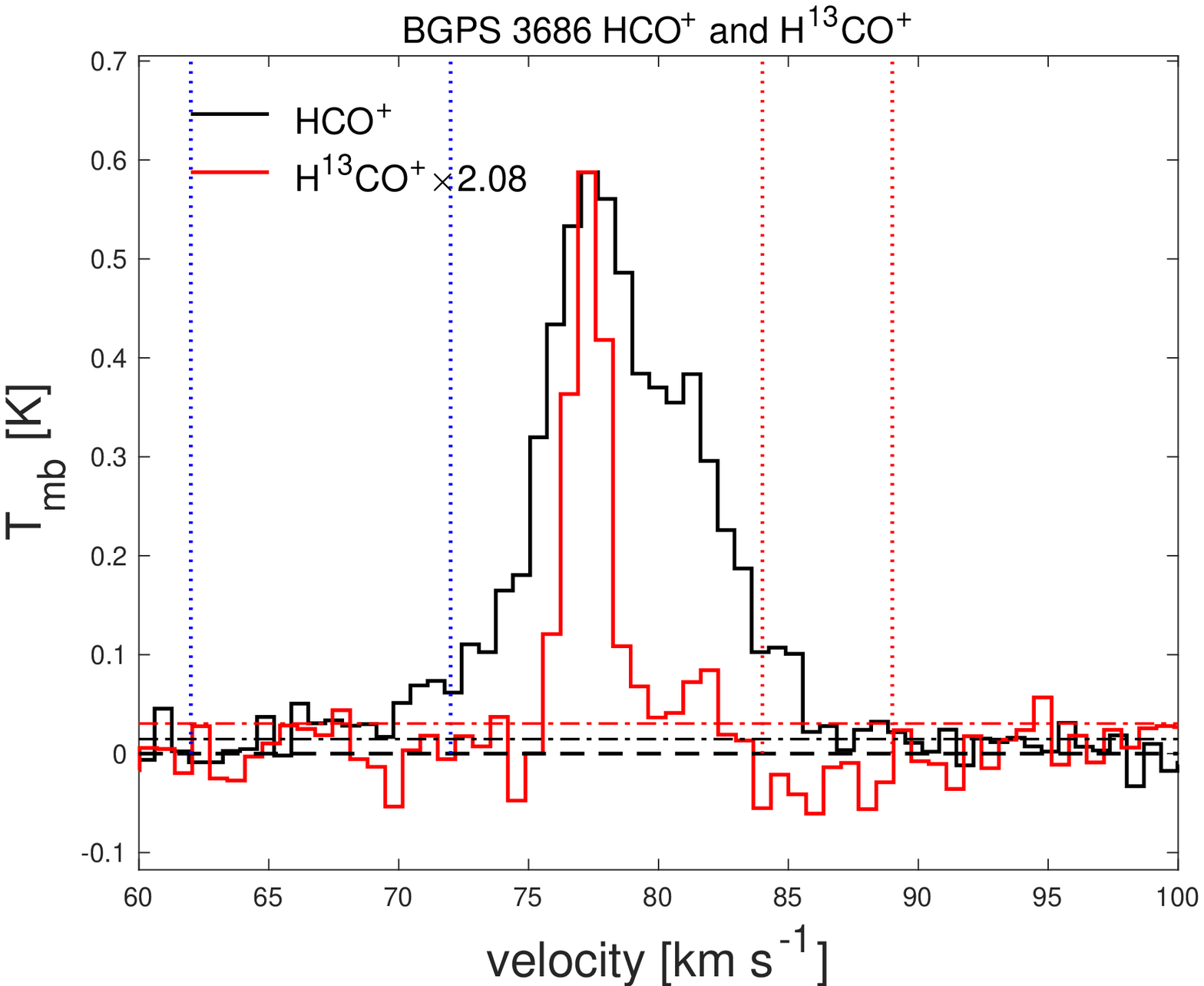}
  \includegraphics[scale=0.45]{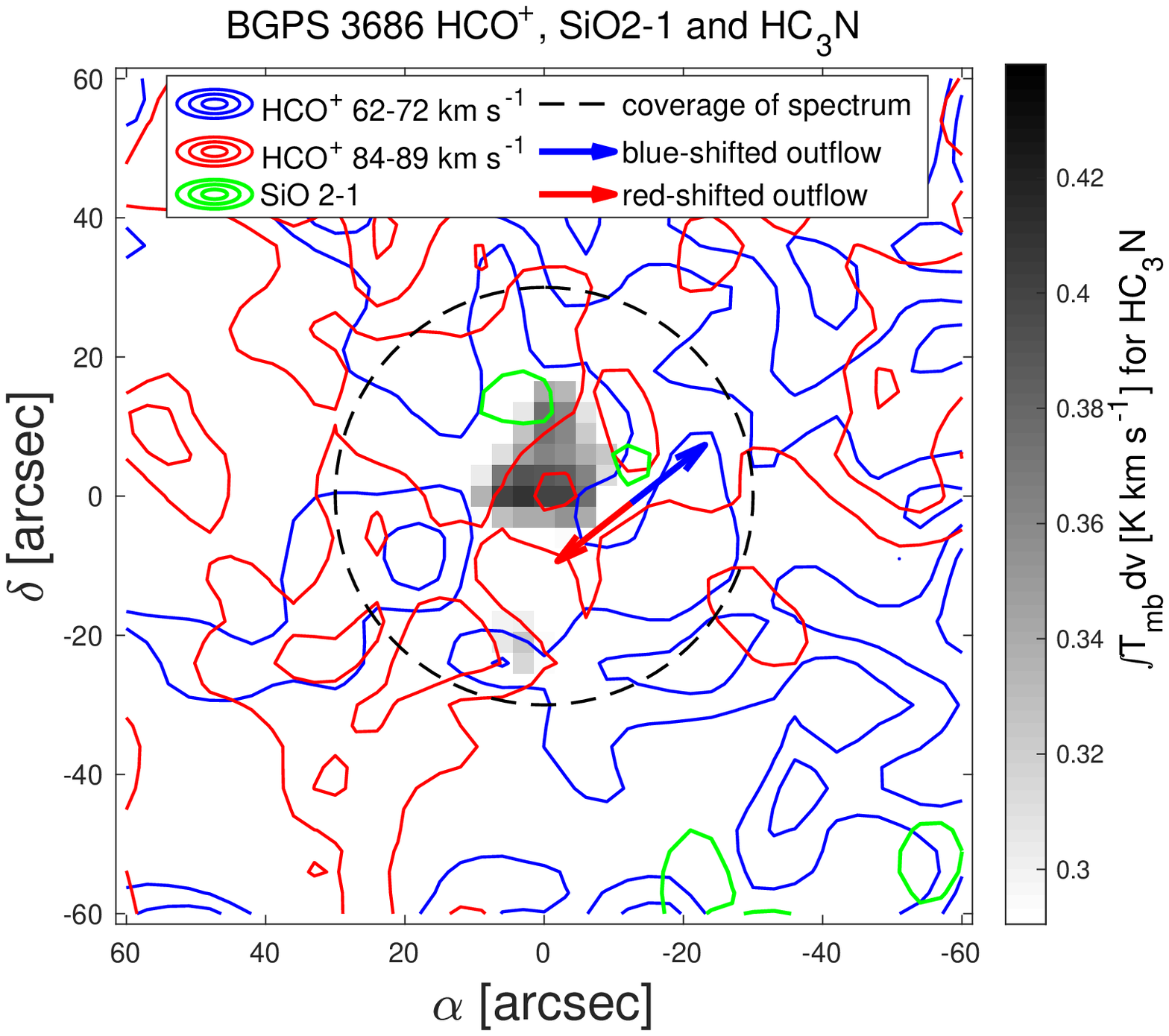}
  \caption{The HCO$^+$ and H$^{13}$CO$^+$ 1-0 spectra toward BGPS 3686 are plotted in the top panel. The blue and red vertical dotted lines indicate the velocity ranges of blue- and red-shifted high-velocity HCO$^+$ components, respectively. The black and red dash-dotted lines indicate the rms noise levels at the velocity resolution in the HCO$^+$ and H$^{13}$CO$^+$ 1-0 spectra. In the bottom panel, the distributions of the 62-72 and 84-89 km s$^{-1}$ high-velocity components of the HCO$^+$ 1-0 line are indicated by the blue and red contours, respectively. The HC$_3$N 10-9 line is shown by the gray-scale image. The contour levels start at $3\sigma$ in steps of $2\sigma$ of the HCO$^+$ 1-0 velocity-integrated intensities. The SiO 2-1 line is presented by the green contours which start at $3\sigma$ in steps of $1\sigma$. The red and blue arrows and their connection point indicate the directions of bipolar outflows and the dense core detected in the CO 2-1 line and the 230 GHz continuum \citep{svo19}. The black dashed circle is the coverage of the spectra shown in the top panel.}\label{fig:BGPS3686_HCO+_HC3N_blrd}
\end{figure}

\begin{figure}
  \centering
  \includegraphics[scale=0.45]{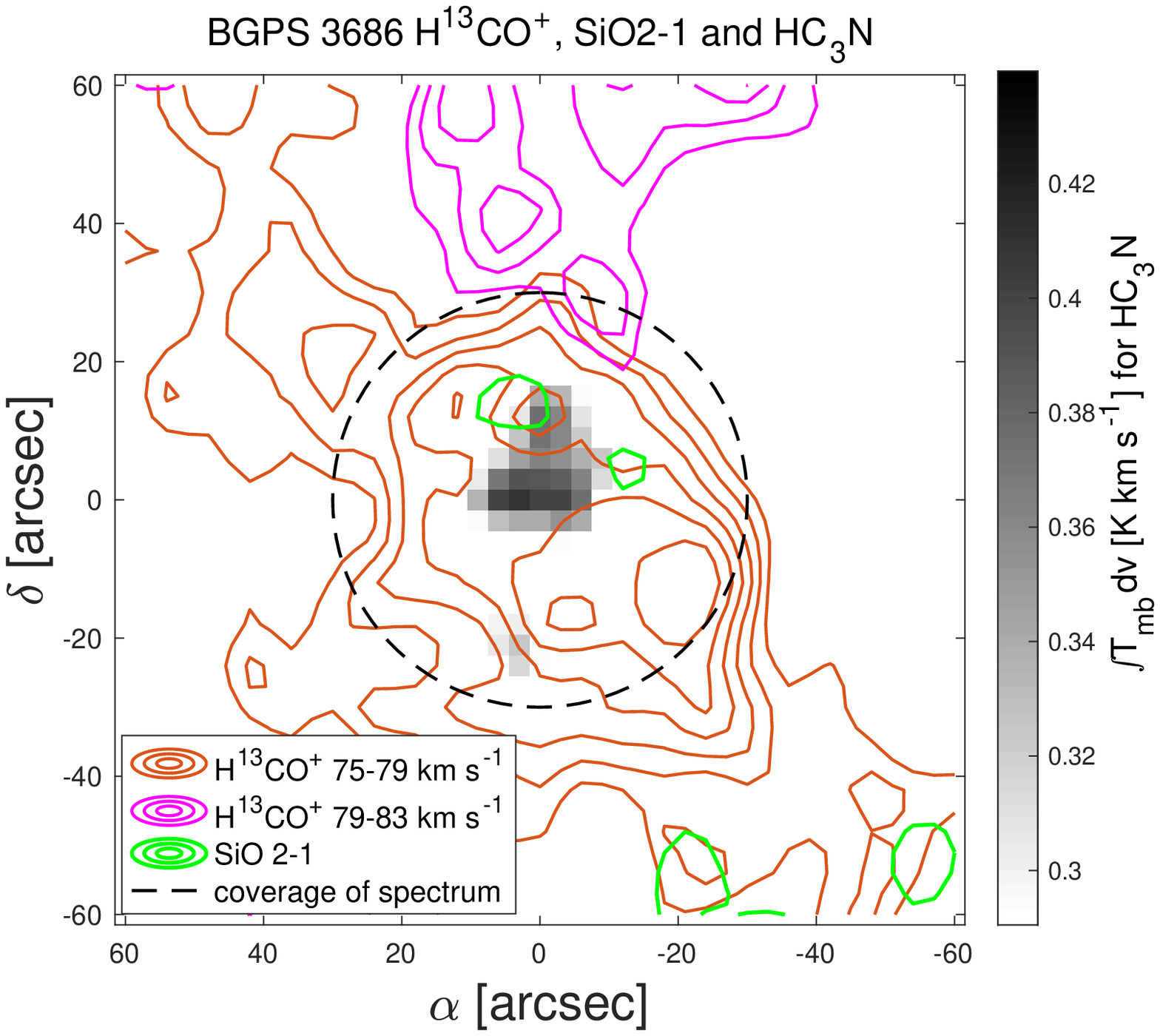}
  \caption{The distributions of the 75-79 and 79-83 km s$^{-1}$ components of the H$^{13}$CO$^+$ line 1-0 is indicated by the brown and magenta contours, respectively. The SiO 2-1 line is presented by the green contours, and the HC$_3$N 10-9 line is shown by the gray-scale image. The contour levels start at $5\sigma$ in steps of $1\sigma$ of the velocity-integrated intensities for the 75-79 km s$^{-1}$ H$^{13}$CO$^+$ component. The contour levels for the 79-83 km s$^{-1}$ H$^{13}$CO$^+$ component start at $4\sigma$ in steps of $1\sigma$. The contour levels for the SiO 2-1 emission start at $3\sigma$ in steps of $1\sigma$. The black dashed circle is the coverage of the spectra shown in the top panel of Figure \ref{fig:BGPS3686_HCO+_HC3N_blrd}.}\label{fig:BGPS3686_HCO+_HC3N}
\end{figure}

%There is no hydrogen recombination lines detected in the other sources.
The HCO$^+$ and H$^{13}$CO$^+$ 1-0 line profiles toward BGPS 3686 are presented in the top panel of Figure \ref{fig:BGPS3686_HCO+_HC3N_blrd}. The region from which the spectra are extracted is indicated by the black dashed circle in the bottom panel. Two velocity components with centers at about 77 and 82 km s$^{-1}$ are shown in both the HCO$^+$ and H$^{13}$CO$^+$ 1-0 spectra. In Figure \ref{fig:BGPS3686_HCO+_HC3N}, the distributions of the dense gas components with corresponding central velocities are shown by the H$^{13}$CO$^+$ 1-0 emission. The $\sim77$ km s$^{-1}$ component spreads widely from the northeast to the southwest, and the brightest part of this component is at the center of observational field. The $\sim82$ km s$^{-1}$ component is concentrated in the north. %The interaction between these two clumps of dense gas is not clear.

The distribution of the HC$_3$N 10-9 emission is also presented in Figure \ref{fig:BGPS3686_HCO+_HC3N}. The HC$_3$N 10-9 emission indicates the compact part of the clump in which a protostellar core could be embedded.  In BGPS 3686, the densest part of gas is distributed at the center of the observational field corresponding to the brightest part of the $\sim77$ km s$^{-1}$ H$^{13}$CO$^+$ component. According to the locations of the HC$_3$N 10-9 and H$^{13}$CO$^+$ 1-0 emissions, the compact gas seems to be only related with the $\sim77$ km s$^{-1}$ clump. The SiO 2-1 emission shows the distribution of the shocked gas. However, the distribution of the SiO 2-1 line is relatively unclear. Since the SiO 2-1 line was detected in previous single-dish observations \citep{zhu20}, the SiO 2-1 emission may be widespread so that the intensity is close to the rms in the current mapping observation. Although the SiO intensity is close to the rms, a part of shocked gas indicated by the SiO 2-1 emission looks distributed near the HC$_3$N 10-9 emission.

The HCO$^+$ 1-0 spectrum shows both red- and blue-shift line wings. The distributions of the fluxes located at these line wings are displayed in the bottom panel of Figure \ref{fig:BGPS3686_HCO+_HC3N_blrd}. Both the high-velocity components in red- and blue-shift line wings are widespread. The feature of outflow cannot be found from the distribution of high-velocity HCO$^+$ components based on our observations in BGPS 3686. %If there is protostellar activity, the high-velocity component caused by outflows might be concealed by the violent gas. The relation between these widespread high-velocity components and the compact region indicated by the HC$_3$N 10-9 line is not clear. And the relation between the high-velocity components shown in the HCO$^+$ 1-0 line and the shocked gas can also not be concluded.

%The HCO$^+$ 1-0 spectrum shows both red- and blue-shift line wings. The distributions of the flux located at these line wings are displayed in Figure \ref{fig:BGPS3686_HCO+_HC3N_blrd}. Both the high-velocity components in red- and blue-shift line wings are widespread. The movement of the molecular gas in BGPS 3686 seems to be violent. If there is protostellar activity, the high-velocity component caused by outflows might be concealed by the violent gas. The relation between these widespread high-velocity components and the compact region indicated by the HC$_3$N 10-9 line is not clear. And the relation between the high-velocity components shown in the HCO$^+$ 1-0 line and the shocked gas can also not be concluded.

\subsubsection{BGPS 4029}

\begin{figure}
  \centering
  \includegraphics[scale=0.40]{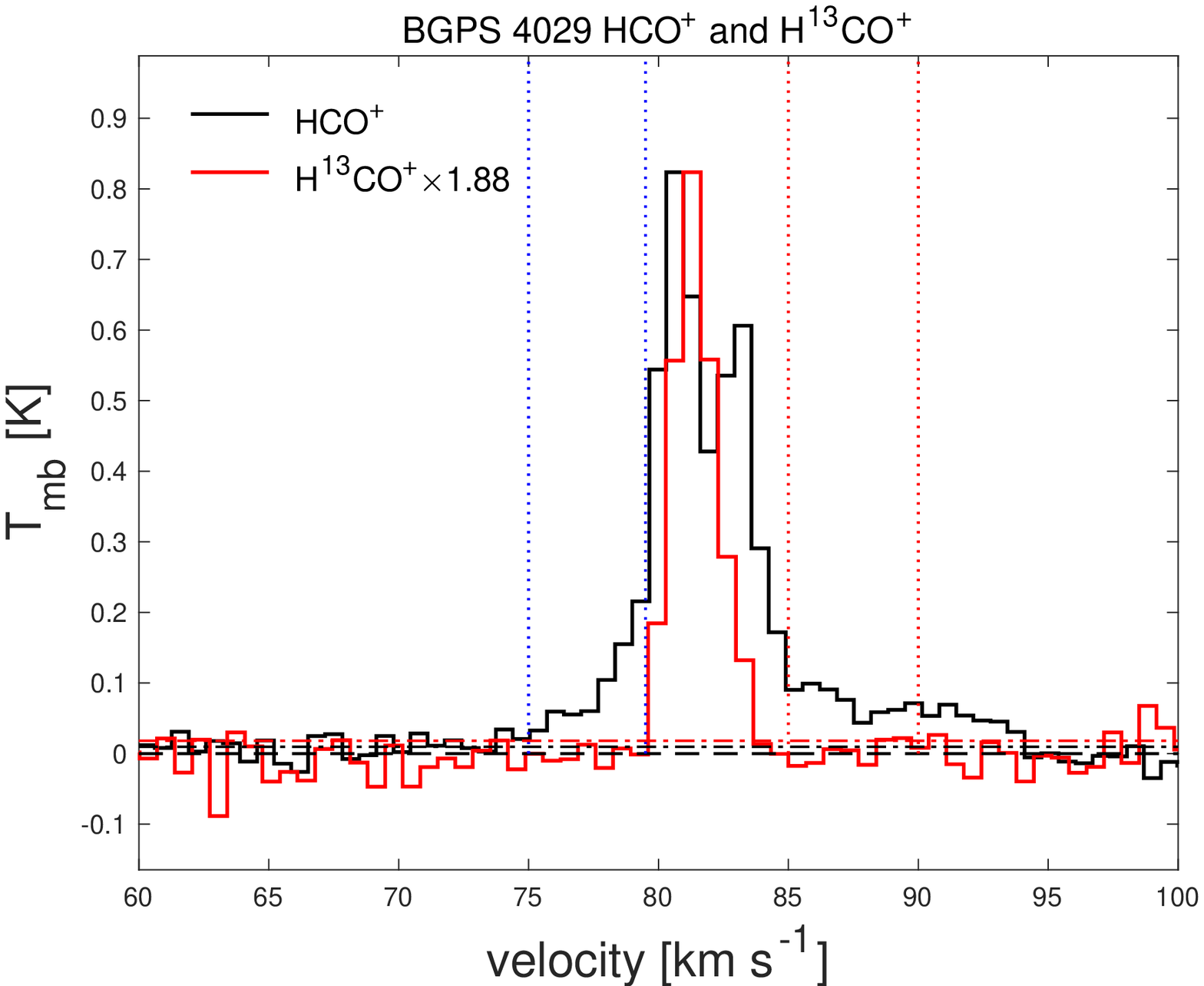}
  \includegraphics[scale=0.45]{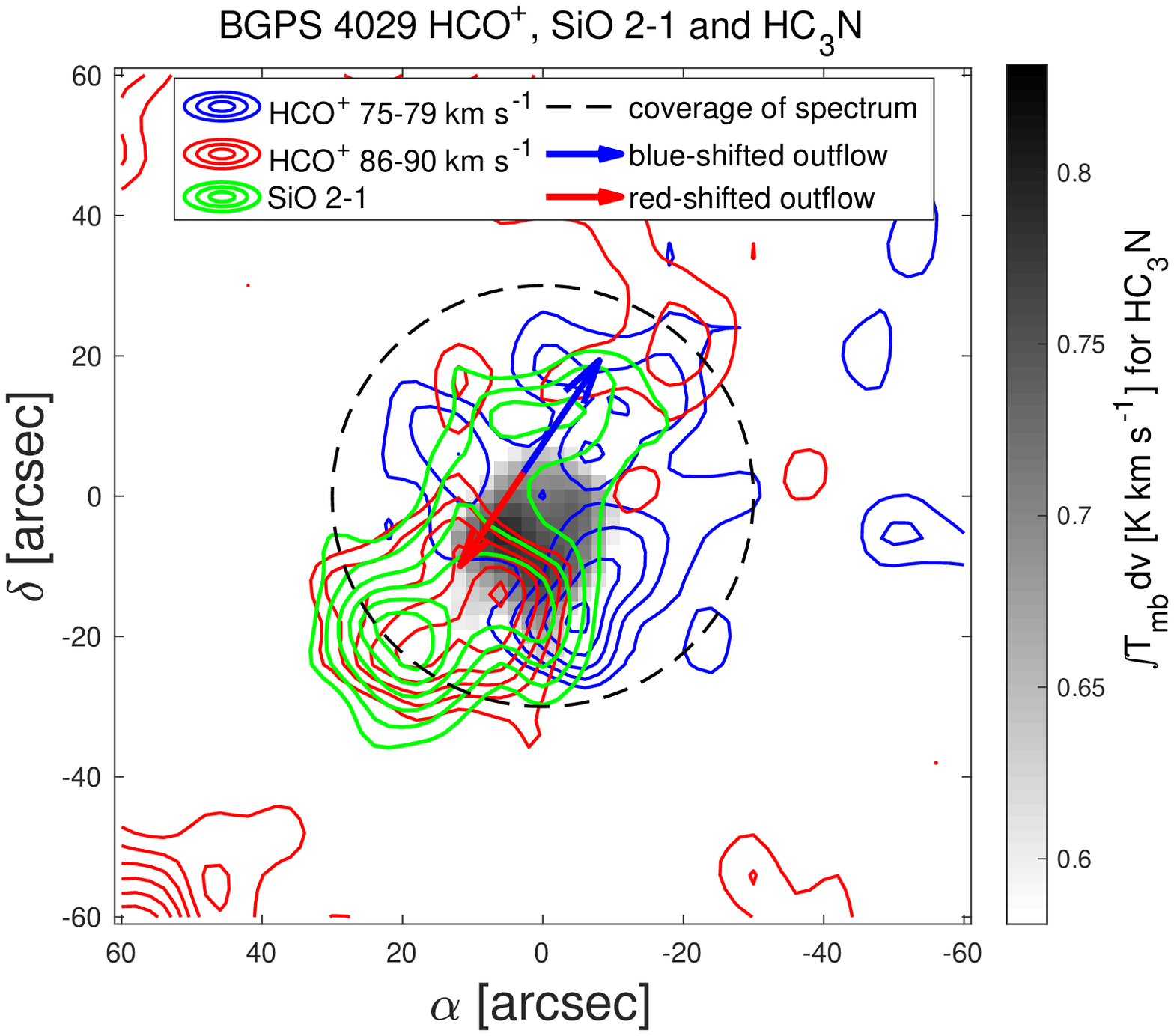}
  \caption{The HCO$^+$ and H$^{13}$CO$^+$ 1-0 spectra toward BGPS 4029 are plotted in the top panel. The blue and red vertical dotted lines indicate the velocity ranges of blue- and red-shifted high-velocity components, respectively. The black and red dash-dotted lines indicate the rms noise levels at the velocity resolution in the HCO$^+$ and H$^{13}$CO$^+$ 1-0 spectra. In the bottom panel, the distributions of the blue- and red-shift velocity components of the HCO$^+$ 1-0 line are shown by the blue and red contours. The SiO 2-1 line is presented by the green contours, and The HC$_3$N 10-9 line is shown by the gray-scale image. The contour levels start at $6\sigma$ in steps of $1\sigma$ of the velocity-integrated intensities for the HCO$^+$ components, and the contour levels corresponding to the SiO 2-1 line start at $3\sigma$. The black dashed circle is the coverage of the spectra shown in the top panel. The red and blue arrows and their connection point indicate the directions of bipolar outflows and the dense core detected in the CO 2-1 line and the 230 GHz continuum \citep{svo19}.}\label{fig:BGPS4029_HCO+_blrd}
\end{figure}

The spectra of the HCO$^+$ and H$^{13}$CO$^+$ 1-0 lines toward BGPS 4029 are plotted in the top panel of Figure \ref{fig:BGPS4029_HCO+_blrd}. A red-shifted absorption dip is shown in the HCO$^+$ 1-0 spectrum, and this feature is also shown in \citet{cal18} as a potential indicator of a global inflow in the clump. According to the comparison between the two spectra, the blue- and red-shifted components of the HCO$^+$ 1-0 line are defined as the components in the velocity ranges of $75-79$ and $86-90$ km s$^{-1}$.  The distributions of the velocity-integrated intensities of the SiO 2-1 and HC$_3$N 10-9 lines toward BGPS 4029 are displayed in the bottom panel of Figure \ref{fig:BGPS4029_HCO+_blrd}. The distributions of the blue- and red-shift components of the HCO$^+$ 1-0 line are also shown. The spreading area of the SiO 2-1 emission shows the distribution of the shocked gas which mainly spreads in the center of observational field. The blue- and red-shift components of the HCO$^+$ 1-0 line are located at the opposite sides of the HC$_3$N 10-9 emission. The coverages of the SiO 2-1 emission and the high-velocity HCO$^+$ components are roughly overlapped. This suggests the relation between the shocked gas and the high-velocity components. %Therefore, we suggest that the shocked gas in BGPS 4029 is originated from the protostellar outflows.

\subsubsection{BGPS 4472}

\begin{figure}
  \centering
  \includegraphics[scale=0.40]{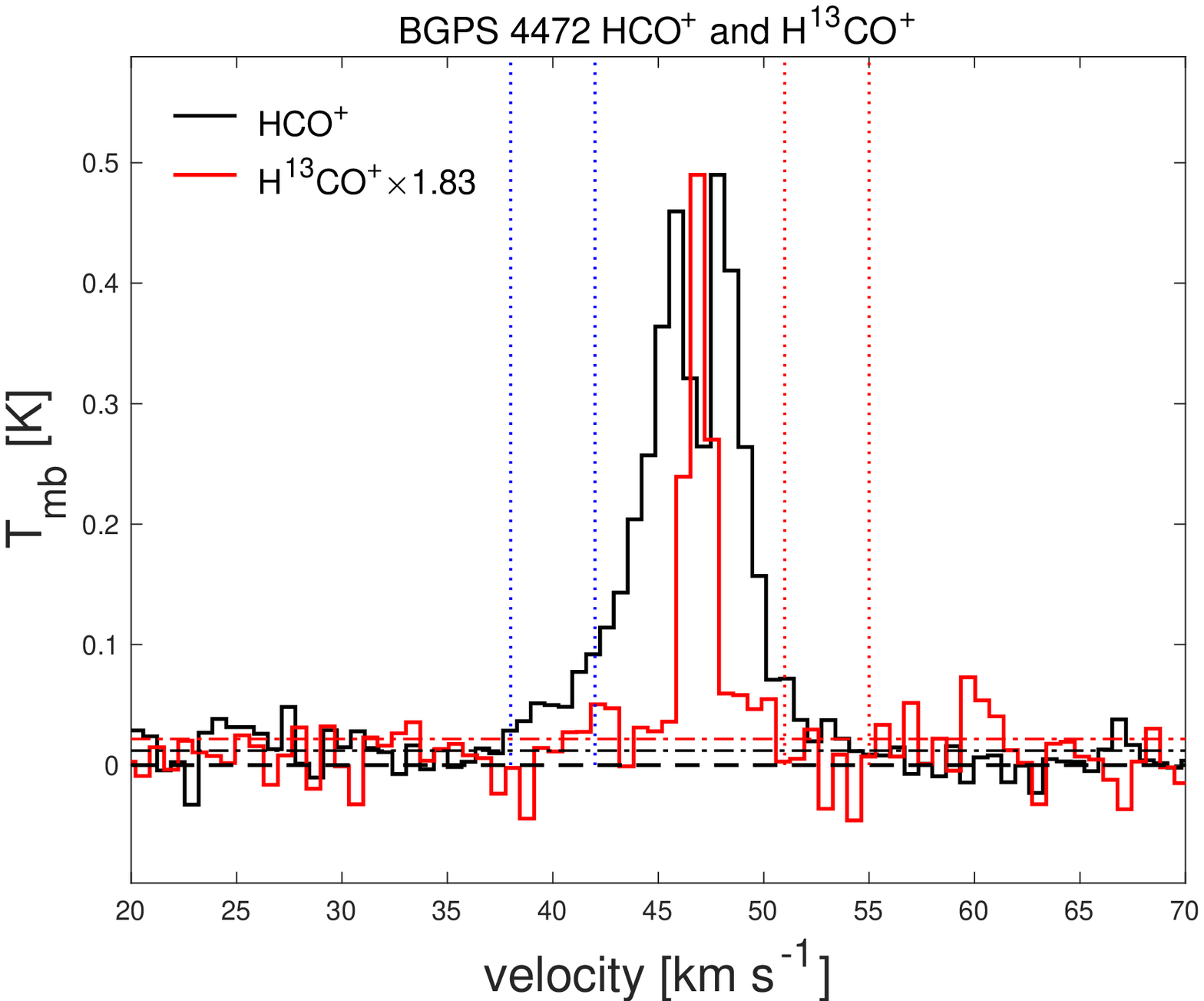}
  \includegraphics[scale=0.45]{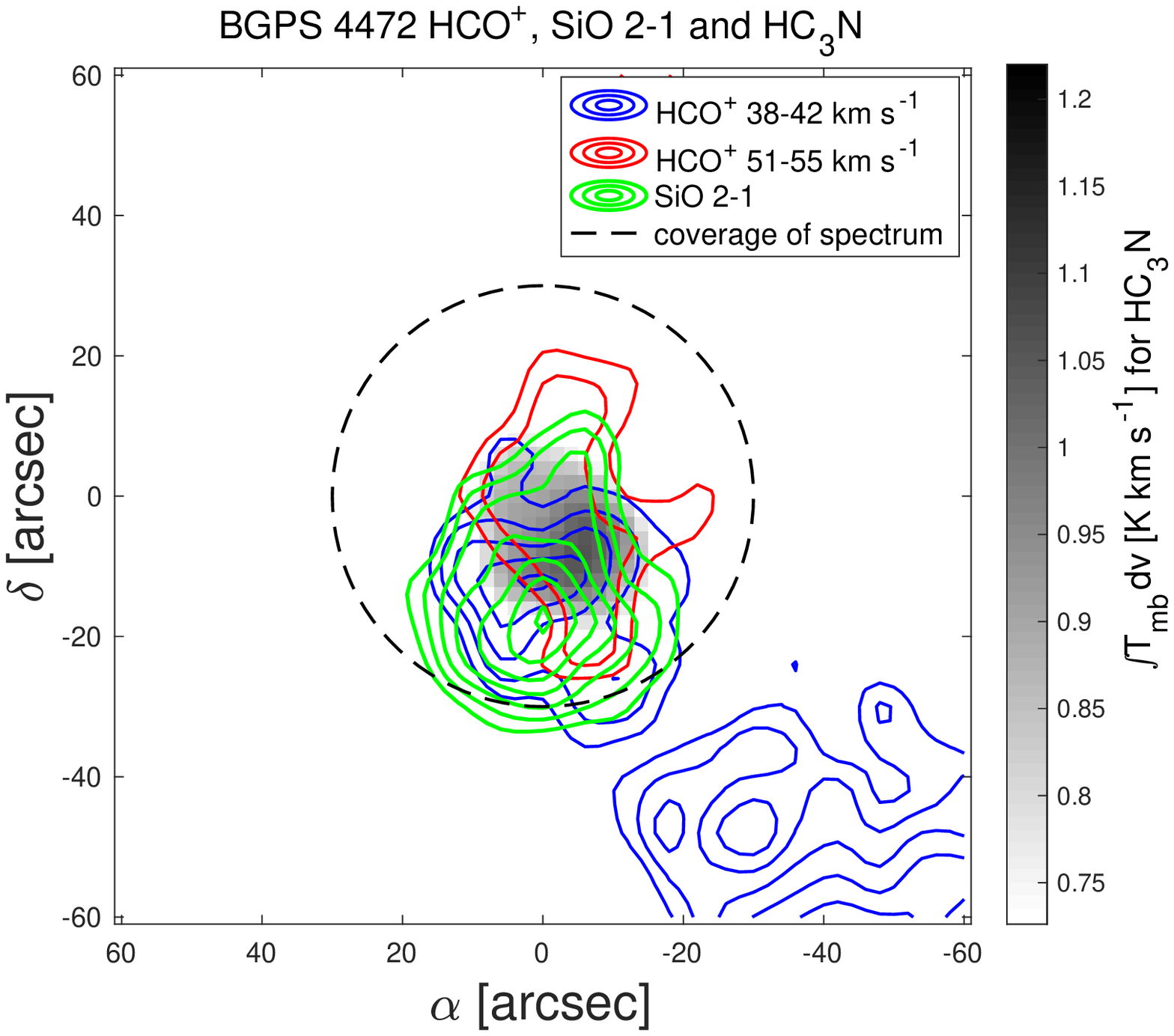}
  \caption{The HCO$^+$ and H$^{13}$CO$^+$ 1-0 spectra toward BGPS 4472 are plotted in the top panel. The blue and red vertical dotted lines indicate the velocity ranges of blue- and red-shifted high-velocity components, respectively. The black and red dash-dotted lines indicate the rms noise levels at the velocity resolution in the HCO$^+$ and H$^{13}$CO$^+$ 1-0 spectra. In the bottom panel, the distributions of the blue- and red-shift velocity components of the HCO$^+$ 1-0 line are shown by the blue and red contours. The SiO 2-1 line is presented by the green contours, and The HC$_3$N 10-9 line is shown by the gray-scale image. The contour levels start at $5\sigma$ and $3\sigma$ in steps of $1\sigma$ of the velocity-integrated intensities for the blue- and red-shifted HCO$^+$ components, respectively. The contour levels corresponding to the SiO 2-1 line start at $5\sigma$. The black dashed circle is the coverage of the spectra shown in the top panel.}\label{fig:BGPS4472_HCO+_blrd}
\end{figure}

The comparison between the HCO$^+$ and H$^{13}$CO$^+$ 1-0 spectra toward BGPS 4472 is displayed in the top panel of Figure \ref{fig:BGPS4472_HCO+_blrd}. As is marked in the panel, the high-velocity components of the HCO$^+$ 1-0 line are defined to be $38-42$ and $51-55$ km s$^{-1}$, respectively. Unlike the HCO$^+$ 1-0 spectrum for BGPS 4029, the signature of inflow is not found in BGPS 4472 since the self-absorption dip is not red-shifted relative to the central velocity of the H$^{13}$CO$^+$ 1-0 line. In the bottom panel of Figure \ref{fig:BGPS4472_HCO+_blrd}, the distributions of the SiO 2-1 and HC$_3$N 10-9 emissions, and the blue- and red-shift HCO$^+$ 1-0 components are presented.  The two high-velocity HCO$^+$ components can be spatially resolved. The compact region of the clump indicated by the HC$_3$N 10-9 line is located at the overlap region of the blue- and red-shift HCO$^+$ components. The spreading region of the shocked gas shown by the SiO 2-1 emission is mostly overlapped with the high-velocity HCO$^+$ components. %So the shock in BGPS 4472 should be caused by the protostellar outflows as in BGPS 4029.

%The SiO 2-1 and 3-2 emissions toward BGPS 4472 are concentrated near the massive clump in the projected image. The brightest point is at the position of (0, -20). The flux weighted central velocity distribution of the SiO 2-1 emission is presetned in Figure \ref{fig:BGPS4472_SiO2-1fwcv}. In this case, the central velocity of the shocked gas roughly decreases from about 52 km s$^{-1}$ to 42 km s$^{-1}$ along the northwest to southeast. From the overall SiO 2-1 spectrum in Figure \ref{fig:BGPS4472_SiO2-1_specall}, the FWHM of the fitted Gaussian profile is 20.48 km s$^{-1}$. This SiO line profile is obviously broad.

\subsubsection{BGPS 5064}

\begin{figure}
  \centering
  \includegraphics[scale=0.40]{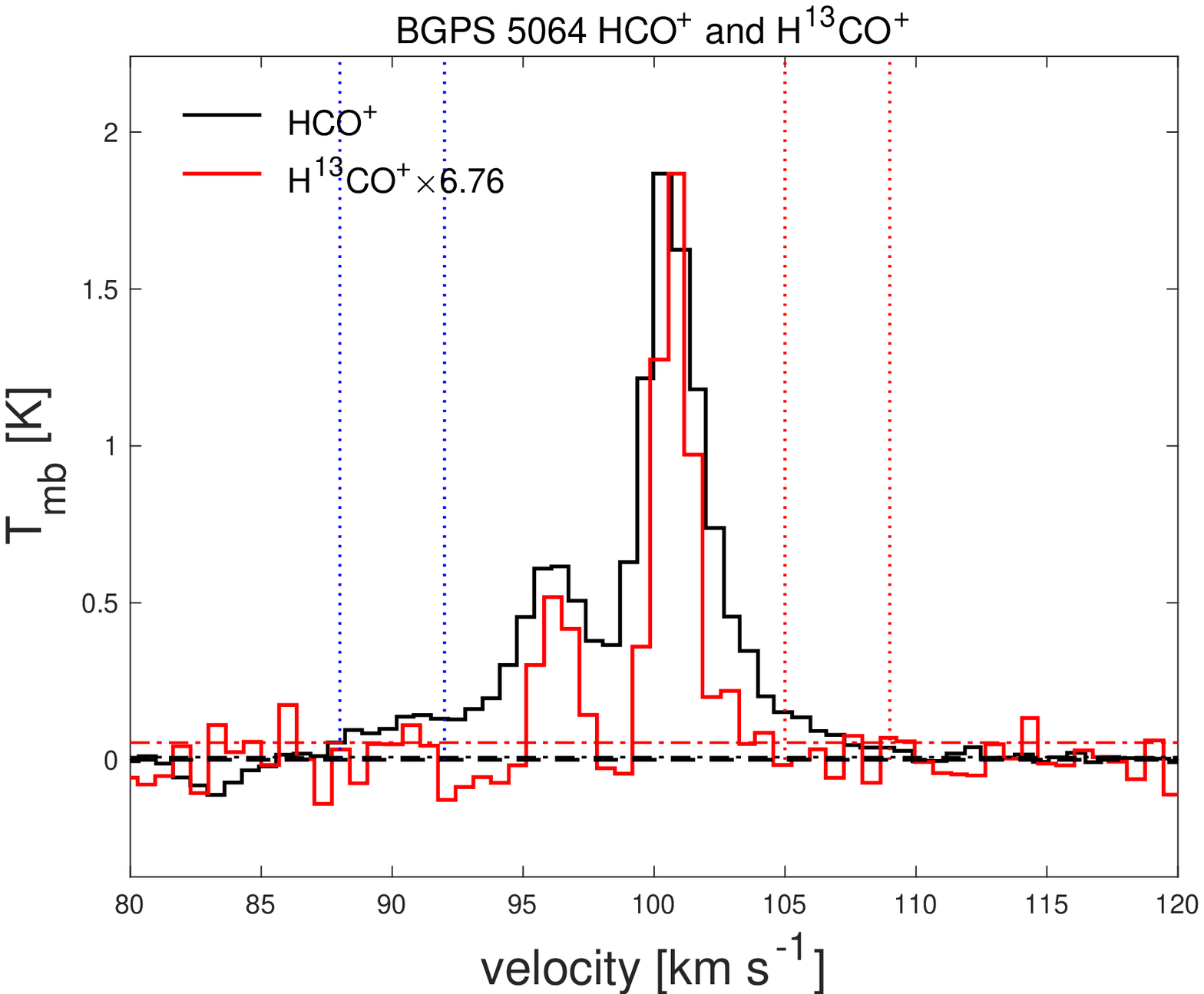}
  \includegraphics[scale=0.45]{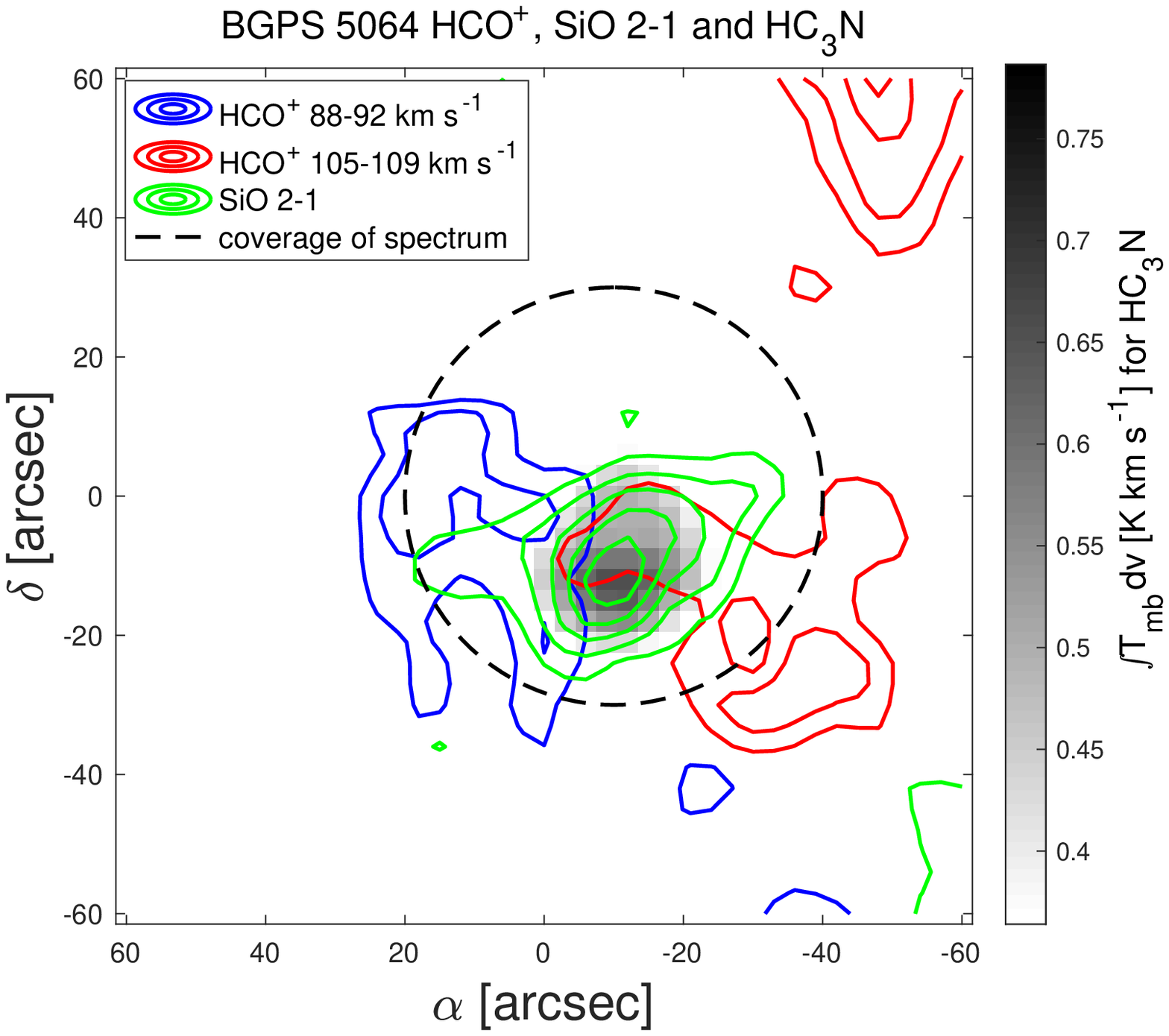}
  \caption{The HCO$^+$ and H$^{13}$CO$^+$ 1-0 spectra toward BGPS 5064 are plotted in the top panel. The blue and red vertical dotted lines indicate the velocity ranges of blue- and red-shifted high-velocity components, respectively. The black and red dash-dotted lines indicate the rms noise levels at the velocity resolution in the HCO$^+$ and H$^{13}$CO$^+$ 1-0 spectra. In the bottom panel, the distributions of the blue- and red-shift velocity components of the HCO$^+$ 1-0 line are shown by the blue and red contours. The SiO 2-1 line is presented by the green contours, and The HC$_3$N 10-9 line is shown by the gray-scale image. The contour levels start at $7\sigma$ in steps of $1\sigma$ of the velocity-integrated intensities for the HCO$^+$ components, and the contour levels corresponding to the SiO 2-1 line start at $3\sigma$. The black dashed circle is the coverage of the spectra shown in the top panel.}\label{fig:BGPS5064_HCO+_blrd}
\end{figure}

\begin{figure}
  \centering
  \includegraphics[scale=0.45]{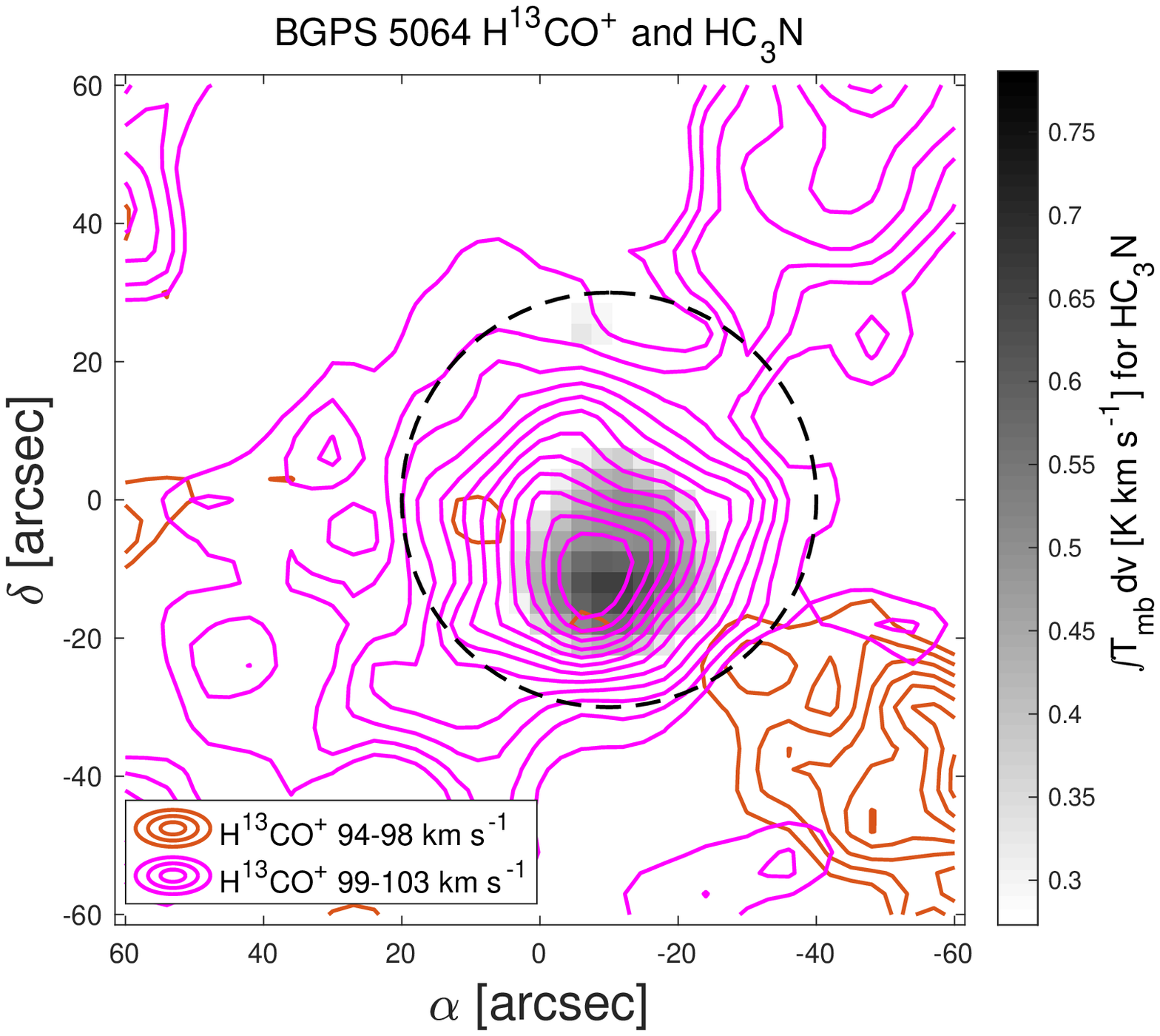}
  \caption{The velocity-integrated intensity distributions of two H$^{13}$CO$^+$ components in the line-of-sight velocity ranges of 94-98 and 99-103 km s$^{-1}$ are presented by brown and magenta contours, respectively. The contour levels start at $5\sigma$ in steps of $1\sigma$. The gray-scale image shows the distribution of the HC$_3$N velocity-integrated intensity. The black dashed circle is the coverage of the spectra shown in the top panel of Figure \ref{fig:BGPS5064_HCO+_blrd}.}\label{fig:BGPS5064_H13CO+_com2}
\end{figure}

%As listed in Table \ref{table_property1}, it is different from the cases of BGPS 4029 and 4472 that the widths (FWHMs) of the SiO 2-1 and 3-2 lines toward BGPS 5064 are narrow.
The HCO$^+$ and H$^{13}$CO$^+$ 1-0 spectra in BGPS 5064 are plotted in the top panel of Figure \ref{fig:BGPS5064_HCO+_blrd}. There are two velocity components in the H$^{13}$CO$^+$ 1-0 spectrum. The spatial distributions of these two components are shown in Figure \ref{fig:BGPS5064_H13CO+_com2}. The component for the velocity range of 94-98 km s$^{-1}$ is mainly distributed in the southwest, and the component for the velocity range of 99-103 km s$^{-1}$ is distributed in the center region and the northwest.

The spectrum of the HCO$^+$ 1-0 line can be divided into three components. Two narrow components correspond to the two velocity components of the H$^{13}$CO$^+$ 1-0 line. The broad component with FWHM$=12.30$ km s$^{-1}$ seems to be related to the shocked gas. These three velocity components of the HCO$^+$ 1-0 line are plotted in Figure \ref{fig:BGPS5064_HCO+_spec3}. The HCO$^+$ components in the velocity ranges of 88-92 and 105-109 km s$^{-1}$ are used to present the distributions of the high-velocity components in the bottom panel of Figure \ref{fig:BGPS5064_HCO+_blrd}. These two high-velocity components can be spatially resolved. The spreading areas of the SiO 2-1 and HC$_3$N 10-9 emissions are also shown in the bottom panel. They are located near the connection point of the blue- and red-shifted HCO$^+$ components. By comparing the center velocity and the spatial distribution, it suggests that the dense clump indicated by the H$^{13}$CO$^+$ 1-0 component in the velocity range of 99-103 km s$^{-1}$ is associated with the shock in BGPS 5064. And the dense gas shown by the other velocity component of the H$^{13}$CO$^+$ 1-0 line is probably not related to the shock.

%\begin{figure}
%  \centering
%  \includegraphics[scale=0.40]{BGPS5064_HCO+_spec3.eps}
%  \caption{The HCO$^+$ 1-0 line profile toward BGPS 5064 is plotted. The coverage is the same as that of the HCO$^+$ spectrum shown in the top panel of Figure \ref{fig:BGPS5064_HCO+_blrd}. The broad velocity component shown in the yellow curve indicates the presence of protostellar outflows.}\label{fig:BGPS5064_HCO+_spec3}
%\end{figure}

\subsubsection{BGPS 5114}

\begin{figure}
  \centering
  \includegraphics[scale=0.40]{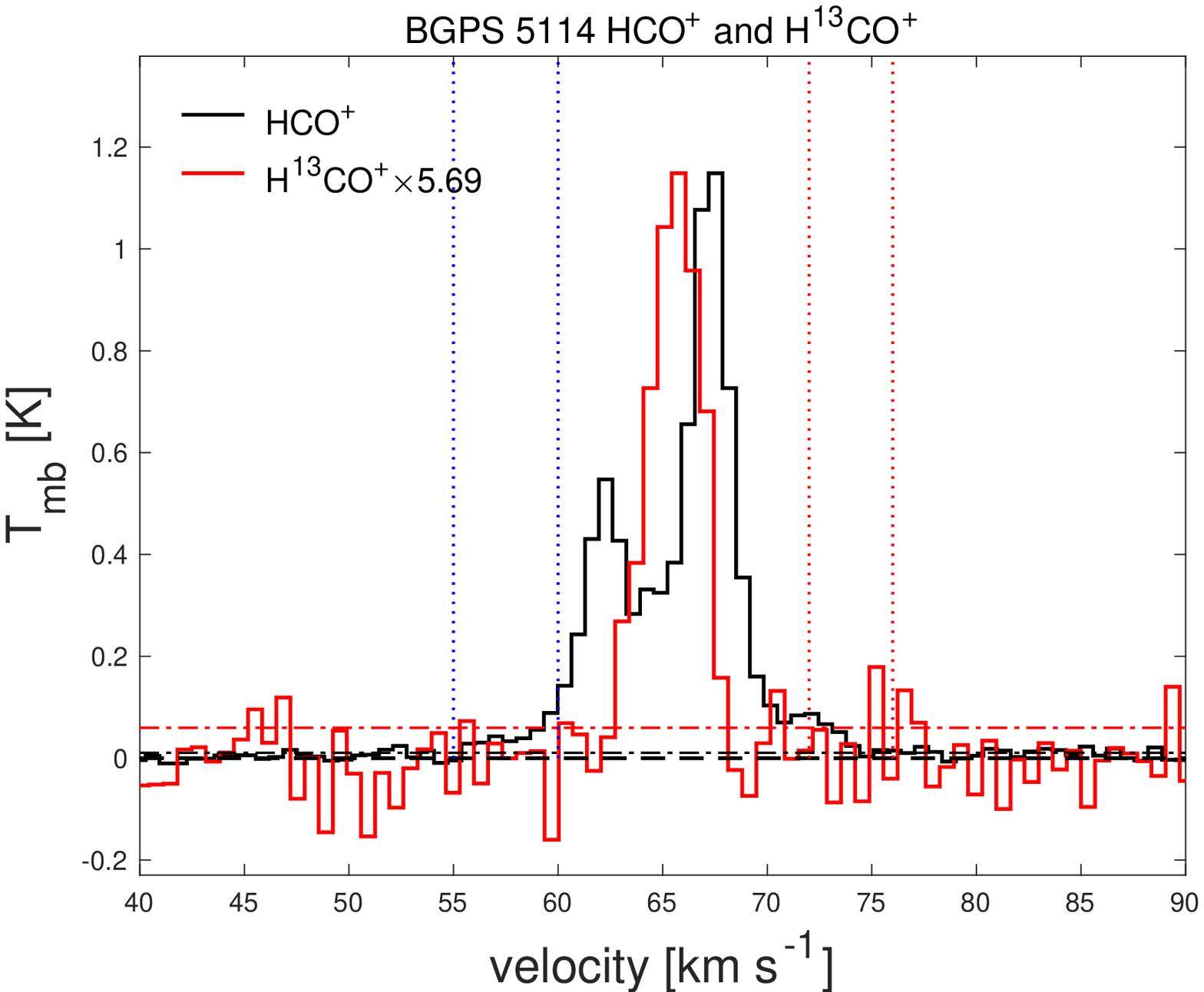}
  \includegraphics[scale=0.45]{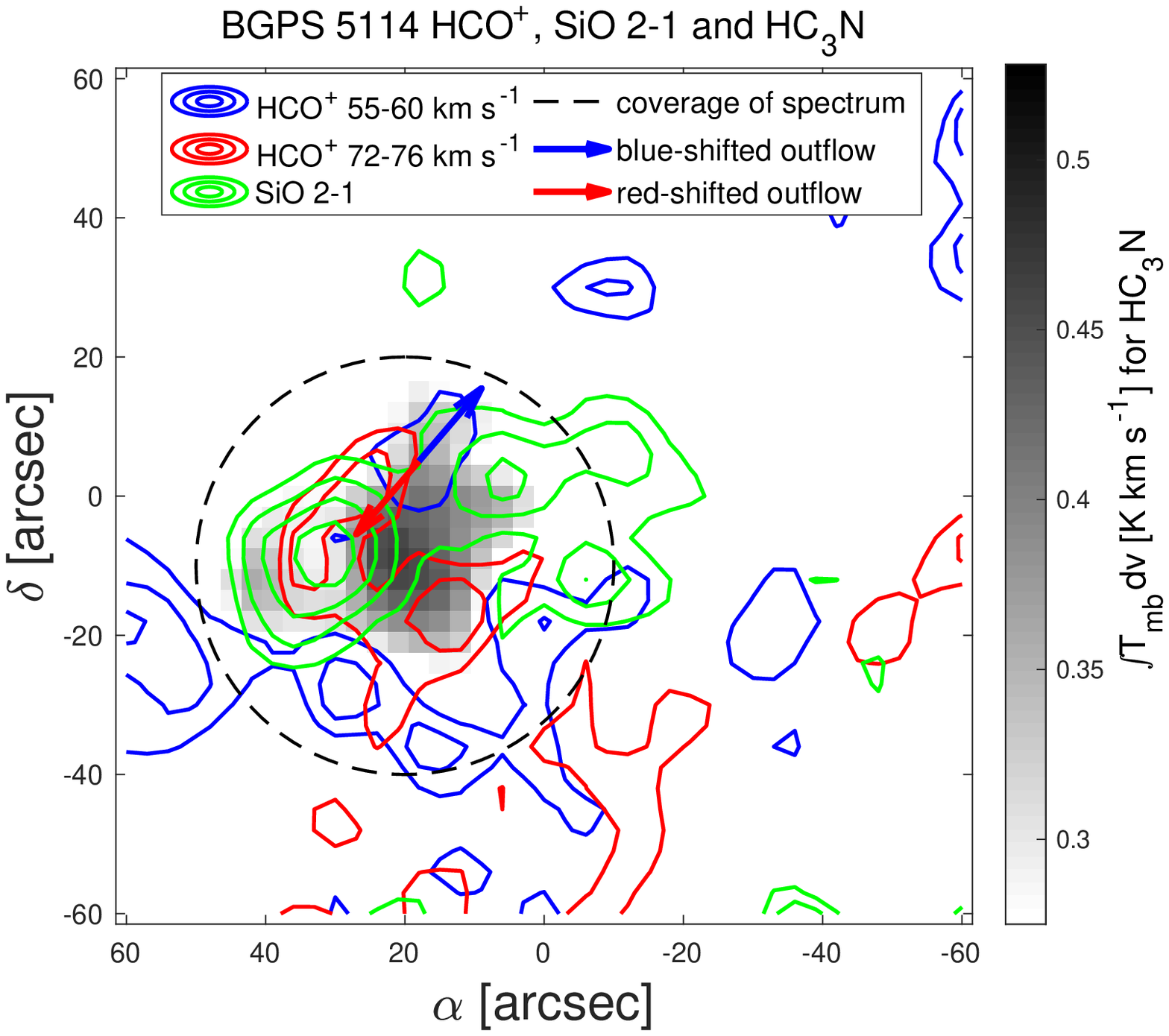}
  \caption{The HCO$^+$ and H$^{13}$CO$^+$ 1-0 spectra toward BGPS 5114 are plotted in the top panel. The blue and red vertical dotted lines indicate the velocity ranges of blue- and red-shifted high-velocity components, respectively. The black and red dash-dotted lines indicate the rms noise levels at the velocity resolution in the HCO$^+$ and H$^{13}$CO$^+$ 1-0 spectra. In the bottom panel, the distributions of the blue- and red-shift velocity components of the HCO$^+$ 1-0 line are shown by the blue and red contours. The SiO 2-1 line is presented by the green contours, and The HC$_3$N 10-9 line is shown by the gray-scale image. The contour levels start at $4\sigma$ in steps of $1\sigma$ of the velocity-integrated intensities for the HCO$^+$ components, and the contour levels corresponding to the SiO 2-1 line start at $3\sigma$. The black dashed circle is the coverage of the spectra shown in the top panel. The red and blue arrows and their connection point indicate the directions of bipolar outflows and the dense core detected in the CO 2-1 line and the 230 GHz continuum \citep{svo19}.}\label{fig:BGPS5114_HCO+_blrd}
\end{figure}

The HCO$^+$ and H$^{13}$CO$^+$ 1-0 spectra toward BGPS 5114 are plotted in the top panel of Figure \ref{fig:BGPS5114_HCO+_blrd}. There is an absorption dip in the HCO$^+$ 1-0 spectrum because of the thick optical depth. The H$^{13}$CO$^+$ 1-0 spectrum is single-peaked. From the comparison between the HCO$^+$ and H$^{13}$CO$^+$ 1-0 spectra, the HCO$^+$ components in velocity ranges of 55-60 and 72-76 km s$^{-1}$ are regarded as the indicators of the high-velocity gas. The distributions of these high-velocity components are shown in the bottom panel of Figure \ref{fig:BGPS5114_HCO+_blrd}. The intensity distributions of the SiO 2-1 and HC$_3$N 10-9 lines are also presented. However, the blue- and red-shift HCO$^+$ components are chaotically distributed unlike those in BGPS 4029, 4472 and 5064. The HC$_3$N 10-9 emission is concentrated near the position of (20, -10). The shocked gas shown by the SiO 2-1 line is located near the HC$_3$N 10-9 emission. The association of the HC$_3$N 10-9 and SiO 2-1 lines is implied by the spatial distributions of them. %If the compact gas indicated by the HC$_3$N 10-9 emission contained protostellar core, the shock could be caused by protostellar outflows.

\subsubsection{BGPS 5243}

%The HCO$^+$ and H$^{13}$CO$^+$ spectra toward BGPS 5243 are plotted in the top panel of Figure \ref{fig:BGPS5243_HCO+_blrd}.

\begin{figure}
  \centering
  \includegraphics[scale=0.45]{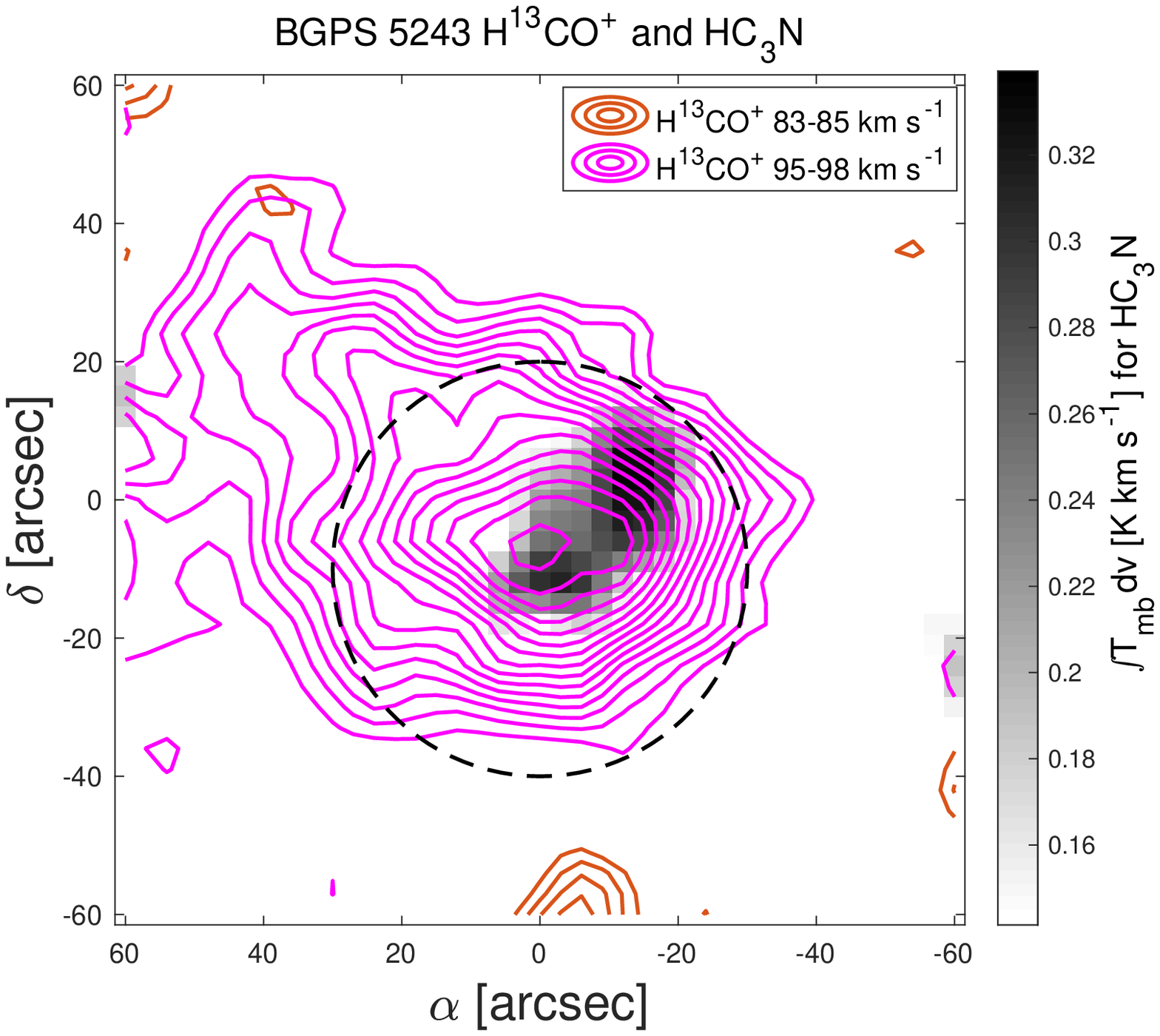}
  \caption{The velocity-integrated intensity distributions of two H$^{13}$CO$^+$ 1-0 components in the line-of-sight velocity ranges of 83-85 and 95-98 km s$^{-1}$ are presented by brown and magenta contours, respectively. The contour levels start at $3\sigma$ in steps of $1\sigma$. The gray-scale image shows the distribution of the HC$_3$N 10-9 velocity-integrated intensity. The black dashed circle is the coverage of the spectra shown in the top panel of Figure \ref{fig:BGPS5243_HCO+_blrd}.}\label{fig:BGPS5243_H13CO+_com2}
\end{figure}

There are two H$^{13}$CO$^+$ 1-0 components with the velocity ranges of 83-85 and 95-98 km s$^{-1}$, respectively. The locations of the two components are presented in Figure \ref{fig:BGPS5243_H13CO+_com2}. One clump of dense gas with the velocity range of 95-98 km s$^{-1}$ is mainly distributed from the center to the east side of observational field while the other dense clump is located at the center of the south side. The central velocities of the two clumps are 85 and 97 km s$^{-1}$, respectively.

\begin{figure}
  \centering
  \includegraphics[scale=0.40]{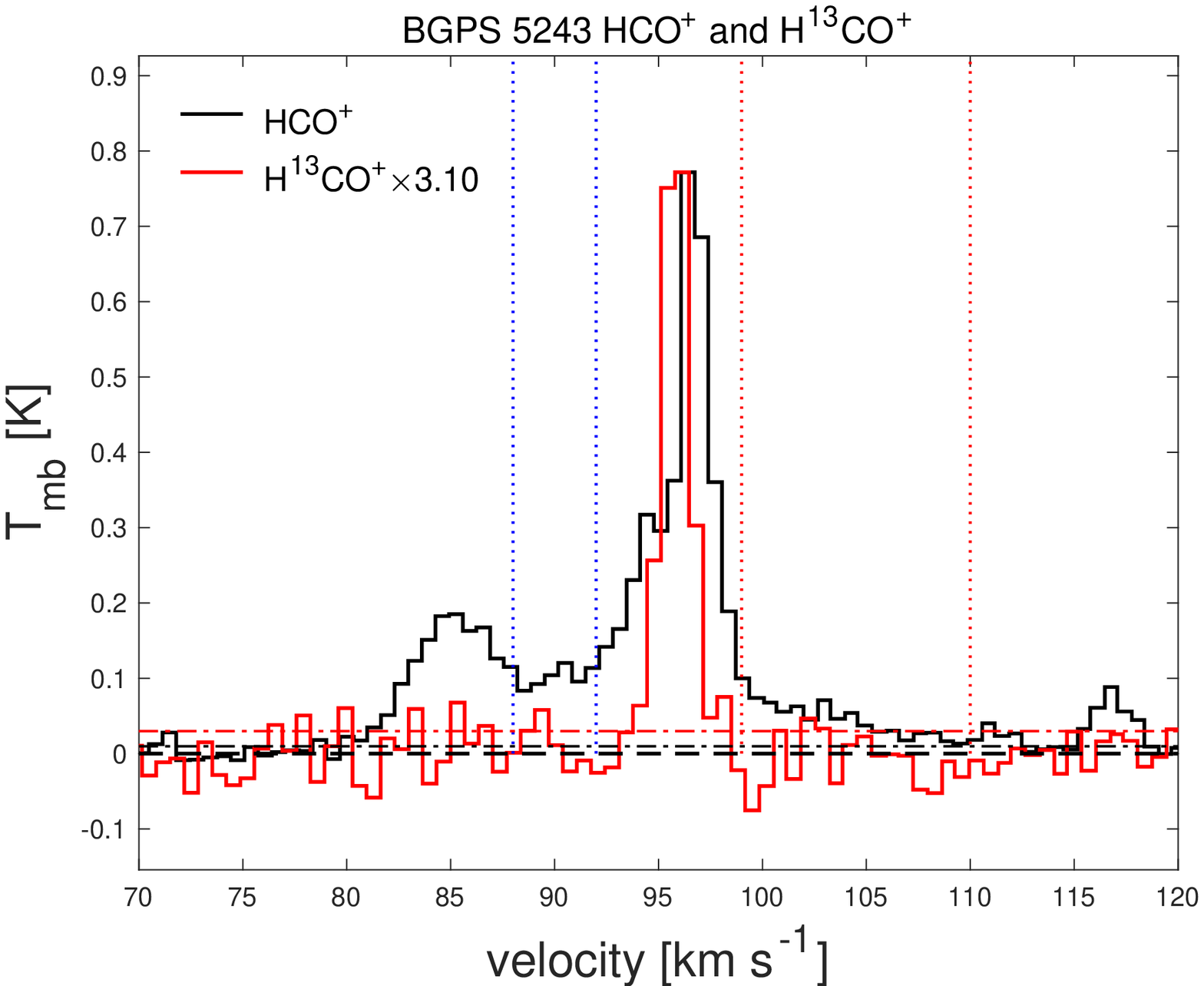}
  \includegraphics[scale=0.45]{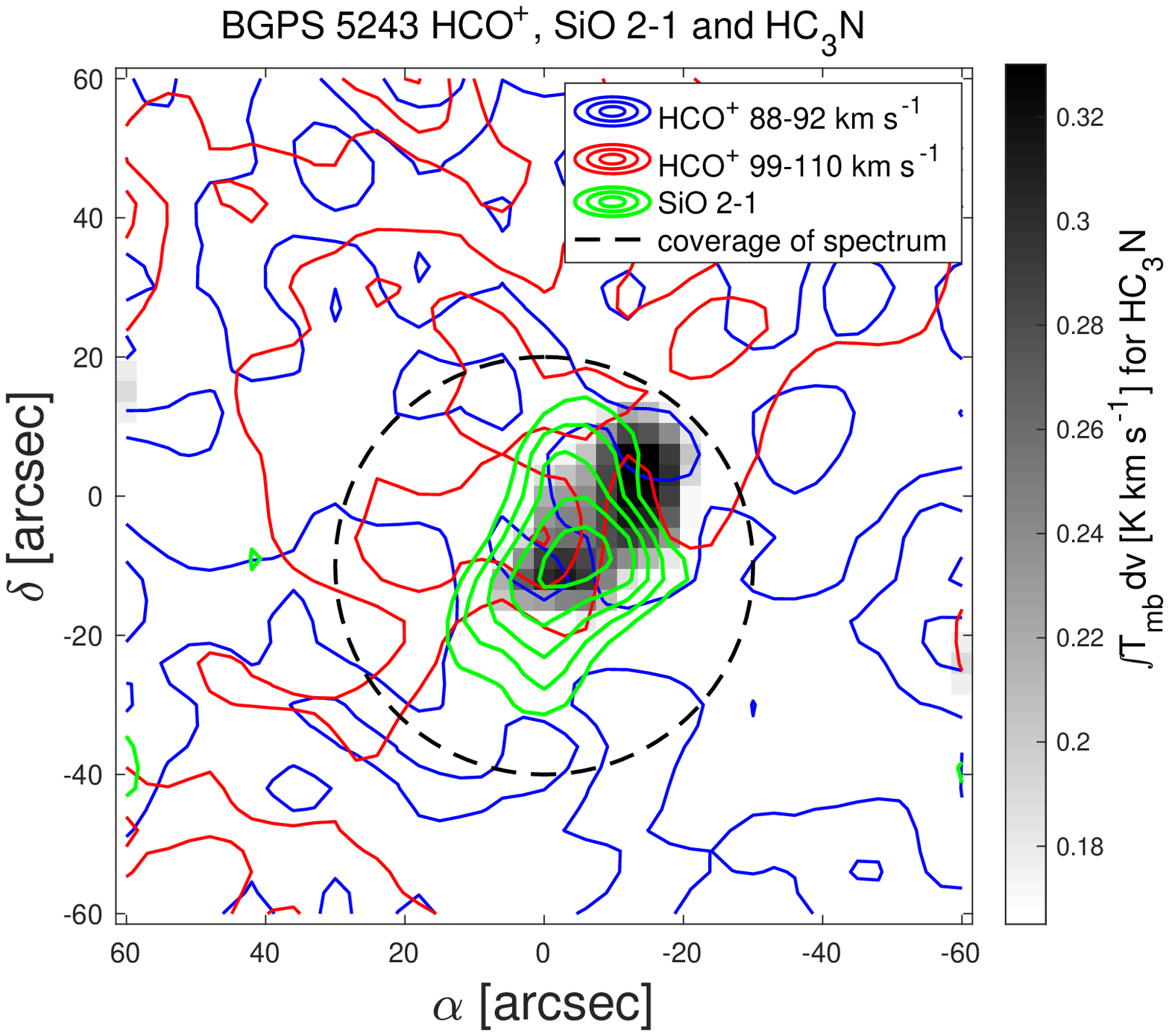}
  \caption{The HCO$^+$ and H$^{13}$CO$^+$ 1-0 spectra toward BGPS 5114 are plotted in the top panel. The blue and red vertical dotted lines indicate the velocity ranges of blue- and red-shifted high-velocity components, respectively. The black and red dash-dotted lines indicate the rms noise levels at the velocity resolution in the HCO$^+$ and H$^{13}$CO$^+$ 1-0 spectra. In the bottom panel, the distributions of the blue- and red-shift velocity components of the HCO$^+$ 1-0 line are shown by the blue and red contours. The SiO line is presented by the green contours, and The HC$_3$N 10-9 line is shown by the gray-scale image. The contour levels start at $5\sigma$ in steps of $2\sigma$ of the velocity-integrated intensities for the HCO$^+$ components, and the contour levels corresponding to the SiO 2-1 line start at $3\sigma$ in steps of $1\sigma$. The black dashed circle is the coverage of the spectra shown in the top panel.}\label{fig:BGPS5243_HCO+_blrd}
\end{figure}

The HCO$^+$ and H$^{13}$CO$^+$ 1-0 spectra toward BGPS 5243 are plotted in the top panel of Figure \ref{fig:BGPS5243_HCO+_blrd}. The two dense clumps with different central velocities mentioned above also lead to the two velocity components in the HCO$^+$ spectrum. The detected distribution of the HCO$^+$ emission from the south clump is wider than that of H$^{13}$CO$^+$ emission. The two velocity components are clear in the HCO$^+$ spectrum, but the H$^{13}$CO$^+$ spectrum only shows the emission from the 97 km s$^{-1}$ clump. The HCO$^+$ 1-0 emission in the velocity range between 85-90 km s$^{-1}$ could be the overlap of line emission from the two clumps with different central velocities. Moreover, the HCO$^+$ 1-0 intensity is also clear in the velocity range between $100-110$ km s$^{-1}$.

The distributions of the velocity-integrated HCO$^+$ intensities in the ranges of 88-92 and 100-110 km s$^{-1}$ are presented in the bottom panel of Figure \ref{fig:BGPS5243_HCO+_blrd}. Both of these two velocity components seem to be widespread. The 88-92 km s$^{-1}$ component should be a mixture of the line emissions from the 85 and 97 km s$^{-1}$ clumps. And the 100-110 km s$^{-1}$ component is mainly distributed in the northeast of the 97 km s$^{-1}$ clump. The broad spreading of the HCO$^+$ emission in the velocity ranges implies a complex velocity field. The HC$_3$N 10-9 emission is located near the center of observational field, while the SiO 2-1 emission spreads in the same area. The HC$_3$N emission displays two closely-positioned intensity peaks in the bottom panel of Figure \ref{fig:BGPS5243_HCO+_blrd}. Since the distance between these peaks is significantly lower than the 28$''$ beam size, it is possible that they result from oversampling. It is difficult to determine if the HC$_3$N emission indeed traces two components. Similar features in the HC$_3$N emission exist in BGPS 3686 and 5114. However, given uncentainties in their reality, they will not be discussed further in this work.

%It is difficult to determine the relation between the 88-92, 100-110 km s$^{-1}$ HCO$^+$ components and the shocked gas indicated by the SiO 2-1 emission.

%The distributions of the velocity-integrated \textbf{HCO$^+$} intensities in the ranges of 88-92 and 100-110 km s$^{-1}$ are presented in the bottom panel of Figure \ref{fig:BGPS5243_HCO+_blrd}. Both of these two velocity components seem to be widespread unlike the result due to protostellar outflows. \textbf{The 88-92 km s$^{-1}$ component should be a mixture of the line emissions from the 85 and 97 km s$^{-1}$ clumps. And the 100-110 km s$^{-1}$ component is mainly distributed in the northeast of the 97 km s$^{-1}$ clump.} The broad spreading of the HCO$^+$ emission in the velocity range implies a complex velocity field. However, because of the limitation of the spatial resolution in the observations, the origin of this velocity field is unknown. The HC$_3$N 10-9 emission is located near the center of observational field, while the SiO 2-1 emission spreads in the same area. It is difficult to determine the relation between the 88-92, 100-110 km s$^{-1}$ HCO$^+$ components and the shocked gas indicated by the SiO 2-1 emission.

\subsection{The sources with H41$\alpha$ detections}

%\subsubsection{Observations toward the sources near H II region G015.0727-00.6447}th

As mentioned above, BGPS 3110, 3114 and 3118 are located near M17 H II region. In this section, the observational results for BGPS 3110, 3114 and 3118 are individually presented in details. In addition, the distributions of blue- and red-shifted high-velocity HCO$^+$ components are not presented because no outflow features are found in these sources. The distributions of HC$_3$N line are not much different from those of H$^{13}$CO$^+$ line in BGPS 3110 and 3118 while the HC$_3$N line is very weak in BGPS 3114 \citep{zhu23}. So the H$^{13}$CO$^+$ line is used to trace high-density gas in the 3 sources, and the distributions of HC$_3$N 10-9 line are not shown below.

%Each of these sources is observed with a $2'\times2'$ observation field in this set of observations. In the further observations performed in December 2020 to January 2021, the whole H II region is observed to study its effect on the nearby star formation environments. The more forceful evidences of the origin of the shock in the three sources are also obtained. The details of the further observations and the relevant analysis are discussed in the other work \citep{zhu22}. In this work, the details of the observational results for BGPS 3110, 3114 and 3118 are not individually presented as those for the other sources. The concise descriptions of the case in BGPS 3110 are written below as the exemplar of the three sources.

\subsubsection{BGPS 3110}

\begin{figure}
  \centering
  \includegraphics[scale=0.40]{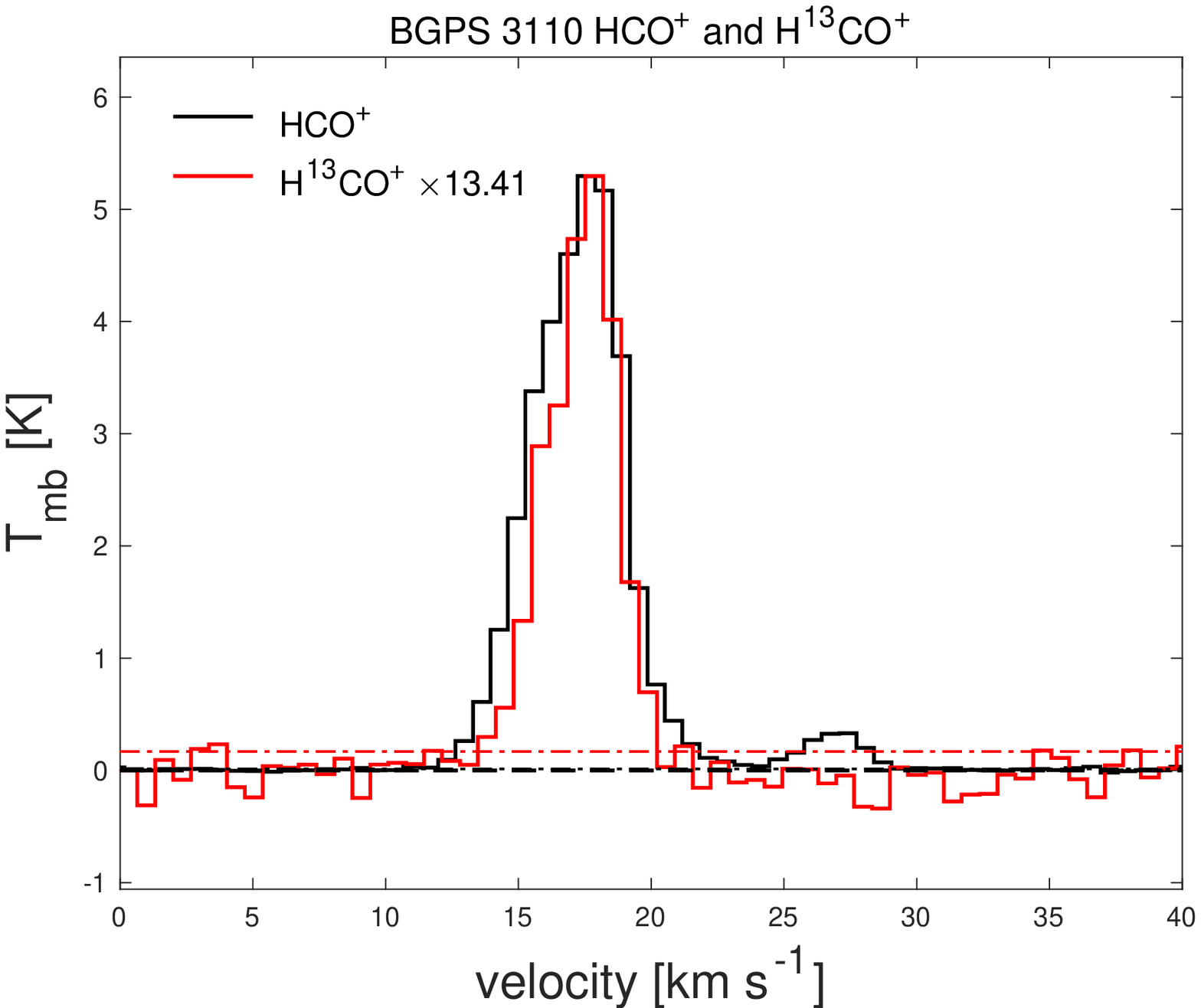}
  \includegraphics[scale=0.45]{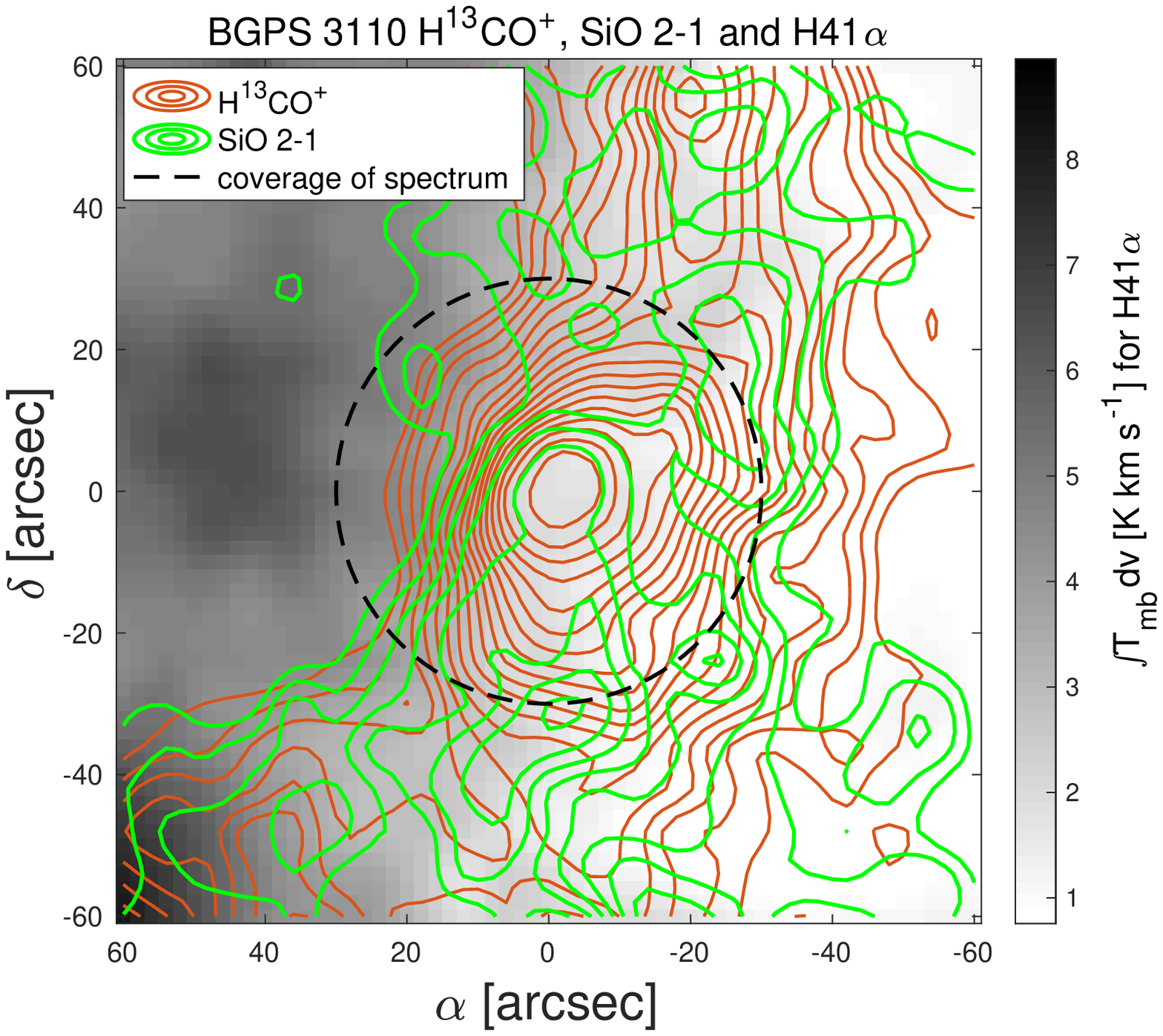}
  \caption{The HCO$^+$ and H$^{13}$CO$^+$ 1-0 spectra toward BGPS 3110 are plotted in the top panel. The black and red lines represent the HCO$^+$ and H$^{13}$CO$^+$ 1-0 lines, respectively. The black and red dash-dotted lines indicate the rms noise levels at the velocity resolution in the HCO$^+$ and H$^{13}$CO$^+$ 1-0 spectra. In the bottom panel, the distributions of the H$^{13}$CO$^+$ 1-0, SiO 2-1 emissions are shown by the blue and green contours. The H41$\alpha$ line is shown by the gray-scale image. The contour levels start at $5\sigma$ in steps of $1\sigma$ of the velocity-integrated intensities for the H$^{13}$CO$^+$ 1-0 line, and the contour levels corresponding to the SiO 2-1 line start at $3\sigma$. The black dashed circle is the coverage of the spectra shown in the top panel.}\label{fig:BGPS3110_HCO+_H41a}
\end{figure}

The HCO$^+$ and H$^{13}$CO$^+$ 1-0 spectra toward BGPS 3110 are presented in the top panel of Figure \ref{fig:BGPS3110_HCO+_H41a}. There is no obvious non-Gaussian and optically thin line wings shown in the HCO$^+$ 1-0 spectrum. And we find no suitable velocity ranges for blue- and red-shifted HCO$^+$ high-velocity components that can indicate the existence of outflows. The distributions of the H$^{13}$CO$^+$ 1-0, H41$\alpha$ and SiO 2-1 lines are shown in the bottom panel of Figure \ref{fig:BGPS3110_HCO+_H41a}. The H41$\alpha$ emission indicates the spreading area of the H II region in the observational field. This line is bright in the east and much weaker in the west. The distribution of the H$^{13}$CO$^+$ 1-0 line represents the spreading area of dense gas distributed from the north to the southeast. The H$^{13}$CO$^+$ line emission is brightest in the center where corresponds to the position of SCC.  In addition, the distribution of the shocked gas indicated by the SiO 2-1 line is approximately overlapped with that of high-density gas. This implies the relation between the shocked and high-density gases. %Since the shell consists of dense gas between the ionization and shock fronts \citep{hos06,zhu15b},  this supports that the layer of dense gas indicated by H$^{13}$CO$^+$ 1-0 emission is a part of the shell of M17 H II region.

%Comparing the distributions of the H$^{13}$CO$^+$ 1-0 and H41$\alpha$ lines in the whole region around M17 H II region shown in \citet{zhu22},  the dense gas indicated by the H$^{13}$CO$^+$ 1-0 emission in BGPS 3110 is probably a part of the shell of the H II region.
%The HC$_3$N spectrum toward BGPS 3110 is compared with the SiO 2-1 spectrum in Figure .
%The velocity component of the HCO$^+$ line located in the velocity range between 25-28 km s$^{-1}$

\subsubsection{BGPS 3114}

\begin{figure}
  \centering
  \includegraphics[scale=0.40]{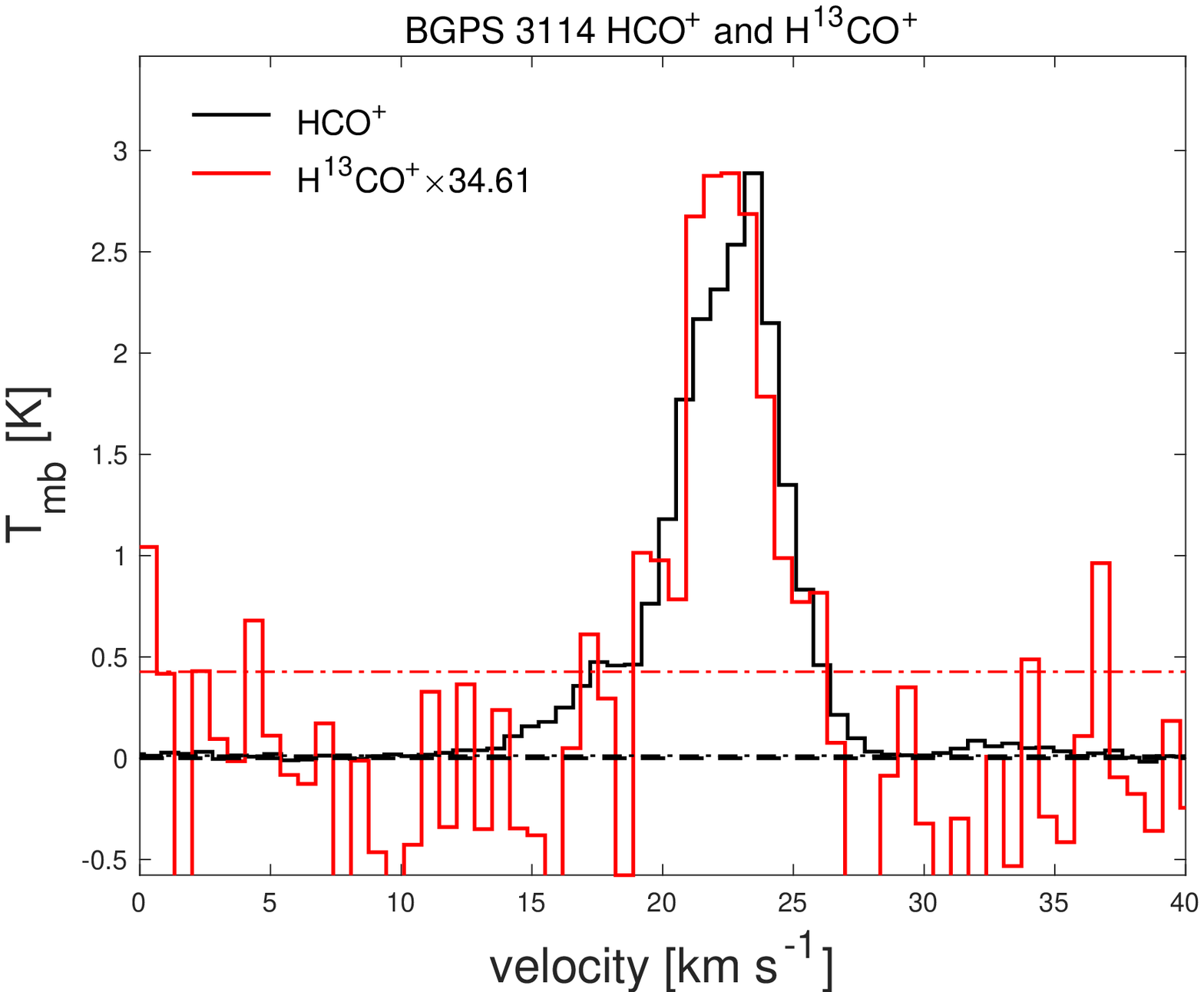}
  \includegraphics[scale=0.45]{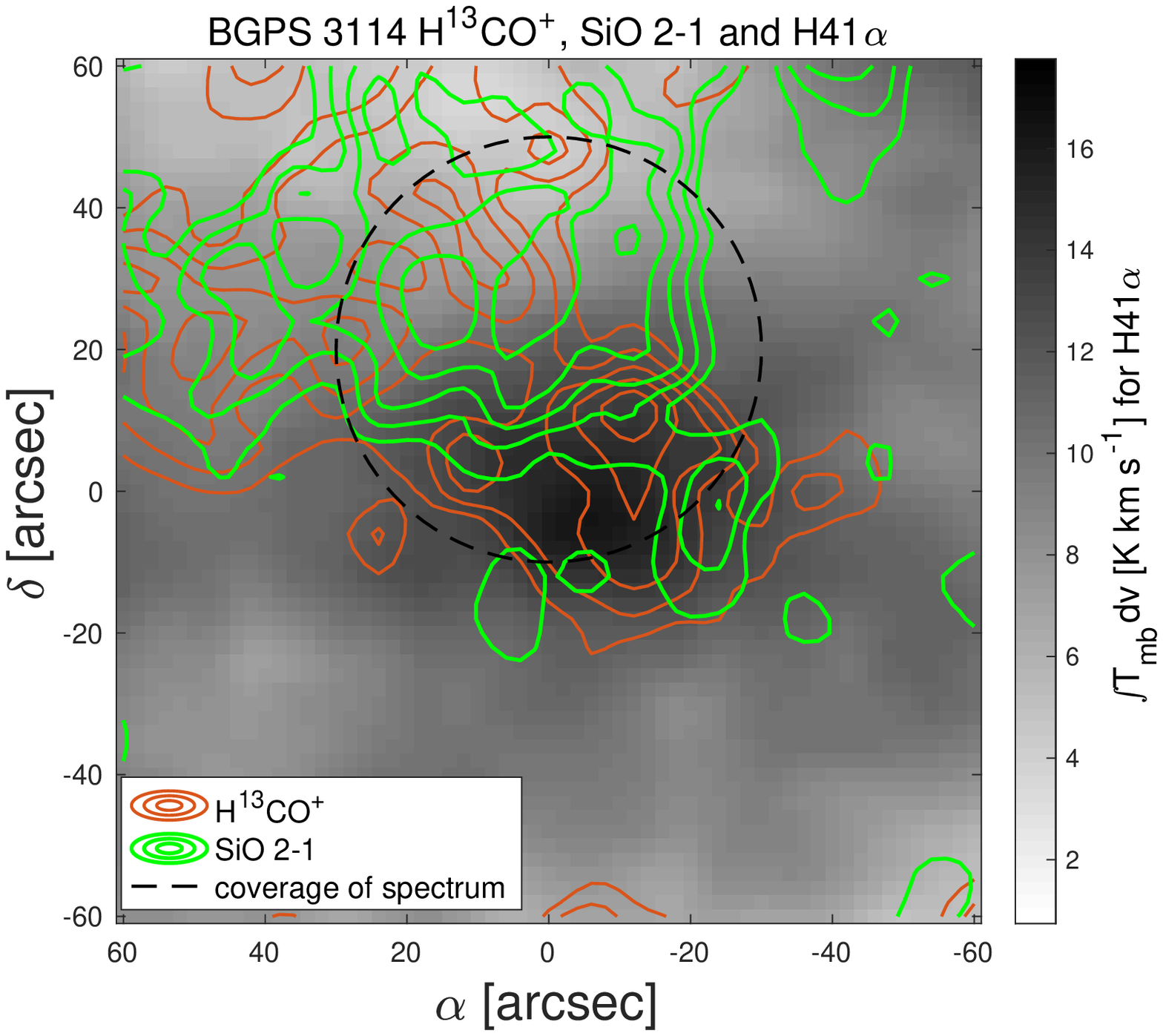}
  \caption{The HCO$^+$ and H$^{13}$CO$^+$ 1-0 spectra toward BGPS 3114 are plotted in the top panel. The black and red lines represent the HCO$^+$ and H$^{13}$CO$^+$ 1-0 lines, respectively. The black and red dash-dotted lines indicate the rms noise levels at the velocity resolution in the HCO$^+$ and H$^{13}$CO$^+$ 1-0 spectra. In the bottom panel, the distributions of the H$^{13}$CO$^+$ 1-0, SiO 2-1 emissions are shown by the brown and green contours. The H41$\alpha$ line is shown by the gray-scale image. The contour levels start at $3\sigma$ in steps of $1\sigma$ of the velocity-integrated intensities for the H$^{13}$CO$^+$ 1-0 and SiO 2-1 lines. The black dashed circle is the coverage of the spectra shown in the top panel.}\label{fig:BGPS3114_HCO+_H41a}
\end{figure}

The HCO$^+$ and H$^{13}$CO$^+$ 1-0 line profiles in BGPS 3114 are plotted in the top panel of Figure \ref{fig:BGPS3114_HCO+_H41a}. There is a slight blue shifted absorption dip in the HCO$^+$ 1-0 line profile. The H$^{13}$CO$^+$ 1-0 line intensity is mainly in the velocity range between 20 and 26 km s$^{-1}$. The spatial distributions of the H$^{13}$CO$^+$ 1-0, SiO 2-1 and H41$\alpha$ lines are presented in the bottom panel of Figure \ref{fig:BGPS3114_HCO+_H41a}. The H41$\alpha$ emission spreads in the whole observational field, and the brightest point is located at the center. Molecular gas can not exist in the H II region because of the processes of ionization and dissociation. According to the location and gas velocity, the dense gas indicated by the H$^{13}$CO$^+$ 1-0 line should be related with the isolated small cloud found by \citet{wil03} which seems to be behind the ionized gas. The distributions of the SiO 2-1 and H$^{13}$CO$^+$ 1-0 lines are also overlapped as those in BGPS 3110. The central velocities of SiO and H$^{13}$CO$^+$ 1-0 lines are similar while they are different from the central velocity of H41$\alpha$ line. %These features suggest the association between the shocked and dense gas.

\subsubsection{BGPS 3118}

\begin{figure}
  \centering
  \includegraphics[scale=0.40]{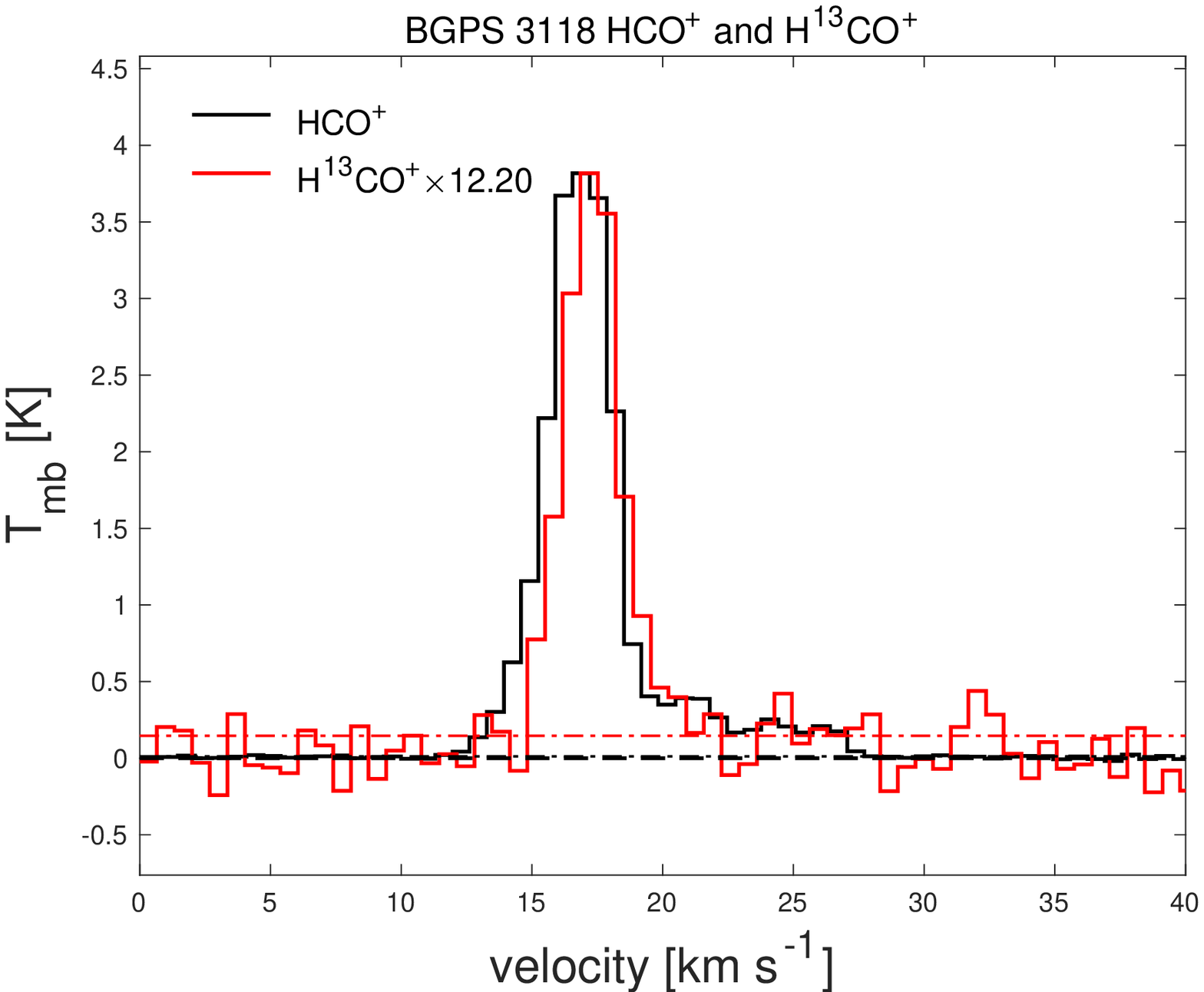}
  \includegraphics[scale=0.45]{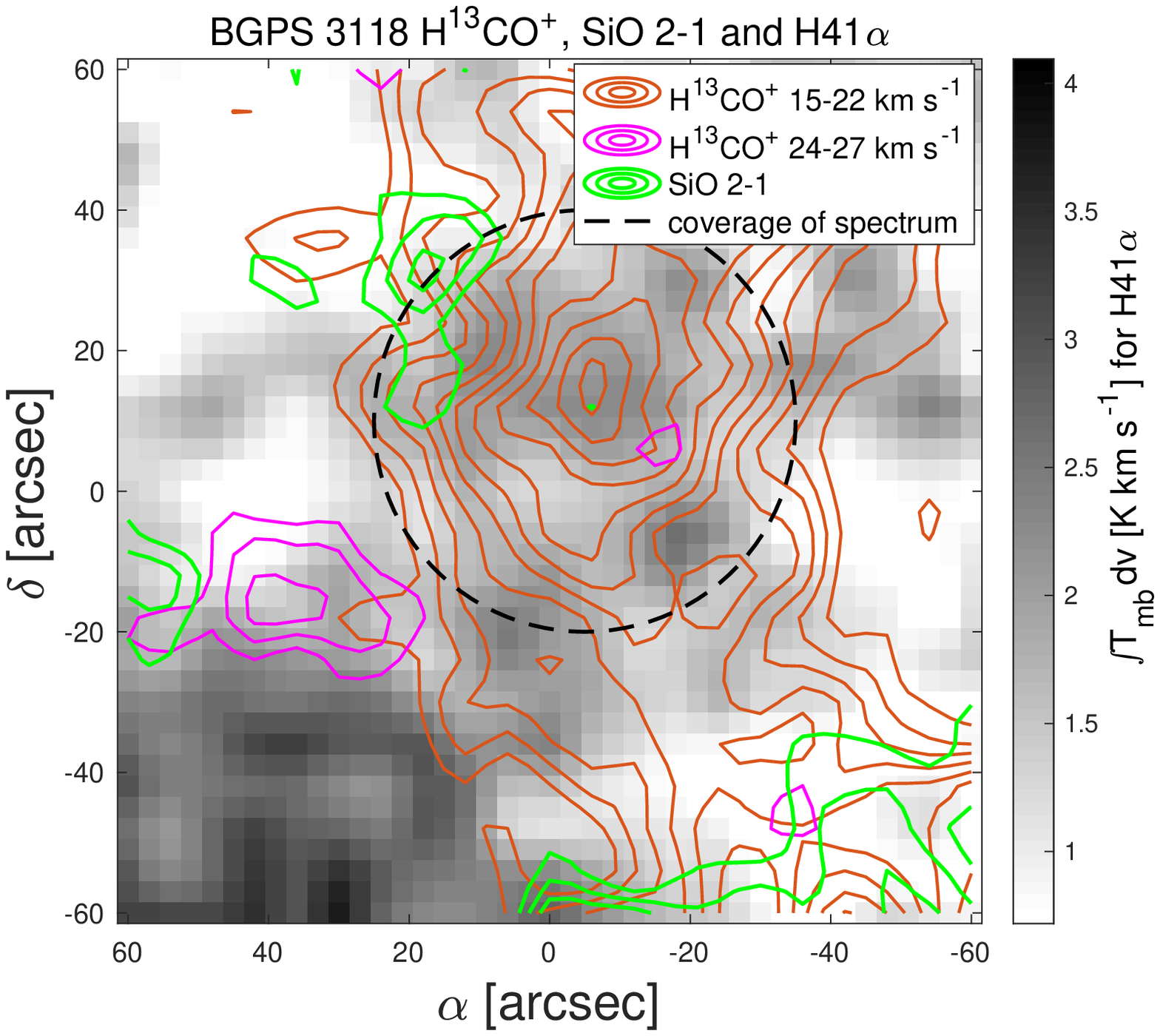}
  \caption{The HCO$^+$ and H$^{13}$CO$^+$ 1-0 spectra toward BGPS 3118 are plotted in the top panel. The black and red lines represent the HCO$^+$ and H$^{13}$CO$^+$ 1-0 lines, respectively. The black and red dash-dotted lines indicate the rms noise levels at the velocity resolution in the HCO$^+$ and H$^{13}$CO$^+$ 1-0 spectra. In the bottom panel, the distributions of the 15-22 and 24-27 km s$^{-1}$ components of the H$^{13}$CO$^+$ line is indicated by the brown and magenta contours, respectively. The SiO 2-1 emission is shown by the green contours. The H41$\alpha$ line is shown by the gray-scale image. The contour levels start at $3\sigma$ in steps of $1\sigma$ of the velocity-integrated intensities for the 24-27 km s$^{-1}$ H$^{13}$CO$^+$ component and the SiO 2-1 line. The contour levels for the 15-22 km s$^{-1}$ H$^{13}$CO$^+$ component start at $5\sigma$ in steps of $1\sigma$. The black dashed circle is the coverage of the spectra shown in the top panel.}\label{fig:BGPS3118_HCO+_H41a}
\end{figure}

The HCO$^+$ and H$^{13}$CO$^+$ 1-0 spectra are shown in the top panel of Figure \ref{fig:BGPS3118_HCO+_H41a}. The profiles of these two lines are similar. The line emissions are mainly in the velocity range between 13 and 22 km s$^{-1}$ with a tail in the range of 22-27 km s$^{-1}$. The spatial distributions of the H$^{13}$CO$^+$ 1-0 emission in the velocity ranges of 15-22 and 24-27 km s$^{-1}$ are shown in the bottom panel of Figure \ref{fig:BGPS3118_HCO+_H41a}. The 15-22 km s$^{-1}$ H$^{13}$CO$^+$ component is bright and spreads widely from the north side to the south side of observational field. The 24-27 km s$^{-1}$ component is much weaker. It is only distributed near the position at offsets (40, -15). The distributions of the SiO 2-1 and H41$\alpha$ lines are also presented. The H41$\alpha$ emission is bright from the southeast to the northwest. The brightest point is in the southeast. In the northeast and southwest of observational field, the H41$\alpha$ emission also exists but is relatively weak. The SiO 2-1 emission is mainly distributed in the southwest. A small part of the SiO flux is distributed elsewhere. According to the positions of BGPS 3110 and 3118, the shocked gas indicated by the SiO 2-1 line in the southwest is likely the extension of the shocked gas in BGPS 3110. %If so, the shock in BGPS 3118 is also caused by the expansion of the H II region.

\subsection{Spectra of the SiO 2-1 and HC$_3$N 10-9 lines}

%\subsubsection{Comparison with the HC$_3$N line}

\begin{figure*}
  \centering
  \includegraphics[scale=0.4]{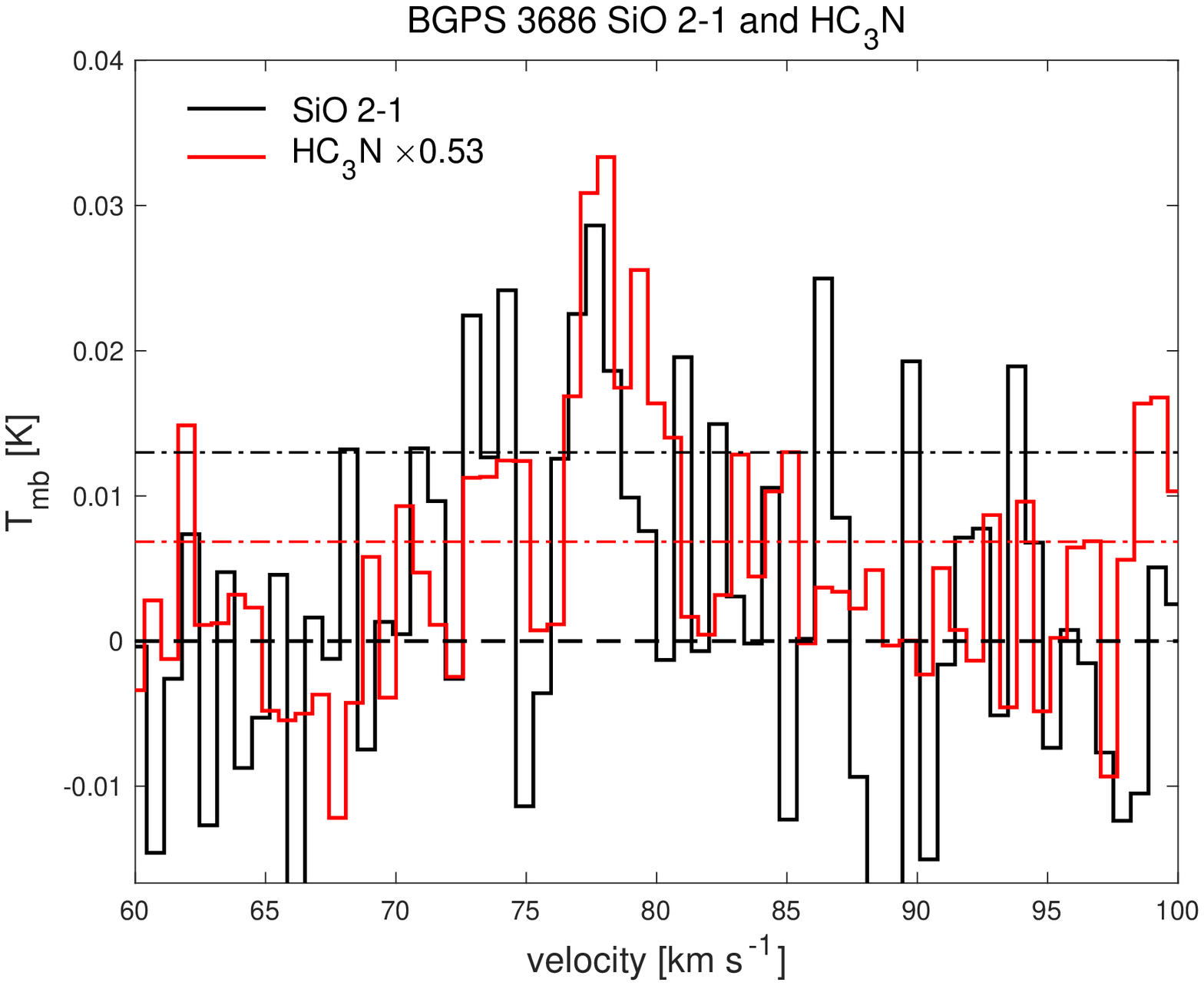}
  \includegraphics[scale=0.4]{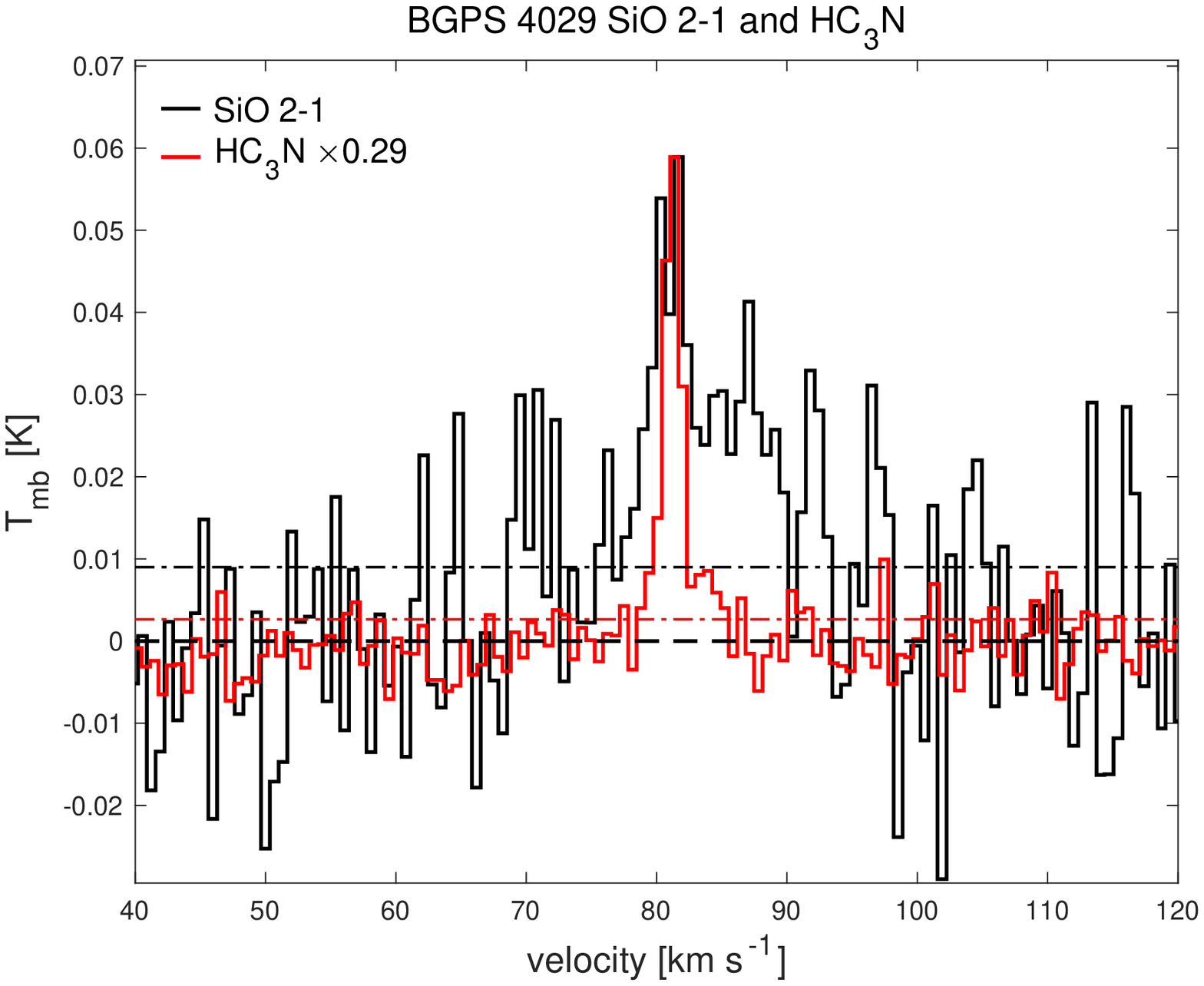}
  \includegraphics[scale=0.4]{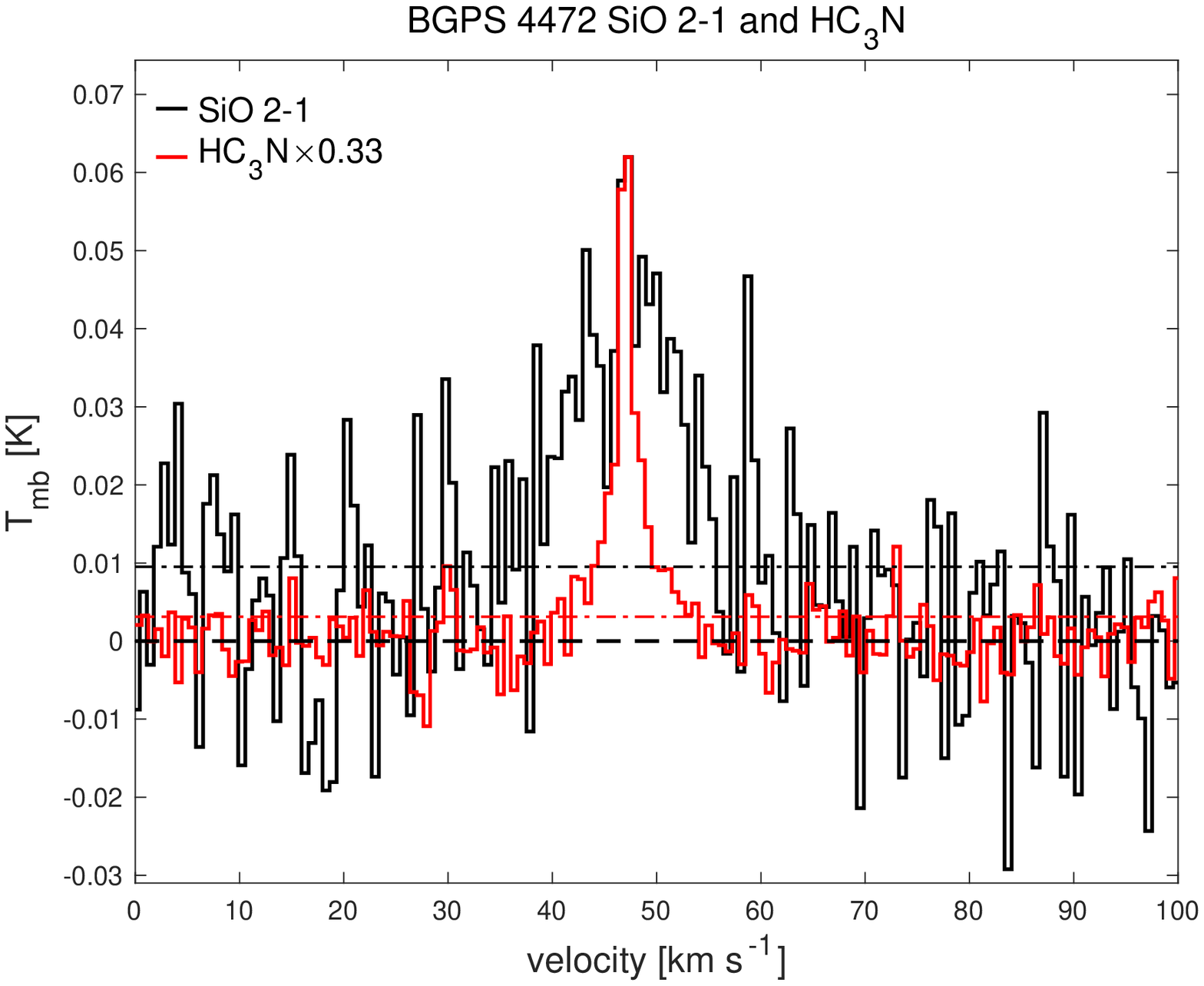}
  \includegraphics[scale=0.4]{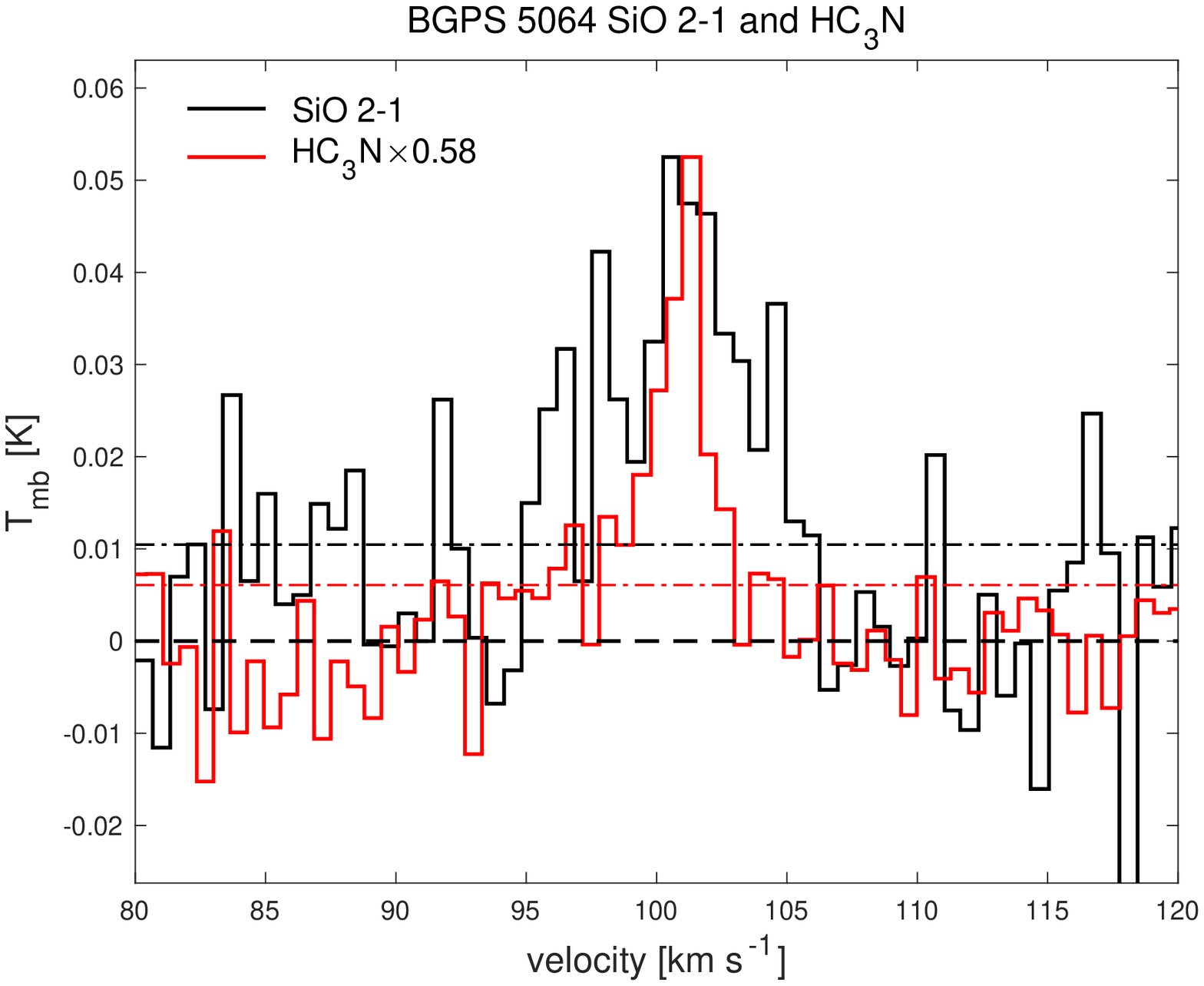}
  \includegraphics[scale=0.4]{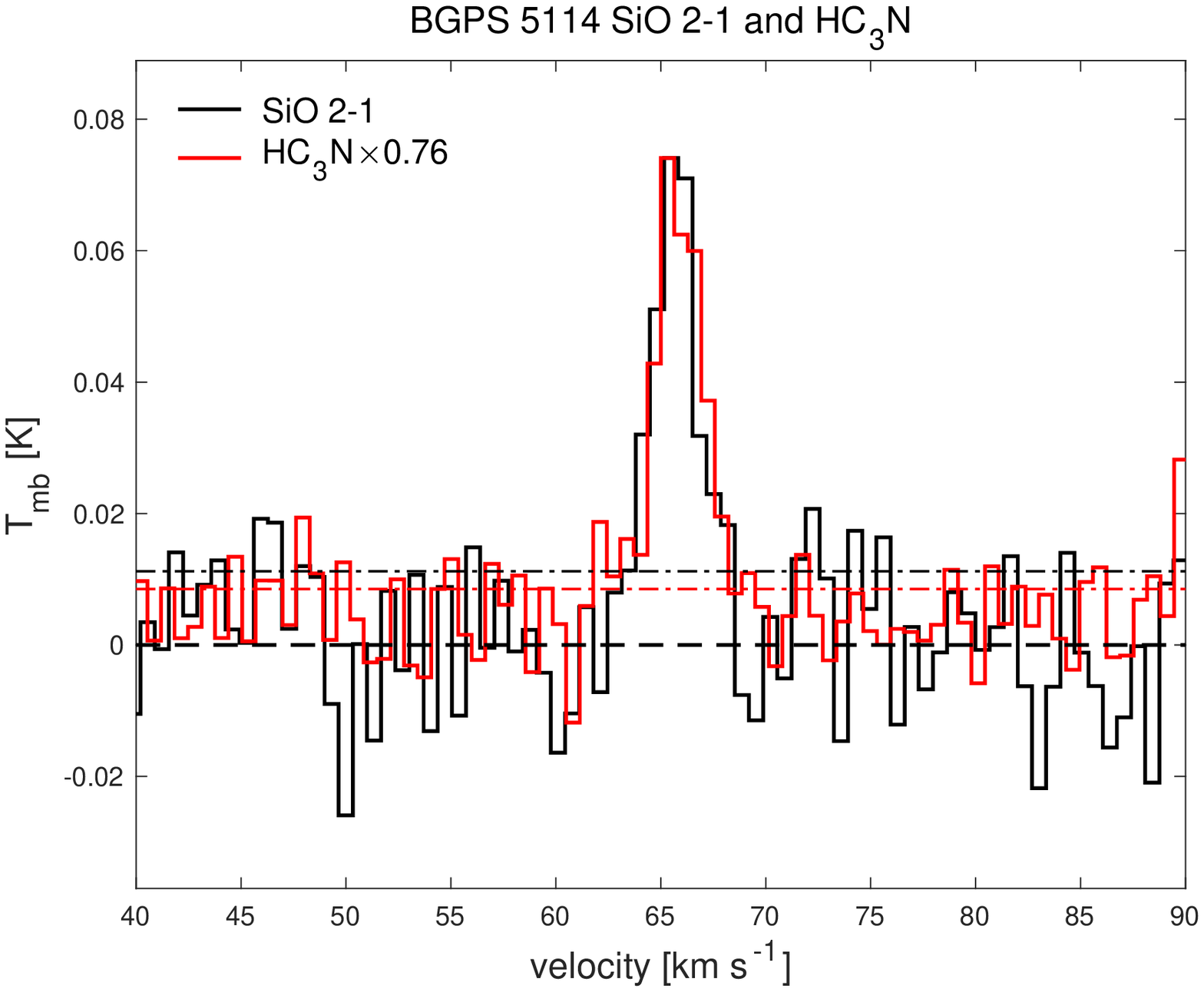}
  \includegraphics[scale=0.4]{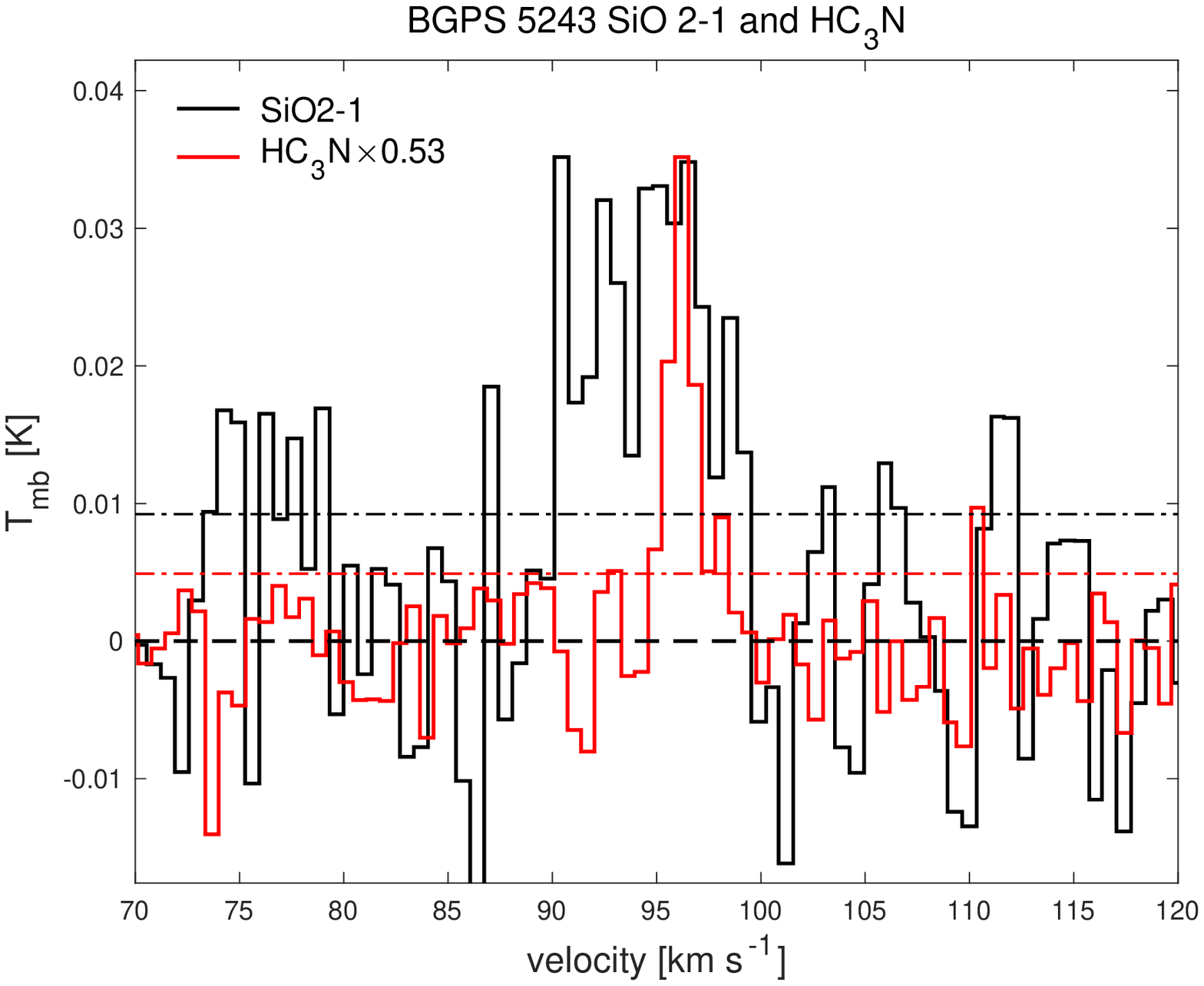}
  \caption{The SiO 2-1 and HC$_3$N spectra toward the SCCs without H41$\alpha$ detections are plotted. The black and red dash-dotted lines indicate the rms noise levels at the velocity resolution in the SiO 2-1 and HC$_3$N spectra, respectively.}\label{fig:SiO2-1_HC3N_com}
\end{figure*}

\begin{figure*}
  \centering
  \includegraphics[scale=0.4]{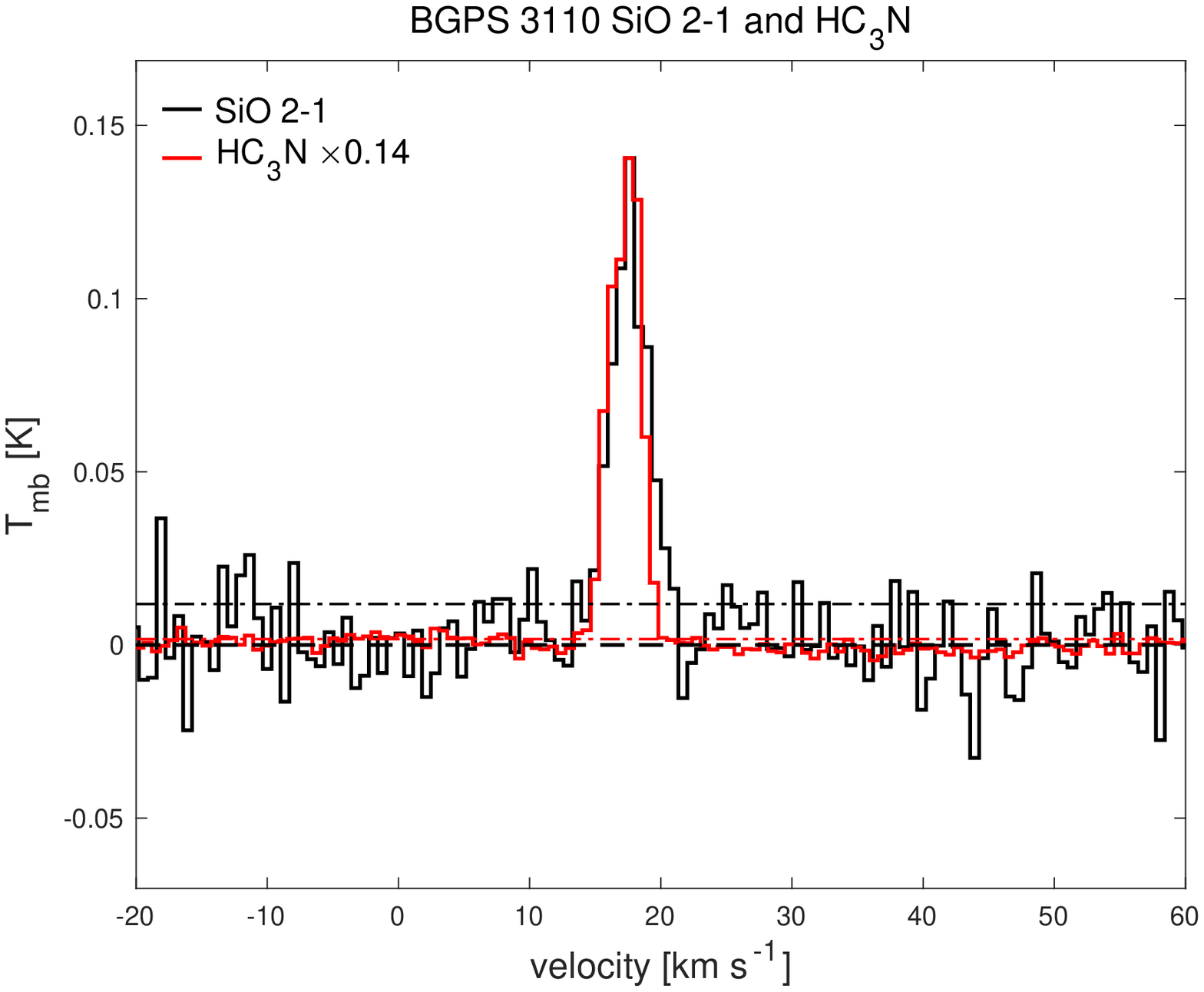}
  \includegraphics[scale=0.4]{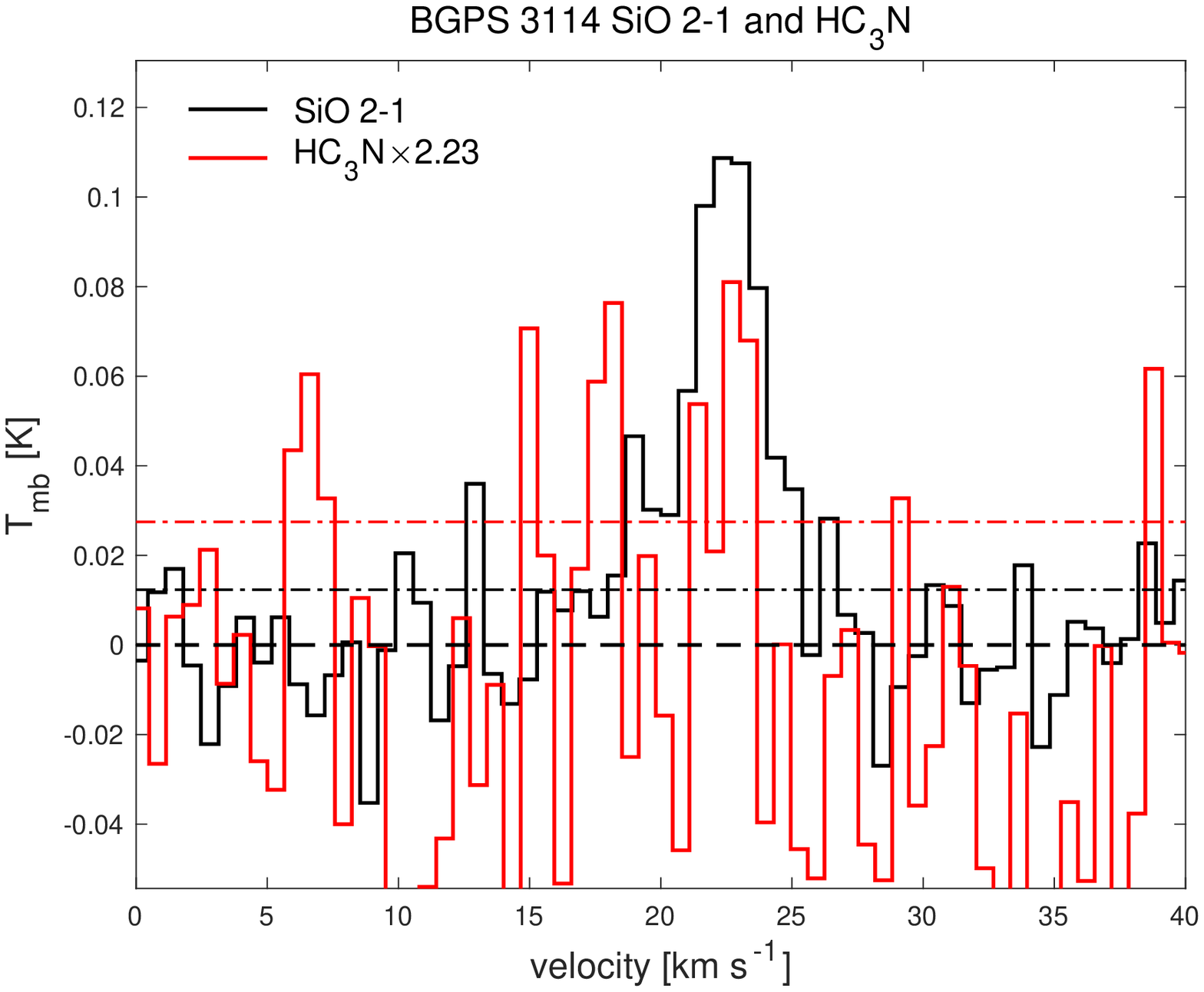}
  \includegraphics[scale=0.4]{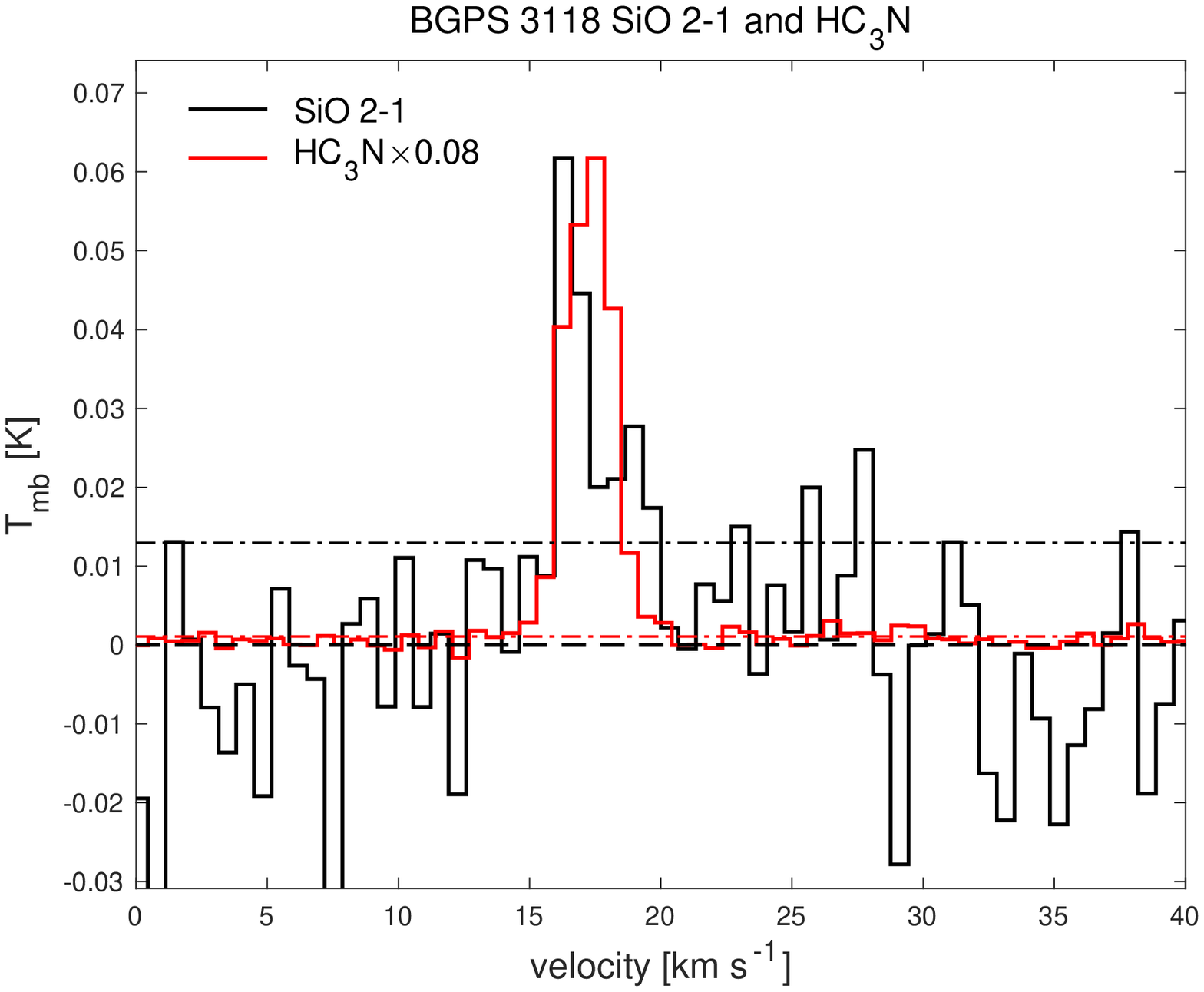}
  \caption{The SiO 2-1 and HC$_3$N spectra toward the SCCs with H41$\alpha$ detections are plotted. The black and red dash-dotted lines indicate the rms noise levels at the velocity resolution in the SiO 2-1 and HC$_3$N spectra, respectively.}\label{fig:SiO2-1_HC3N_com2}
\end{figure*}

The different origins of the SiO 2-1 and HC$_3$N 10-9 lines could lead to a distinction between the two line profiles if the shocked and dense gases have significantly different velocity distributions in the line of sight.  The profiles of the SiO 2-1 and HC$_3$N 10-9 lines toward the centers of the observational fields for the SCCs with and without H41$\alpha$ detections are plotted in Figure \ref{fig:SiO2-1_HC3N_com} and \ref{fig:SiO2-1_HC3N_com2}, respectively. The spatial coverages of the spectra are the same as those of the H$^{13}$CO$^+$ and HCO$^+$ 1-0 spectra mentioned above.

%The central velocities of the SiO 2-1 and HC$_3$N 10-9 lines are always close to each other in each SCCs. The SiO 2-1 line is much wider than the HC$_3$N 10-9 line in BGPS 4029, 4472, 5064 and 5243. \textit{In these sources, the high-velocity component of the SiO 2-1 emission is clear. The HC$_3$N 10-9 line also have line wings in BGPS 4029, 4472 and 5064. In the other sources,} the difference between the SiO 2-1 and HC$_3$N 10-9 line widths is not significant. Additionally, in BGPS 3110 and 3114 neighbouring M17 H II region, the SiO 2-1 and HC$_3$N 10-9 line profiles are very alike. %The peak value ratios of the SiO 2-1 line to the HC$_3$N line for different sources are marked in the corresponding panels.

The central velocities of the SiO 2-1 and HC$_3$N 10-9 lines are always close to each other in each SCCs. However, the widths of these two lines are different in some sources. The SiO 2-1 line is much wider than the HC$_3$N 10-9 line in BGPS 4029, 4472, 5064 and 5243. In the other sources, the difference between the SiO 2-1 and HC$_3$N 10-9 line widths is not significant. Additionally, in BGPS 3110 and 3114 neighbouring M17 H II region, the SiO 2-1 line widths are slightly broader than those of HC$_3$N 10-9 line.

%Additionally, in BGPS 3110 and 3114 neighbouring M17 H II region, the SiO 2-1 and HC$_3$N 10-9 line profiles are very alike. %The peak value ratios of the SiO 2-1 line to the HC$_3$N line for different sources are marked in the corresponding panels.

%The spatial distributions of the SiO 2-1 and HC$_3$N lines presented above also show that the spreading area of the SiO 2-1 line is much larger than that of the HC$_3$ line in these SCCs.

\section{discussions} \label{sec:discussion}

By analyzing the distributions and the spectra of the molecular and hydrogen recombination lines, the origins of the shocks in SCCs are studied. The origins of the shocks in the sources without H41$\alpha$ line detections are discussed in Section \ref{sec:early_stage}. Those in the sources detected in the H41$\alpha$ line including BGPS 3110, 3114, and 3118 are analyzed in Section \ref{sec:later_scc}.

\subsection{Shocks originated from early-stage star formation} \label{sec:early_stage}

\subsubsection{The origins of the shocks in BGPS 4029, 4472, 5064}

In BGPS 4029, 4472 and 5064, the HCO$^+$ 1-0 spectra all show high-velocity components. The distributions of the blue- and red-shifted HCO$^+$ components are distributed in the opposite sides of the location of the HC$_3$N 10-9 emission which possible indicate the positions of protostellar cores. The shocked gas shown by the SiO emission spreads around the location of the HC$_3$N 10-9 emission. In addition, as is presented in Figure \ref{fig:SiO2-1_HC3N_com}, the SiO 2-1 line is clearly broader than the HC$_3$N 10-9 line in the three sources. Especially, the SiO 2-1 and 3-2 line widths are wider than 12 km s$^{-1}$ in BGPS 4029 and 4472. Such broad widths of the SiO lines are thought to be caused by powerful feedbacks from protostars \citep{ngu13,cse16}. These features are the indicators of protostellar outflows. So we suggest that the shocked gas in these sources are originated from protostellar activities. In addition, the bipolar outflows in BGPS 4029 found by the current work were also suggested in \citet{svo19}. In those observations, the CO 2-1 line was used to found outflows, and compact points in 1.3 mm continuum maps were used to point out the positions of protostellar cores. The directions of bipolar outflow presented in \citet{svo19} are indicated in Figure \ref{fig:BGPS4029_HCO+_blrd}.

\subsubsection{The origins of the shocks in BGPS 3686 and 5114}

The SiO 2-1 and 3-2 line widths in BGPS 3686 are broad. However, due to the low signal-to-noise ratios of the SiO lines, the reliability of the measured widths is diminished. The SiO line widths in BGPS 5114 are narrower than those in BGPS 4029, 4472, and 5064. In addition, they are not significantly broader than the HC$_3$N line width which indicates the properties of the clump core. This implies that the shocks in BGPS 5114 are not fast shocks.

In our observations, no obvious outflow features can be found in BGPS 3686 and 5114. The central regions of the observational fields toward BGPS 3686 and 5114 were also observed in \citet{svo19}. Outflow features are found in these two clumps. The directions of bipolar outflows detected in CO 2-1 emission and the dense cores found in the 1.3 mm continuum are presented in Figure \ref{fig:BGPS3686_HCO+_HC3N_blrd} and \ref{fig:BGPS5114_HCO+_blrd}.

In BGPS 3686, the dense core called G22695 S1 is suggested to be associated with a bipolar outflow. This core and corresponding outflows are located near the shocked gas indicated by the SiO 2-1 emission in our observations. This suggests that the shocked gas is caused by protostellar activities in BGPS 3686. The distributions of the CO 2-1 high-velocity components are presented in Figure 7 in \citet{svo19}. There are multiple pairs of blue- and red-shifted gas components located close to each others in BGPS 3686. Because the current observations with $\sim28''$ beam size can not resolve the spatial distributions of these high-velocity components, the outflows can not be identified in the current observations alone.

%For BGPS 3686, some compact sources with bipolar outflows shown by the CO 2-1 line were detected. Their positions are now covered by the HC$_3$N 10-9 emission in our observations. This implies that the gas indicated by the HC$_3$N 10-9 line is influenced by the protostellar activity. And the shocked gas indicated by the SiO 1-0 and 2-1 lines in our previous work \citep{zhu20} could also be originated from the protostellar activity.

A compact target in BGPS 5114, G30120 S1, is associated with strong CO bipolar outflows \citep{svo19}. In the bottom panel of Figure \ref{fig:BGPS5114_HCO+_blrd}, it is presented that this compact source is near the shocked component shown by the SiO 2-1 emission. In addition, a pair of blue- and red-shifted high-velocity HCO$^+$ components near G30120 S1 seem to show a bipolar outflow feature although they are not very clear because of the confusion with other high-velocity gas components. So we suggest that the shocked gas in BGPS 5114 is also caused by the protostellar outflows as that in BGPS 4029. In addition, there are some other high-velocity components shown in the CO 2-1 line in BGPS 5114 \citep{svo19}. Although they may be not related with the bipolar outflows associated with G30120 S1, the spreading areas of these high-velocity components seem to be also overlapped by the distribution region of the SiO 2-1 emission in our observation. These high-velocity components could also be related with the origin of the shocked gas. However, these detailed structures in BGPS 5114 can not be resolved in our observations because of the relatively big beam size.

It is noticed that the peaks of HC$_3$N line emission are a little shifted with respect to the 1.3 mm continuum peaks associated with outflows in \citet{svo19} in BGPS 3686, 4029, and 5114. Two mechanisms could lead to this. First, the beam size of IRAM 30m is much larger than that of ALMA observations. Then dilution effect is important especially when there are multiple dense cores and filamentary structures in these SCCs. Second, the HC$_3$N line traces gas with high volume density, but 1.3 mm continuum emission traces column density. This difference can cause the shift \citep{wil03}. % In addition, because the locations of dense cores associated with outflows are within or close to the spreading areas of HC$_3$N emission in BGPS 3686, 4029, and 5114, we still suggest that the HC$_3$N line can trace dense cores in clumps.}

%\textbf{It is noticed that the peaks of HC$_3$N line emission are a little shifted with respect to the 1.3 mm continuum peaks associated with outflows in \citet{svo19} in BGPS 3686, 4029, and 5114. Several mechanisms could lead to this. First, the beam size of IRAM 30m is much larger than that of ALMA observations. Then dilution effect is important especially when there are multiple dense cores and filamentary structures in these SCCs. Second, the HC$_3$N line trace gas with high volume density, but 1.3 mm continuum emission which indicate the dense cores traces column density. Then this shift is not surprised \citep{wil03}. In addition, because the locations of dense cores associated with outflows are within or close to the spreading areas of HC$_3$N emission in BGPS 3686, 4029, and 5114, we still suggest that the HC$_3$N line can trace dense cores in clumps.}

\subsubsection{The origin of the shock in BGPS 5243}

%In BGPS 3686 and 5243, it is difficult to distinguish the high-velocity HCO$^+$ components that indicate the outflowing gas from the violent ambient gas. So the origins of the shocks in these sources are not easy to conclude. Especially for the shocked gas in BGPS 3686, its distribution is not very clear. But the observation in \citet{svo19} toward BGPS 3686 showed a bipolar outflow in the CO 2-1 line. This suggests the shocked gas is originated from protostellar activities.
The shocked gas indicated by the SiO 2-1 emission in BGPS 5243 is concentrated near the compact region of the clump shown by the HC$_3$N 10-9 line. The central velocity and spatial distribution of the SiO 2-1 line show that the shocked gas should be related with the 97 km s$^{-1}$ clump. The widths of SiO 2-1 and 3-2 lines are 7.1 and 5.4 km s$^{-1}$. According to these FWHMs, it is difficult to judge whether the shock in BGPS 5243 is fast shock \citep{jim10,dua14}. Additionally, the SiO 2-1 line width is wider than that of HC$_3$N 10-9  as those in BGPS 4029, 4472 and 5064. These features are more similar to the result of protostellar activity.

However, the red- and blue-shifted high-velocity HCO$^+$ components shown in Figure \ref{fig:BGPS5243_HCO+_blrd} are widespread unlike the result due to protostellar outflows. Different velocity ranges of the high-velocity components are also checked. It is difficult to distinguish the high-velocity components related with the potential outflowing gas from ambient components. If protostellar outflows actually exist in BGPS 5243, we guess that the case in BGPS 5243 may be similar to that in BGPS 3686. Multiple high-velocity components are closely located so that they cannot be distinguished in the spatial resolution of the IRAM 30m telescope. It is necessary to perform observations with a higher spatial resolution to confirm this assumption.

\subsection{Shocks originated from H II regions} \label{sec:later_scc}

BGPS 3110, 3114 and 3118 are all near M17 H II region. The gas kinetic temperatures calculated from the NH$_3$ lines provided by \citet{svo16} are also given in Tables \ref{table_property1}. As given in Table \ref{table_property1}, the kinetic temperatures of molecular gas in BGPS 3110 and 3118 are 25.17 and 23.68 K, respectively,  which are higher than those of  the sources without H41$\alpha$ detections. %{Delete this$->$} These temperatures are consistent with the conditions of the dense gas associated with H II regions \citep{zha20}.}

%The gas kinetic temperatures calculated from the NH$_3$ lines provided by \citet{svo16} are also given in Tables \ref{table_property1}. The kinetic temperatures for BGPS 3110 and 3118 with the H41$\alpha$ detections are higher than the temperatures for the other sources. The simulations of \citet{hos06} and \citet{zhu15b} show that the FUV radiation from ionizing stars can heat the neutral gas near H II regions. It is also displayed in \citet{zha20} that the dust temperatures of the starless clumps associated with H II regions are higher.

In this work, there are no evidences of outflows indicated by high-velocity components of HCO$^+$ 1-0 emission found in these sources. The SiO 2-1 line widths in BGPS 3110 and 3118 are both narrow although they are slightly wider than the widths of HC$_3$N line. In BGPS 3114,  the SiO 2-1 line width is not broad while the HC$_3$N 10-9 emission is very weak. The shocked gas and high-density gas spread over an elongated zone from the north side to the southeast side of observational field in BGPS 3110. This is not like the distributions of shocked and high-density components in those SCCs with outflow detections.

%In this work, there are no evidences of outflows indicated by high-velocity components of HCO$^+$ 1-0 emission found in these sources. The SiO 2-1 line width is not significantly broader than that of HC$_3$N 10-9 line in BGPS 3110 and 3118 as plotted in Figure \ref{fig:SiO2-1_HC3N_com2}. In BGPS 3114,  the SiO 2-1 line width is not broad while the HC$_3$N 10-9 emission is very weak. The shocked gas and high-density gas spread over an elongated zone from the north side to the southeast side of observational field in BGPS 3110. This is not like the distributions of shocked and high-density components in those SCCs with outflow detections.

In the further observations performed in December 2020 to January 2021, the whole M17 H II region is observed to study its effect on the nearby star formation environments. The more forceful evidences of the origin of the shock in BGPS 3110, 3114, and 3118 are obtained \citep{zhu23}. The collision between  ionized gas in M17 H II region and the surrounding molecular gas leads to the shocked gas distributed in the boundary region of the H II region including BGPS 3110, 3114, and 3118. So we suggest that the shocked gas in BGPS 3110, 3114, and 3118 is originated from the expansion of M17 H II region. Moreover, in \citet{zhu23}, it is found that the 1.1 mm continuum emission in BGPS 3114 is mainly contributed from free-free continuum emission from ionized gas. Since BGPS 3114 was identified as a massive clump because the 1.1 mm continuum was attributed to dust emission \citep{gin13,svo16}, the identification of this source is likely erroneous.

\subsection{Comparison between properties of SiO 2-1 and 3-2 lines}

The distributions of SiO 2-1 and 3-2 lines for the sources with and without H41$\alpha$ detections are shown in Figure \ref{fig:SiO_map1} and \ref{fig:SiO_map2}, respectively. Because the rms noise level of SiO 3-2 line is much higher than that of SiO 2-1 line, the distributions area of SiO 3-2 line are significantly smaller than those of SiO 2-1 line in every source. The SiO 2-1 and 3-2 spectra corresponding to the results in Table \ref{table_property1} are plotted in Figure \ref{fig:SiO_spectrum1} and \ref{fig:SiO_spectrum2} for the sources with and without H41$\alpha$ detections. Because of its higher rms noise level, the signal-to-noise ratio of SiO 3-2 line intensity is lower than that of SiO 2-1 line intensity in most of the sources. %will be vague if they are measured within the spreading areas of SiO 2-1 line. So different regions are chosen for the SiO 2-1 and 3-2 lines to measure the line properties in Table \ref{table_property1}.

\begin{figure}
  \centering
  \includegraphics[scale=0.4]{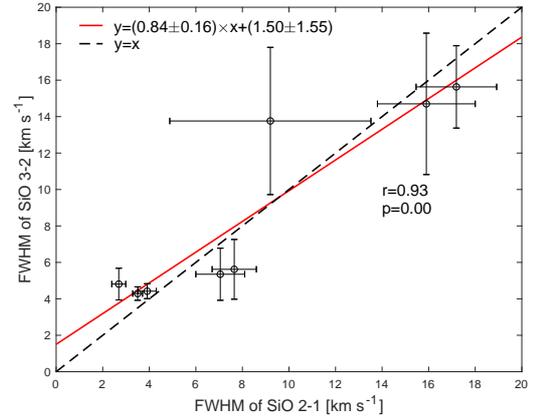}
  \caption{The comparison between the FWHMs of SiO 2-1 and 3-2 lines in SCCs except for BGPS 3118. The red line represents the linear fit of the SiO line widths. The parameters r and p is the Spearman's rank correlation coefficient and the p-value, respectively.}\label{fig:sio_fwhm_com}
\end{figure}

In Table \ref{table_property1}, the differences between central velocities of SiO 2-1 and SiO 3-2 lines for different sources are all less than 2 km s$^{-1}$. In most sources, differences in widths of the two lines are also not significant. The comparison between the SiO 2-1 and 3-2 line widths toward SCCs are plotted in Figure \ref{fig:sio_fwhm_com}. The Spearman test is conducted to estimate the correlation between the widths of the two lines. The Spearman's rank correlation coefficient of 0.93 with a corresponding p-value of 0.22 per cent indicates a strong correlation. These results suggest that the two SiO lines have same origins in these sources. %In addition, the width of SiO 2-1 line is broader in BGPS 5243 while the central velocity of SiO 2-1 line is relatively blue-shifted. This is because the shocked component with slightly blue-shifted velocity in the north side only contributes to the SiO 2-1 emission.

%As presented in Figure \ref{fig:SiO_map1} and \ref{fig:SiO_map2}. In all of the 9 sources, the distribution areas of SiO 3-2 line are significantly smaller than those of SiO 2-1 lines because of the higher rms noise level. Hence, only the distribution of SiO 2-1 line is used to trace the distribution of shocked gas in the text. In addition, the SiO 3-2 spectra are vague when they are measured within the spreading areas of SiO 2-1 line. So different regions are chosen for the SiO 2-1 and 3-2 lines to measure the line properties in Table \ref{table_property1}. The SiO 2-1 and 3-2 spectra corresponding to the results in Table \ref{table_property1} are plotted in Figure \ref{fig:SiO_spectrum1} and \ref{fig:SiO_spectrum2}.

\section{Summary} \label{sec:conclusion}

In this paper, we perform mapping observations of the SiO 2-1, 3-2, HCO$^+$ 1-0, H$^{13}$CO$^+$ 1-0, HC$_3$N 10-9, and H41$\alpha$ lines toward 9 SCCs, which were found to be associated with shocks in our previous single-dish observations \citep{zhu20}. The properties of the observed lines are shown. The spatial distributions of different gas components traced by these lines are investigated.

The origins of the shocks in these SCCs are studied. The H41$\alpha$ line is found in BGPS 3110, 3114 and 3118. The shocked gas in these 3 SCCs is likely caused by the expansion of M17 H II region. This result suggests that the existence of classical H II region should not be neglected in searching for early-stage star-forming targets. The shocks in BGPS 4029, 4472 and 5064 are probably originated from protostellar activities since the features of bipolar outflows are indicated by the HCO$^+$ line. The shocks in BGPS 3686 and 5114 are suggested to be also caused by protostellar activities after comparing the results in our observations with those in previous ALMA observations toward the two sources \citep{svo19}. However, the origin of the shock in BGPS 5243 is still unclear. We acknowledge that further studies with larger samples are needed to obtain statistical results, but our results imply that a large part of shocks in SCCs are originated from protostellar outflows. Therefore, outflow and accretion processes could have occurred in a significant portion of SCCs identified in previous works since shocks are common in these SCCs \citep{zhu20}. In addition, the shocked gas is widespread in BGPS 3110, 3114 and 3118. These SCCs are close to M17 H II region. On the contrary, the distribution of shocked gas is concentrated in the other SCCs.

%2. No evidence for cloud-cloud collision is found in the 9 SCCs. But this does not support that no shocks caused by cloud-cloud collision are formed in SCCs because the selected sample is only a small part of the SiO-detected sources in the previous observations. In addition, the shocks made by cloud-cloud collisions are possibly weaker than those caused by protostellar outflows. In our selected sources with the relatively bright SiO emission, the shocks are more likely produced by protostellar activities.  %The SiO emissions of the sources selected in the current work are relatively strong in the sample observed in the previous observations \citep{zhu20}. This could lead to the non-detections of cloud-cloud collision since the shocks made by cloud-cloud collision are possibly weaker than those caused by protostellar acitivity.

%3. BGPS 3110 and 3114 may be embedded in the dense shell of M17 H II region.

%3. The kinetic temperatures of the neutral gas in BGPS 3110 and 3118, the SCCs associated with an H II region, are obviously higher than those in the SCCs without H II regions. This is consistent with the results in \citet{zha20}, and caused by the heatings from the H II region and its ionizing massive star.

%4. The results from the current observations and our previous single-dish observations by using KVN21-m telescopes are roughly consistent. This suggests the reliability of the detections of the shocks in the sample of the previous observations.

\section*{Acknowledgements}

This work is based on the observations performed using IRAM 30m telescope. F.-Y. Z. thanks Dr. Jiang Xuejian for providing the knowledge about chemical abundances. It is supported by the National Science Foundation of China No. 12003055, and Key Research Project of Zhejiang Lab (No. 2021PE0AC03). Y. T. Y. is a member of the International Max Planck Research School (IMPRS) for Astronomy and Astrophysics at Universities of Bonn and Cologne. Y. T. Y. would like to thank China Scholarship Council (CSC) and the Max-Planck-Institut f\"{u}r Radioastronomie (MPIfR) for the financial support.

\section*{DATA AVAILABILITY STATEMENT}

The data underlying this paper will be shared on reasonable request to the corresponding author.

\clearpage

\appendix

\section{The spectra and distributions of SiO 2-1 and 3-2 lines} \label{sec:siospectrum}

\begin{figure*}
  \centering
  \includegraphics[scale=0.35]{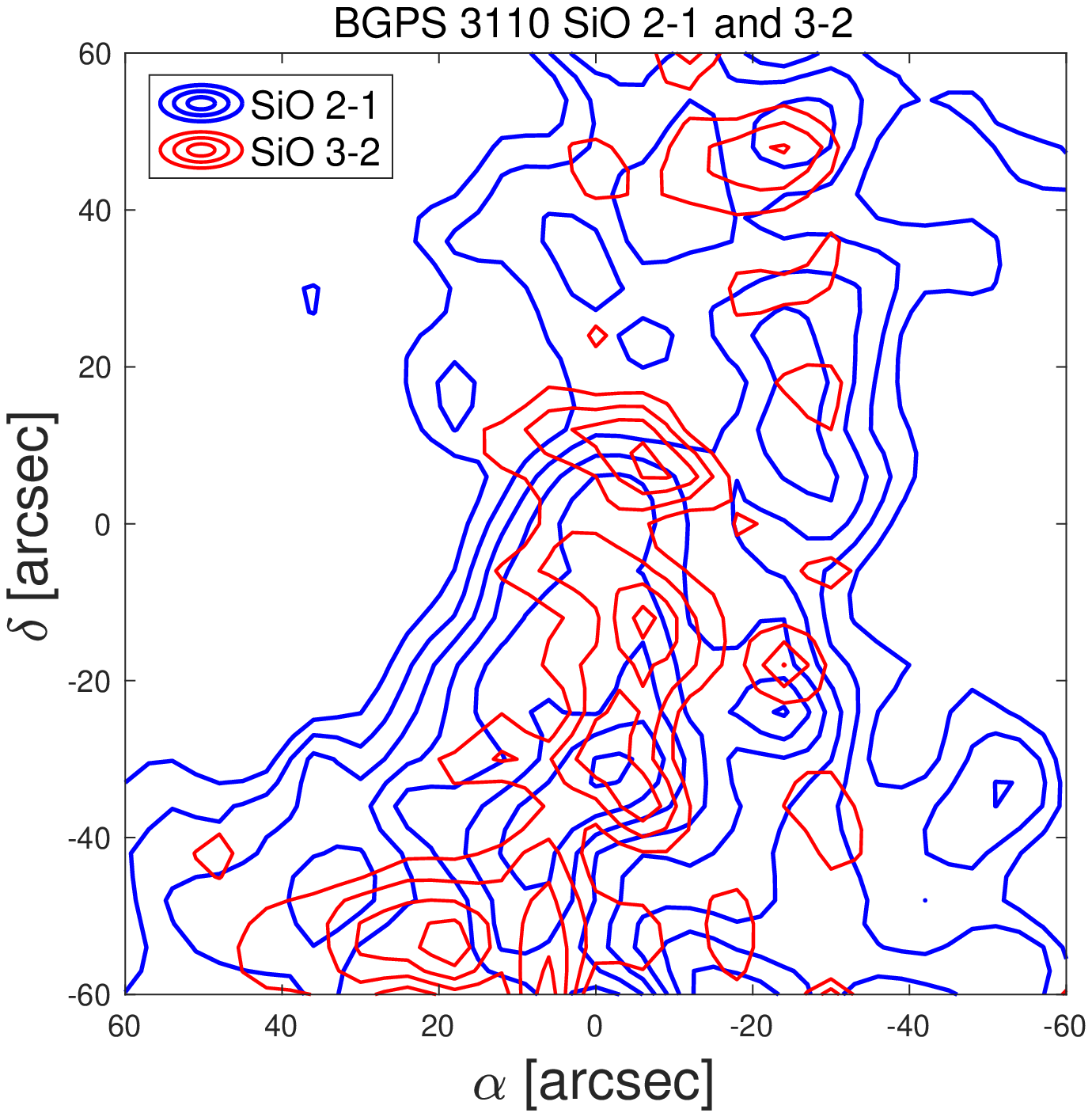}
  \includegraphics[scale=0.35]{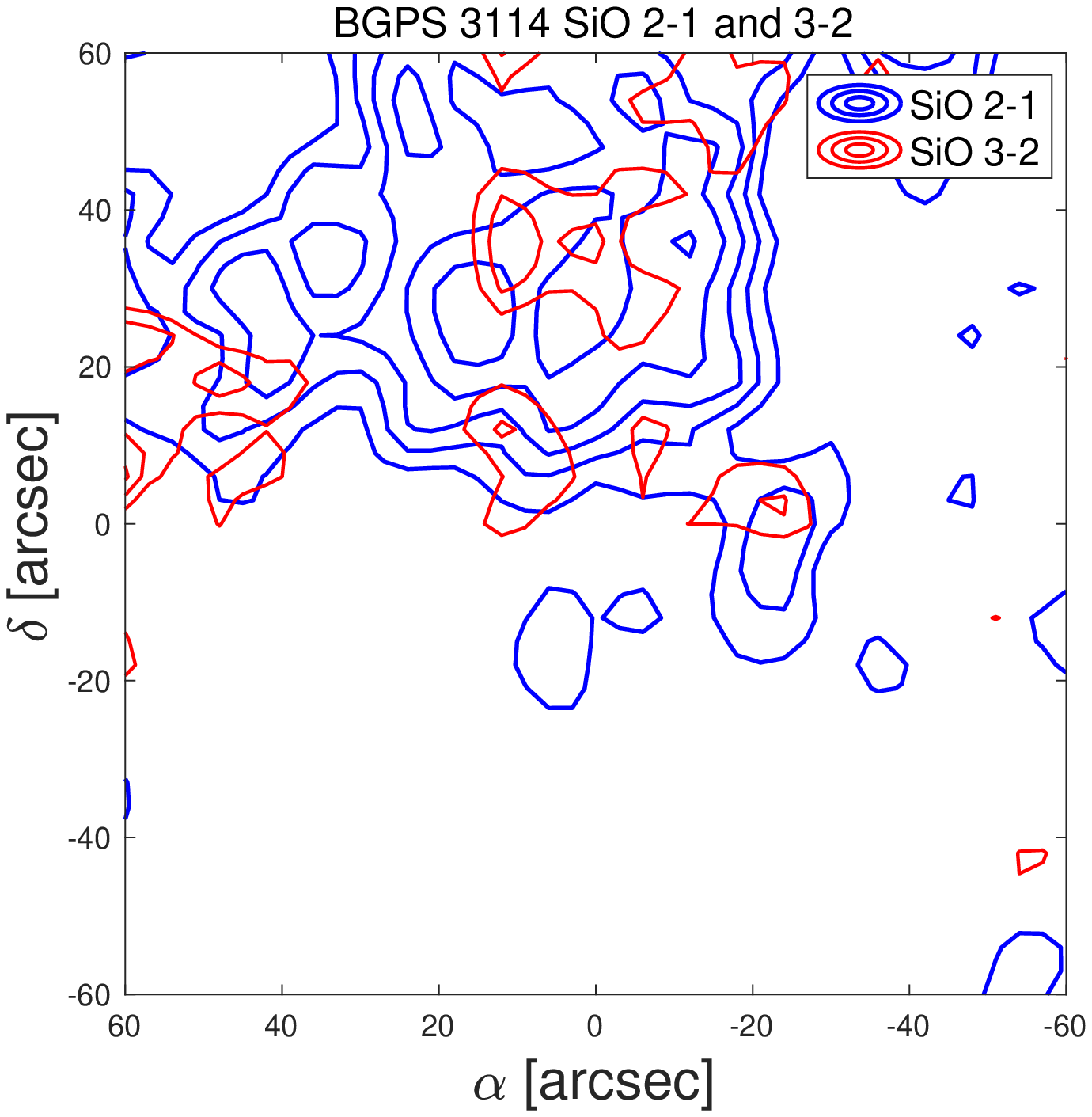}
  \includegraphics[scale=0.35]{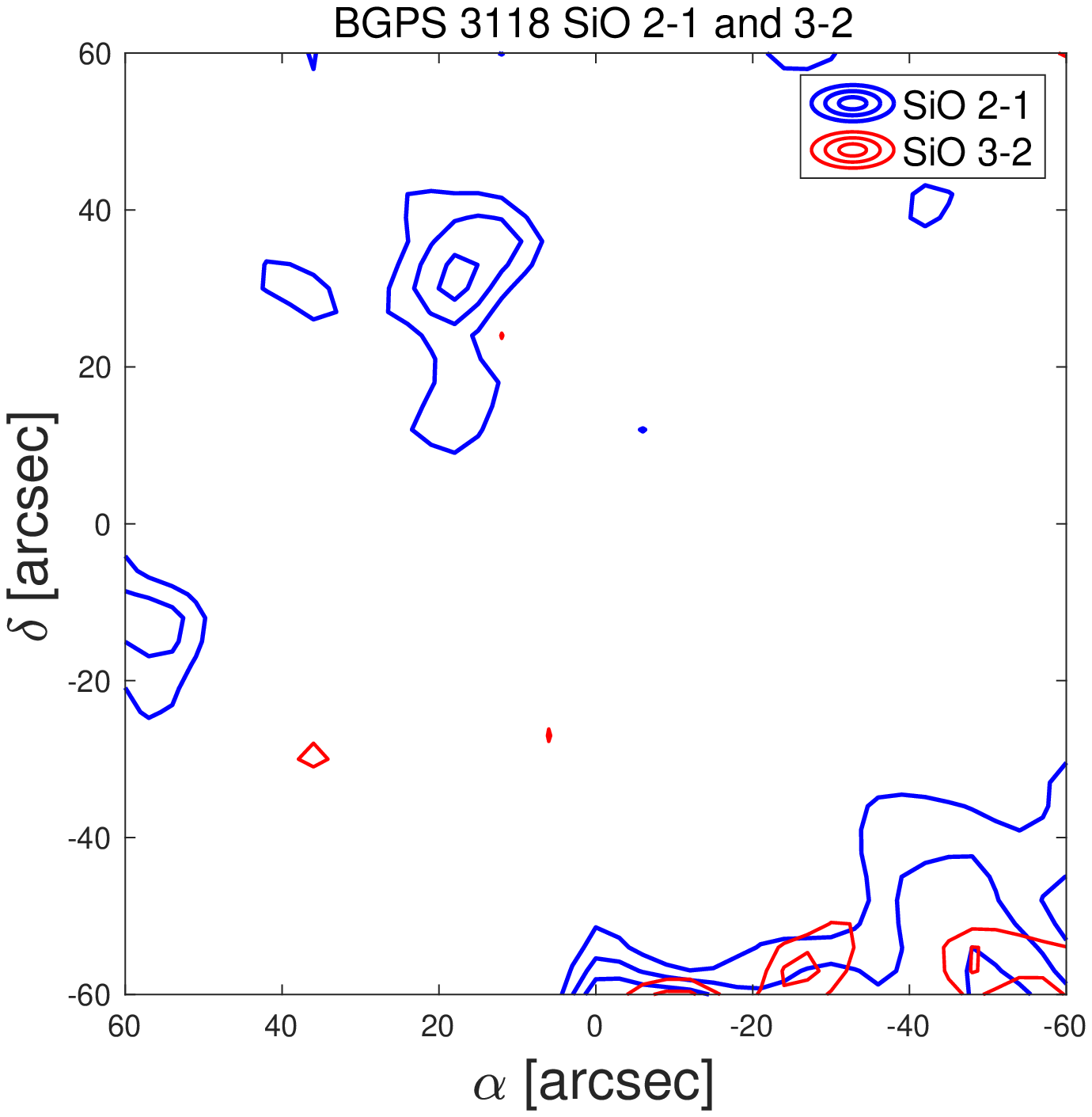}
  \caption{The distributions of SiO 2-1 and 3-2 lines toward the SCCs with H41$\alpha$ detections. The contour levels start at $3\sigma$ in steps of $3\sigma$.}\label{fig:SiO_map1}
\end{figure*}

\begin{figure*}
  \centering
  \includegraphics[scale=0.35]{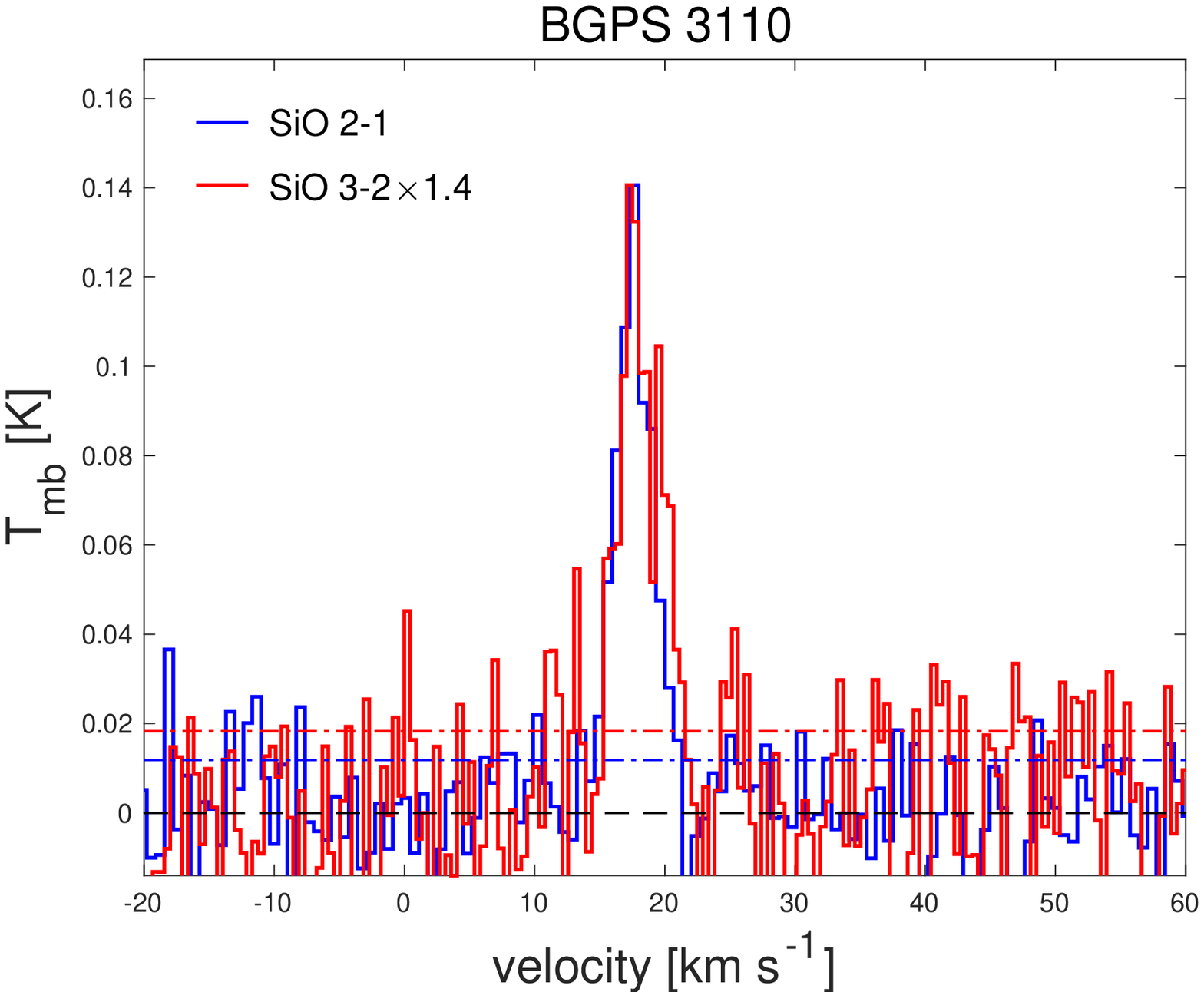}
  \includegraphics[scale=0.35]{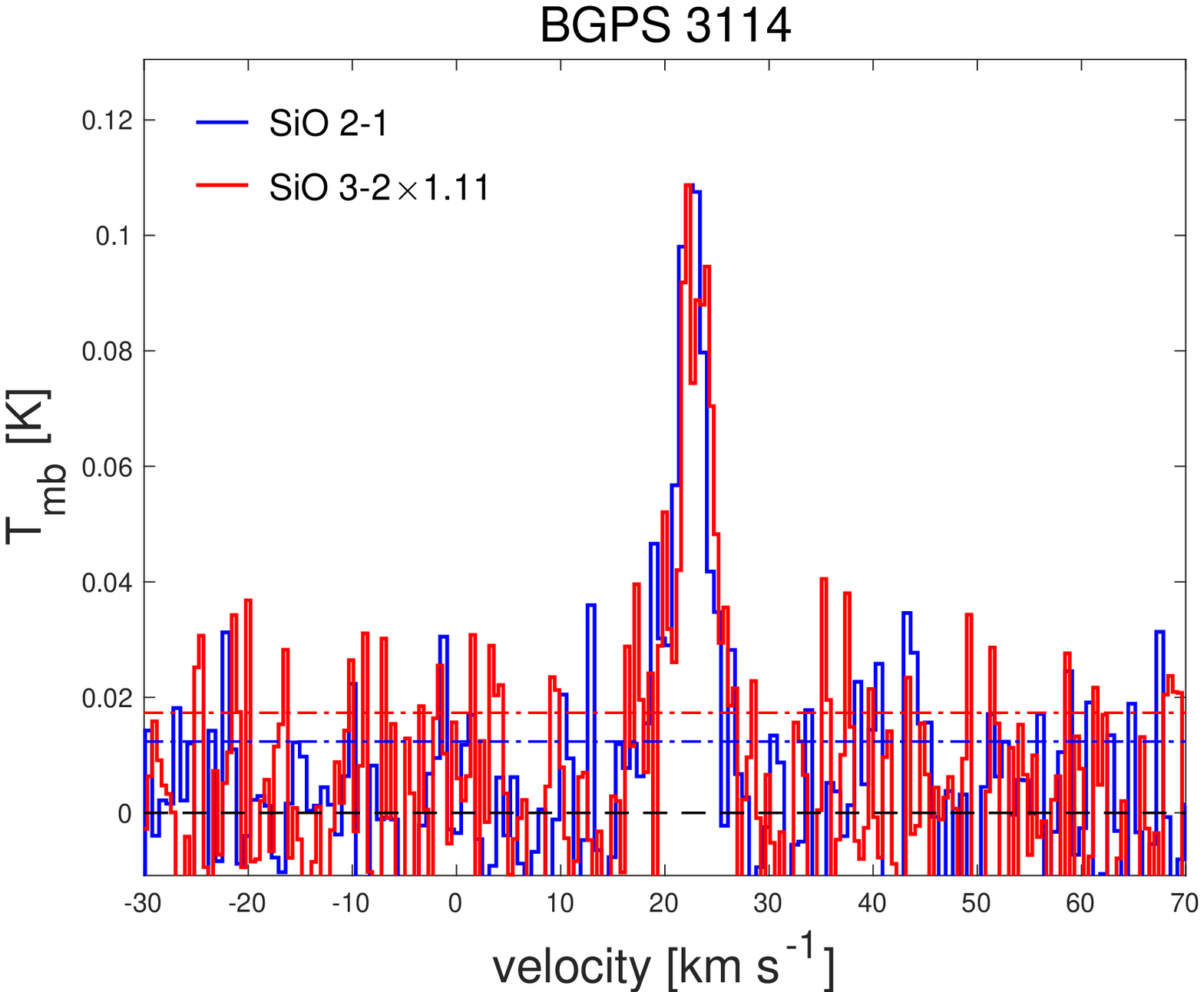}
  \includegraphics[scale=0.35]{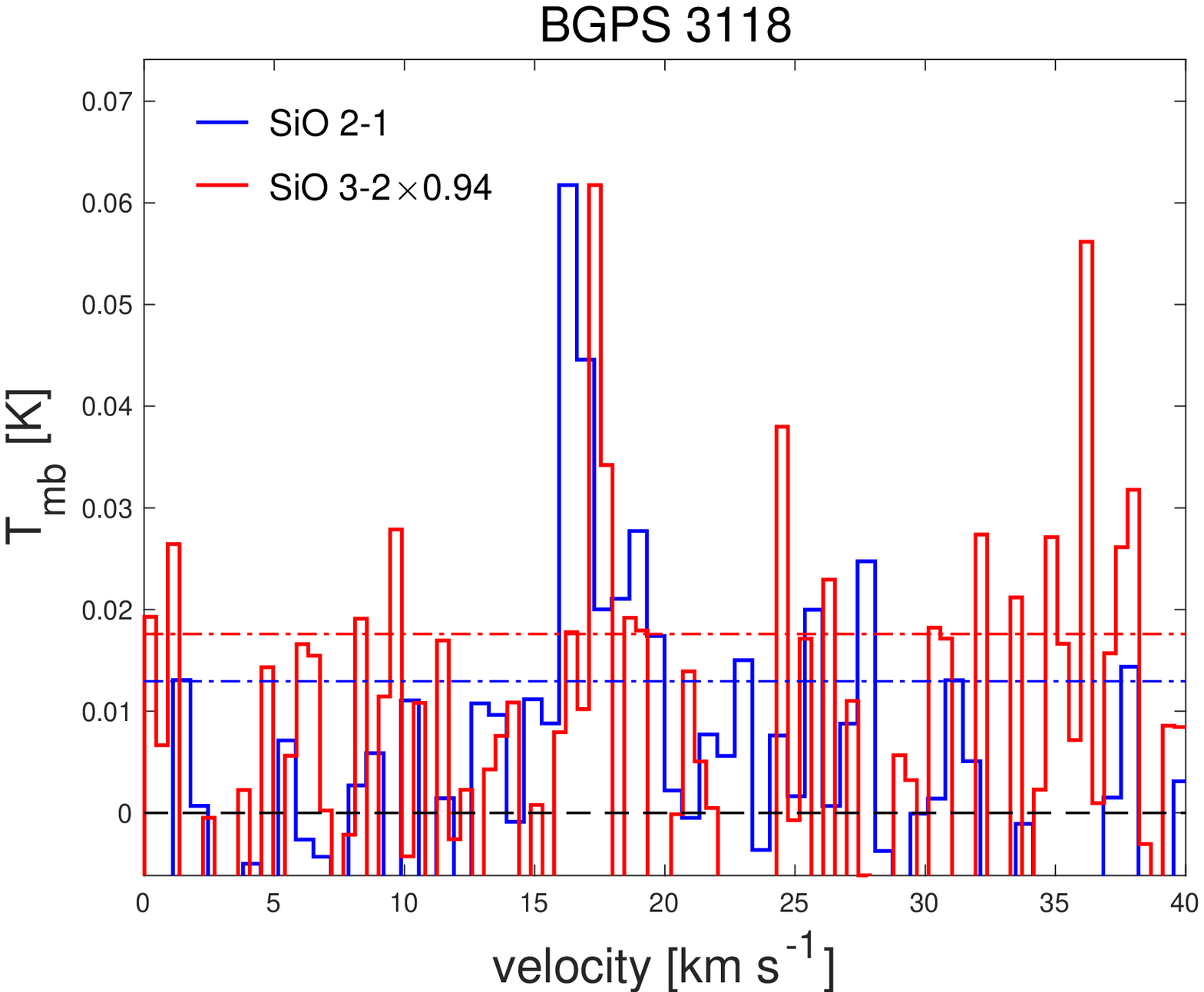}
  \caption{The SiO 2-1 and 3-2 spectra toward the SCCs with H41$\alpha$ detections. The blue and red dash-dotted horizontal lines are the rms noise levels at the velocity resolution for the SiO 2-1 and 3-2 lines, respectively.}\label{fig:SiO_spectrum1}
\end{figure*}

\begin{figure*}
  \centering
  \includegraphics[scale=0.35]{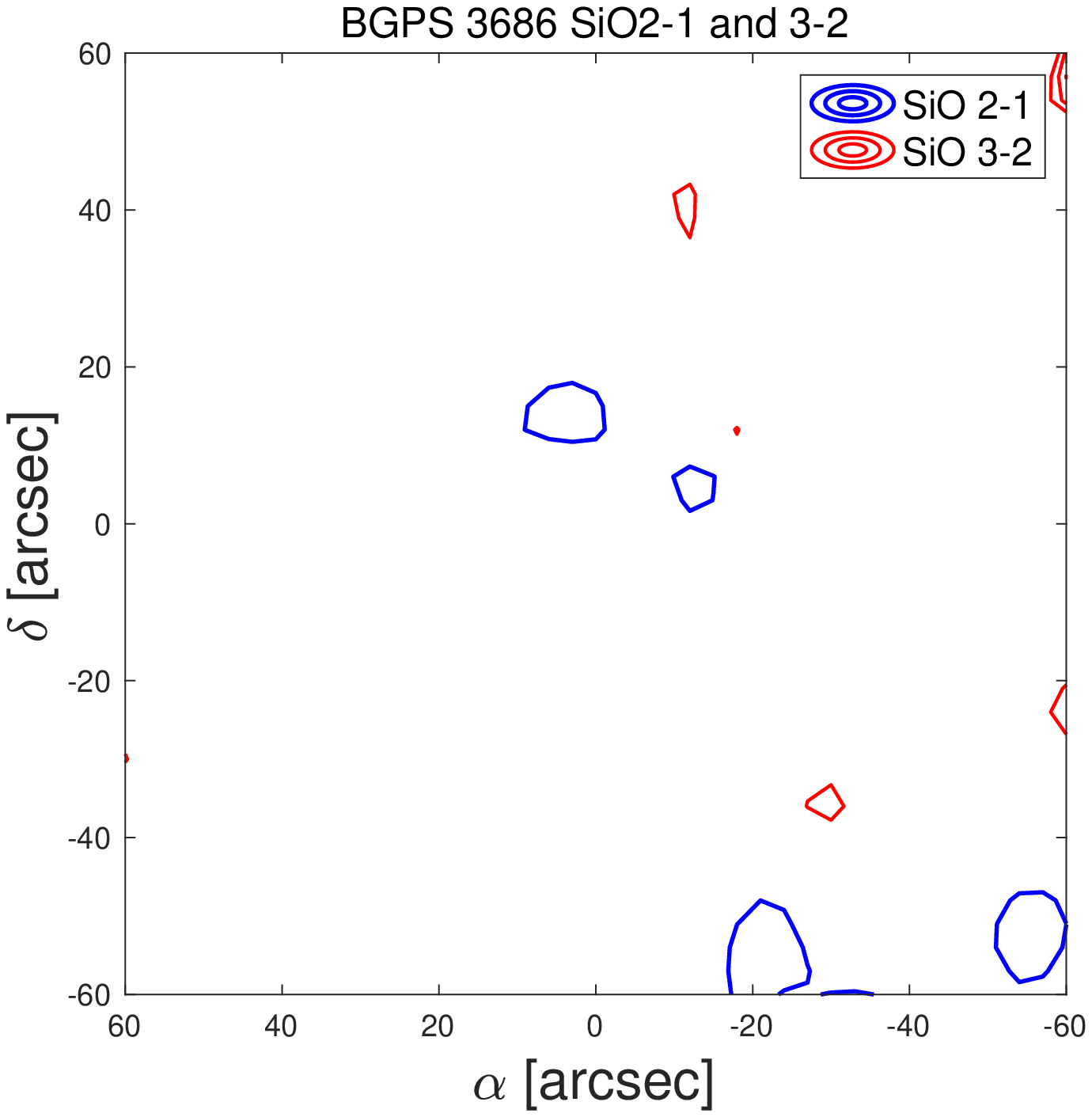}
  \includegraphics[scale=0.35]{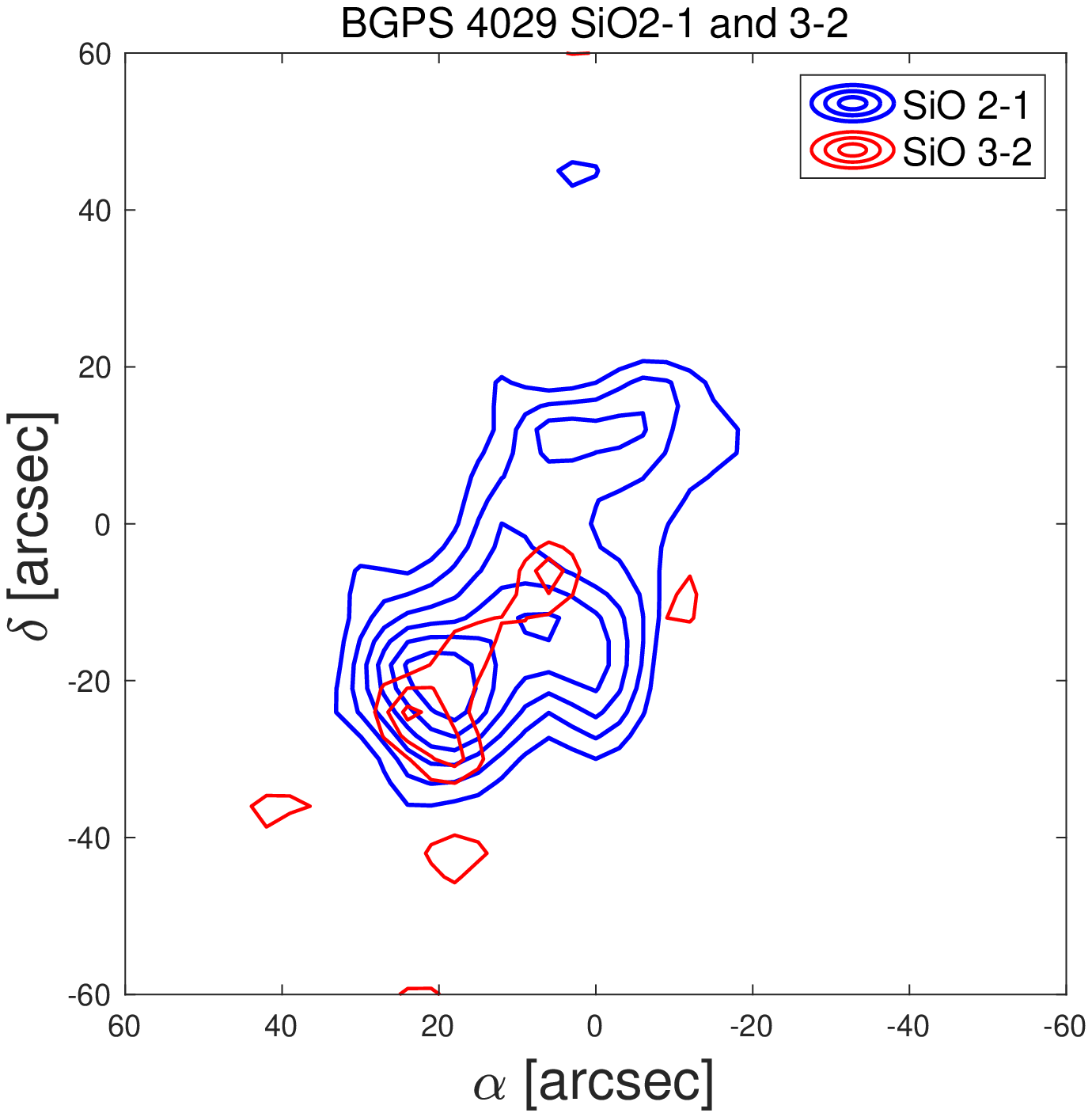}
  \includegraphics[scale=0.35]{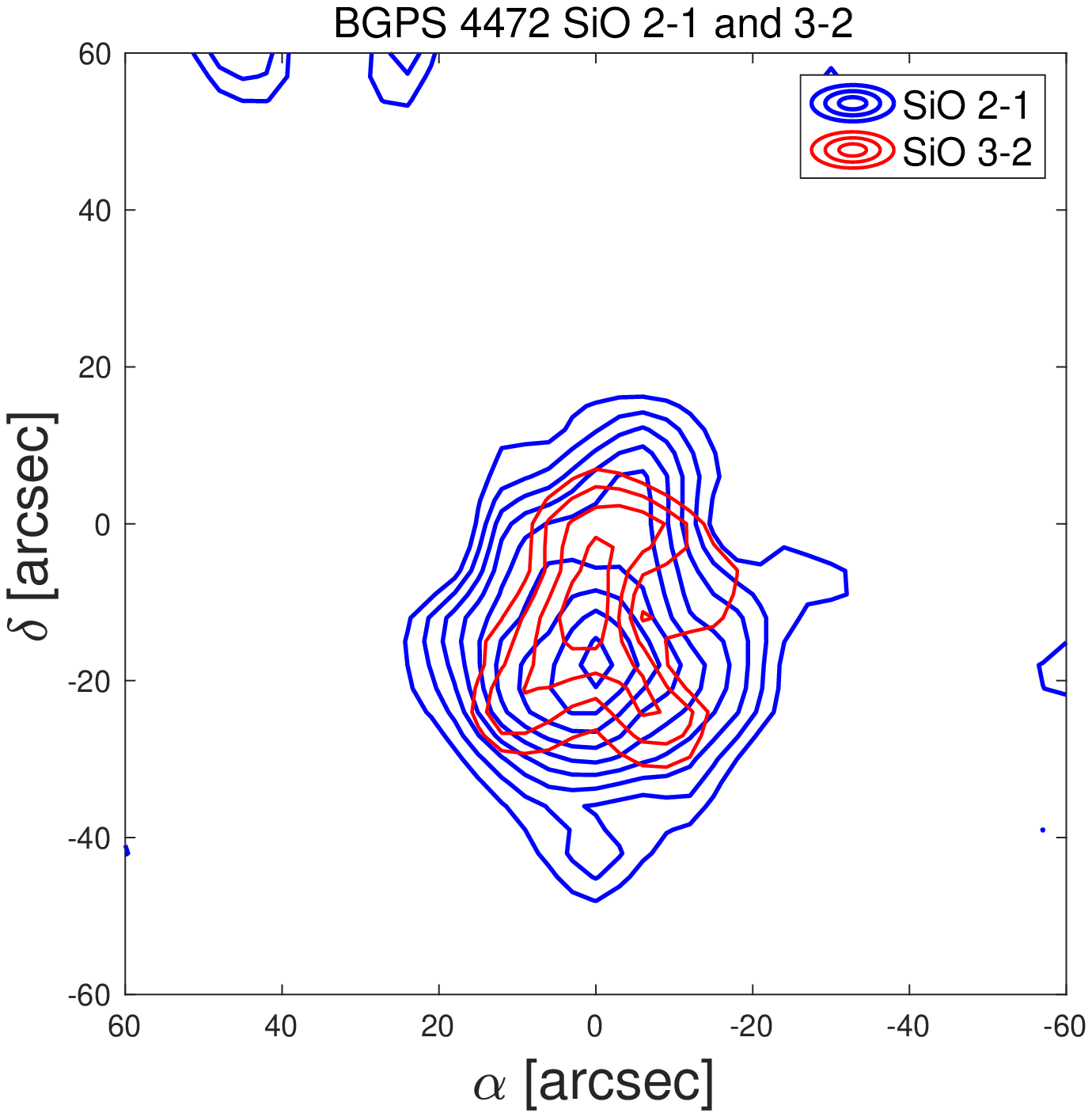}
  \includegraphics[scale=0.35]{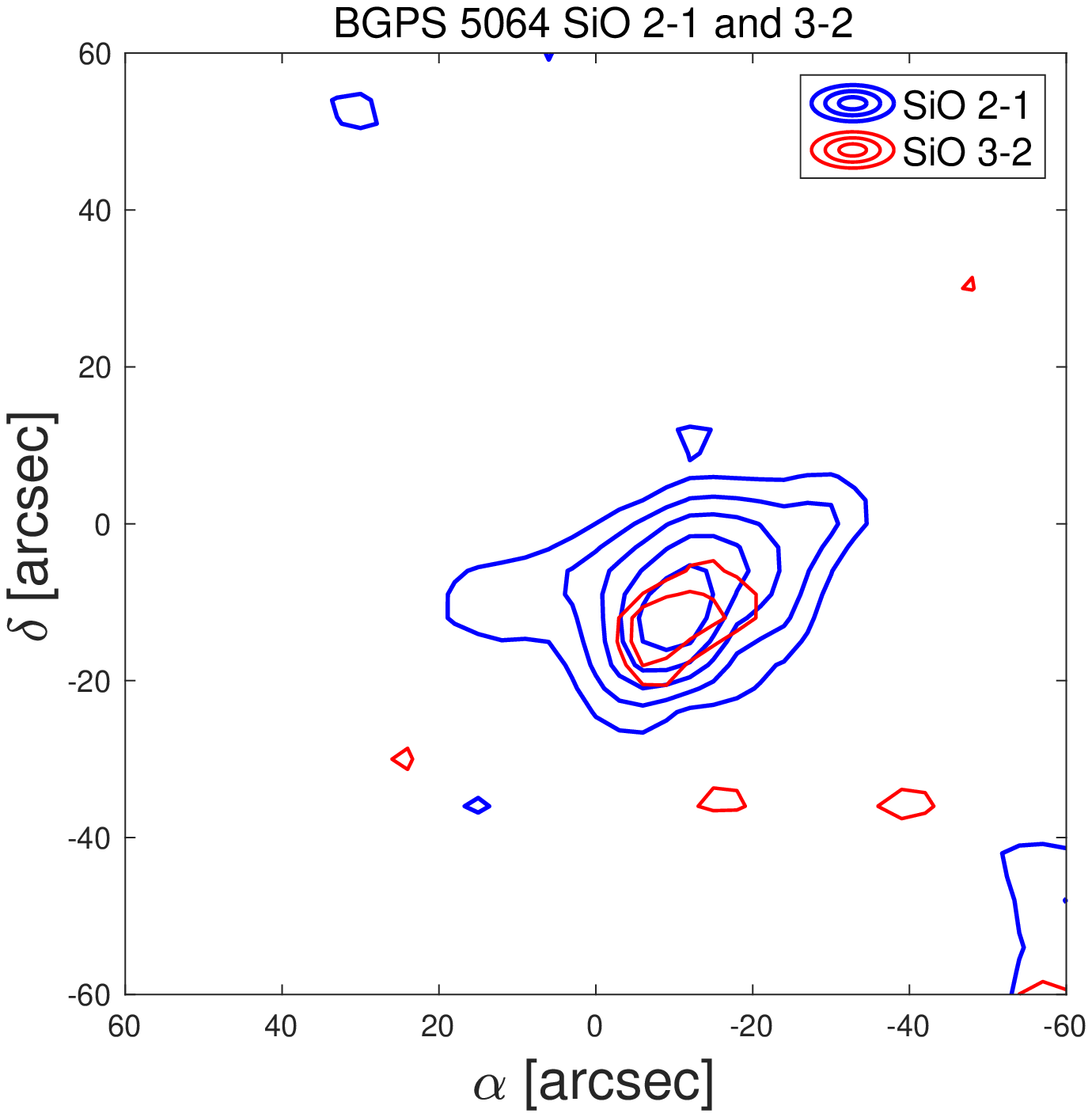}
  \includegraphics[scale=0.35]{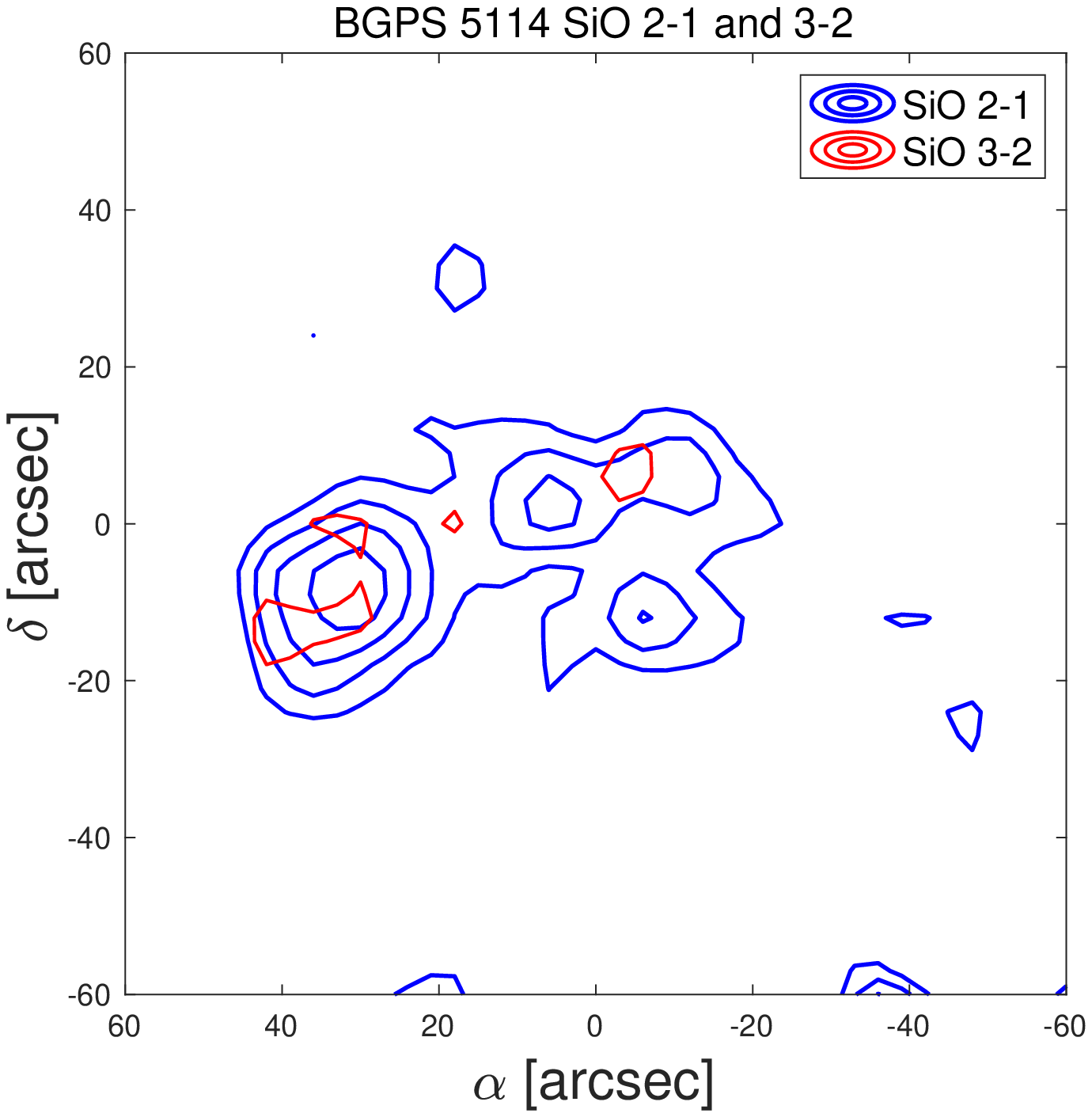}
  \includegraphics[scale=0.35]{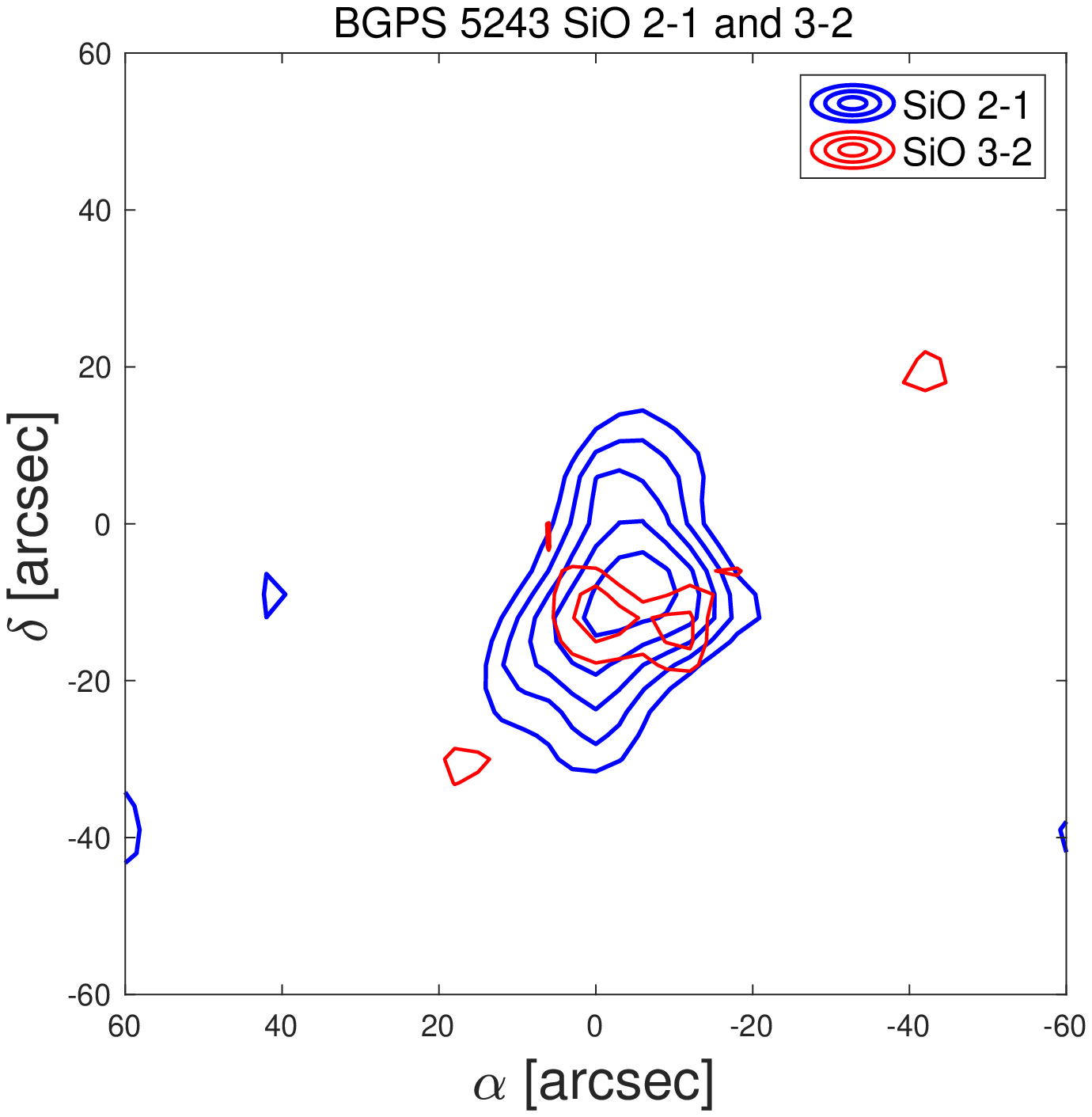}
  \caption{The distributions of SiO 2-1 and 3-2 lines toward the SCCs without H41$\alpha$ detections. The contour levels start at $3\sigma$ in steps of $3\sigma$.}\label{fig:SiO_map2}
\end{figure*}

\begin{figure*}
  \centering
  \includegraphics[scale=0.35]{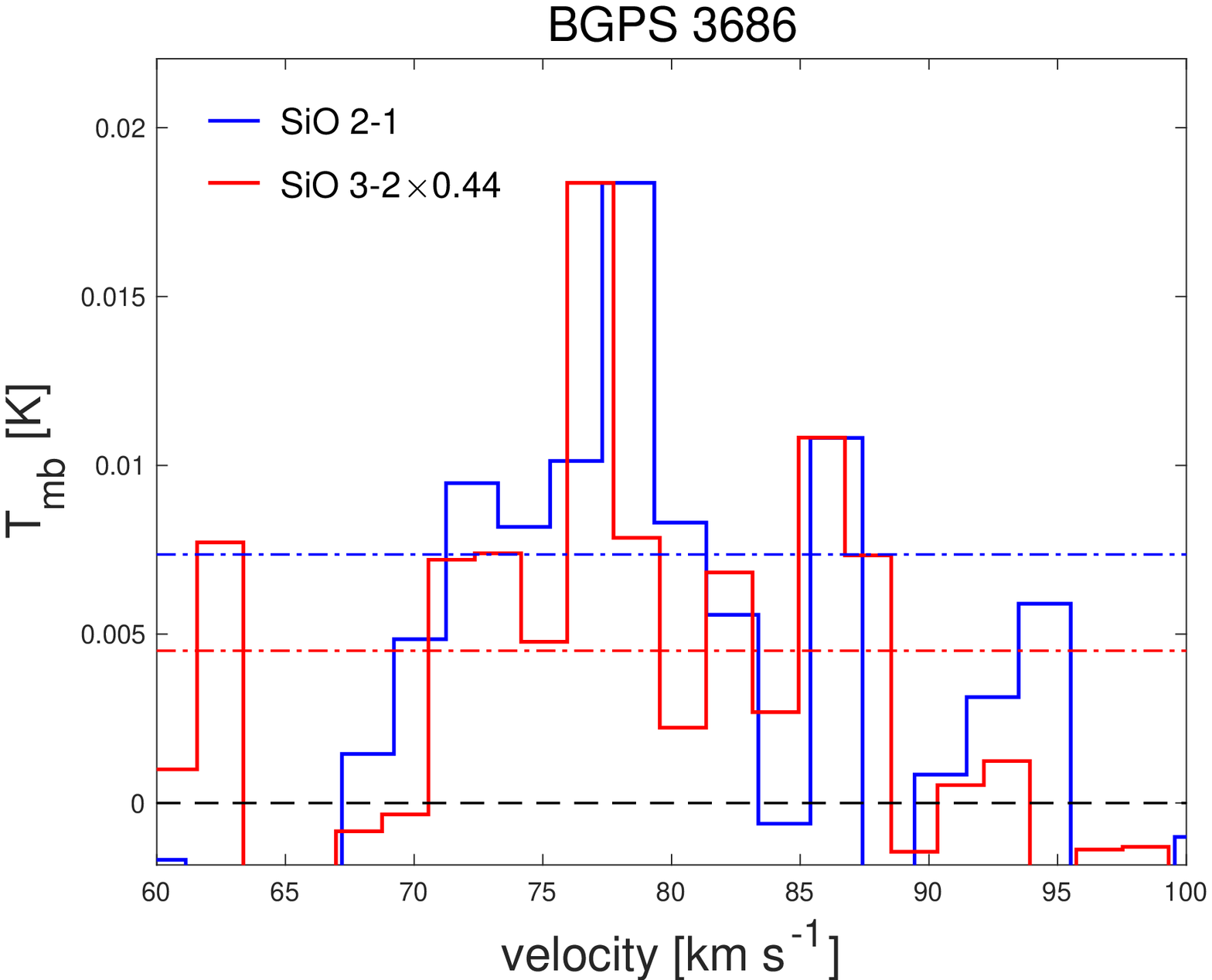}
  \includegraphics[scale=0.35]{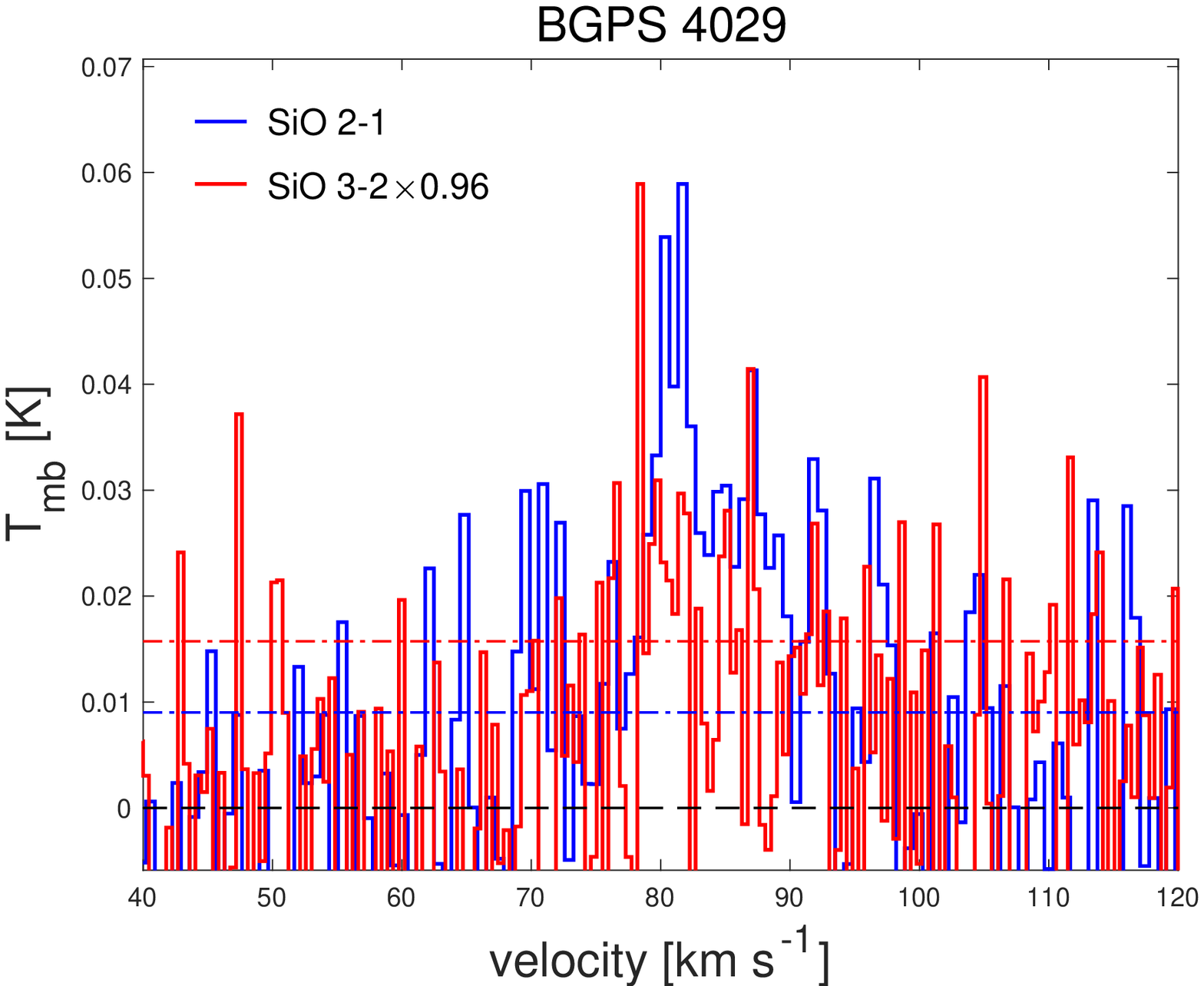}
  \includegraphics[scale=0.35]{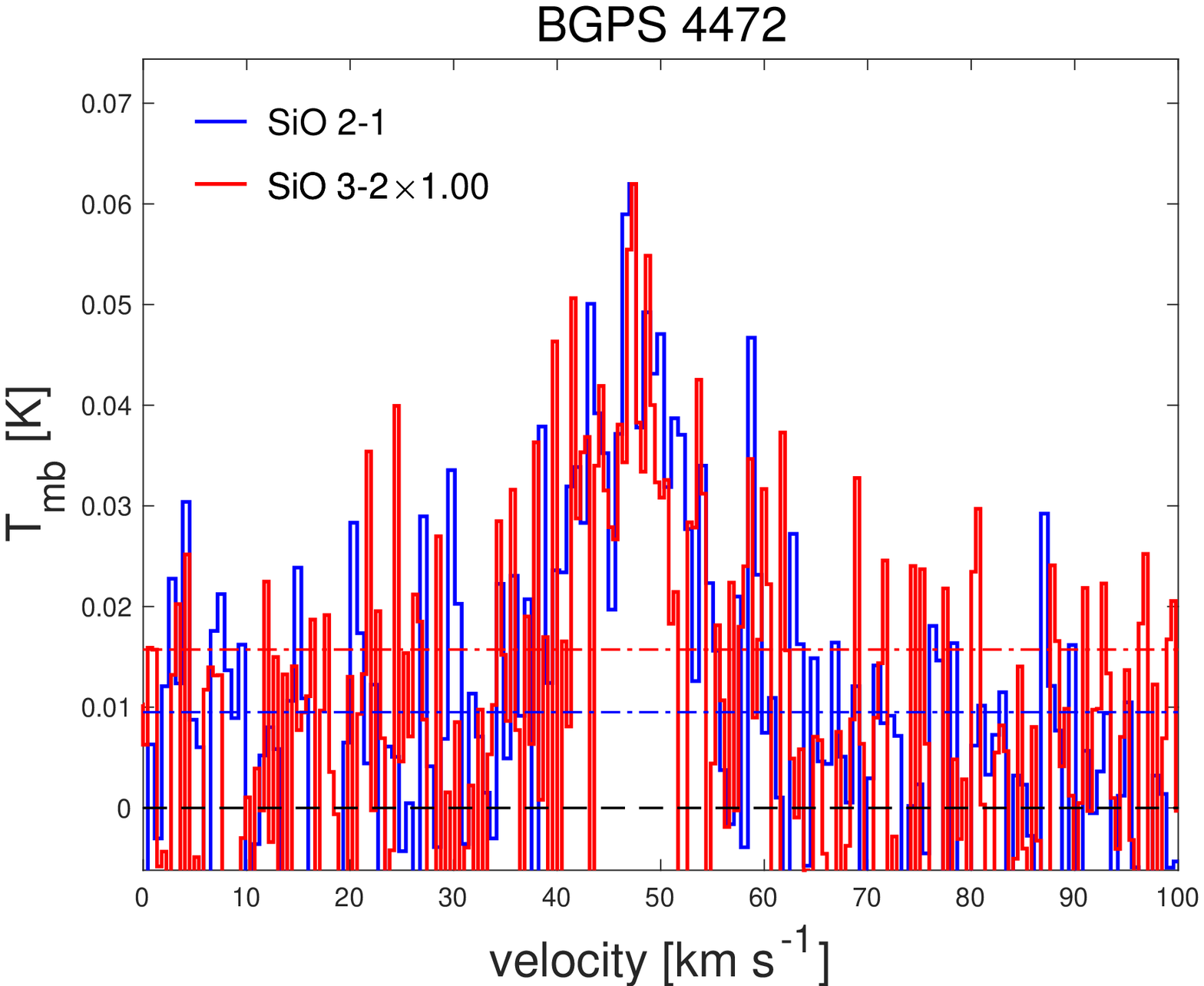}
  \includegraphics[scale=0.35]{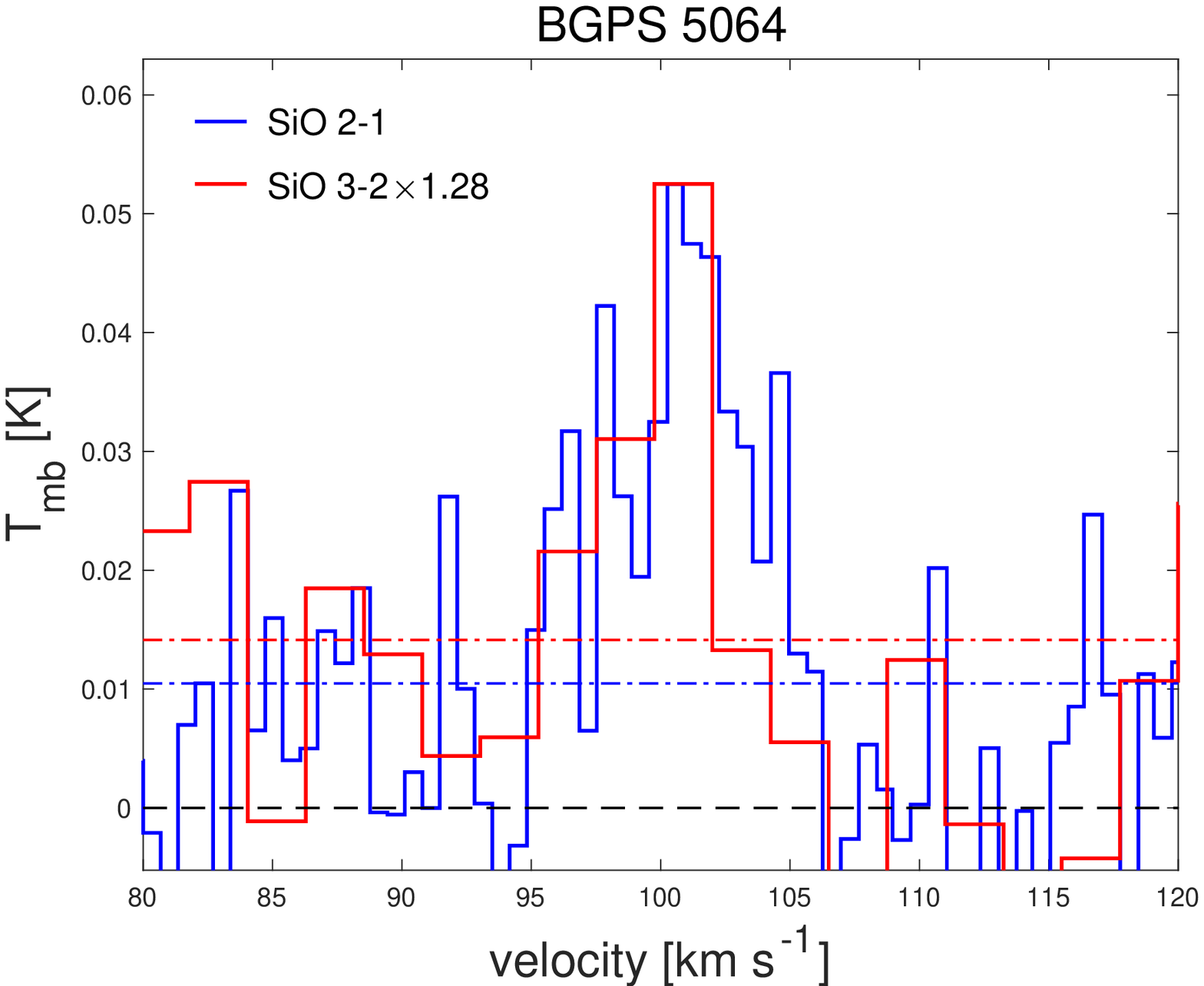}
  \includegraphics[scale=0.35]{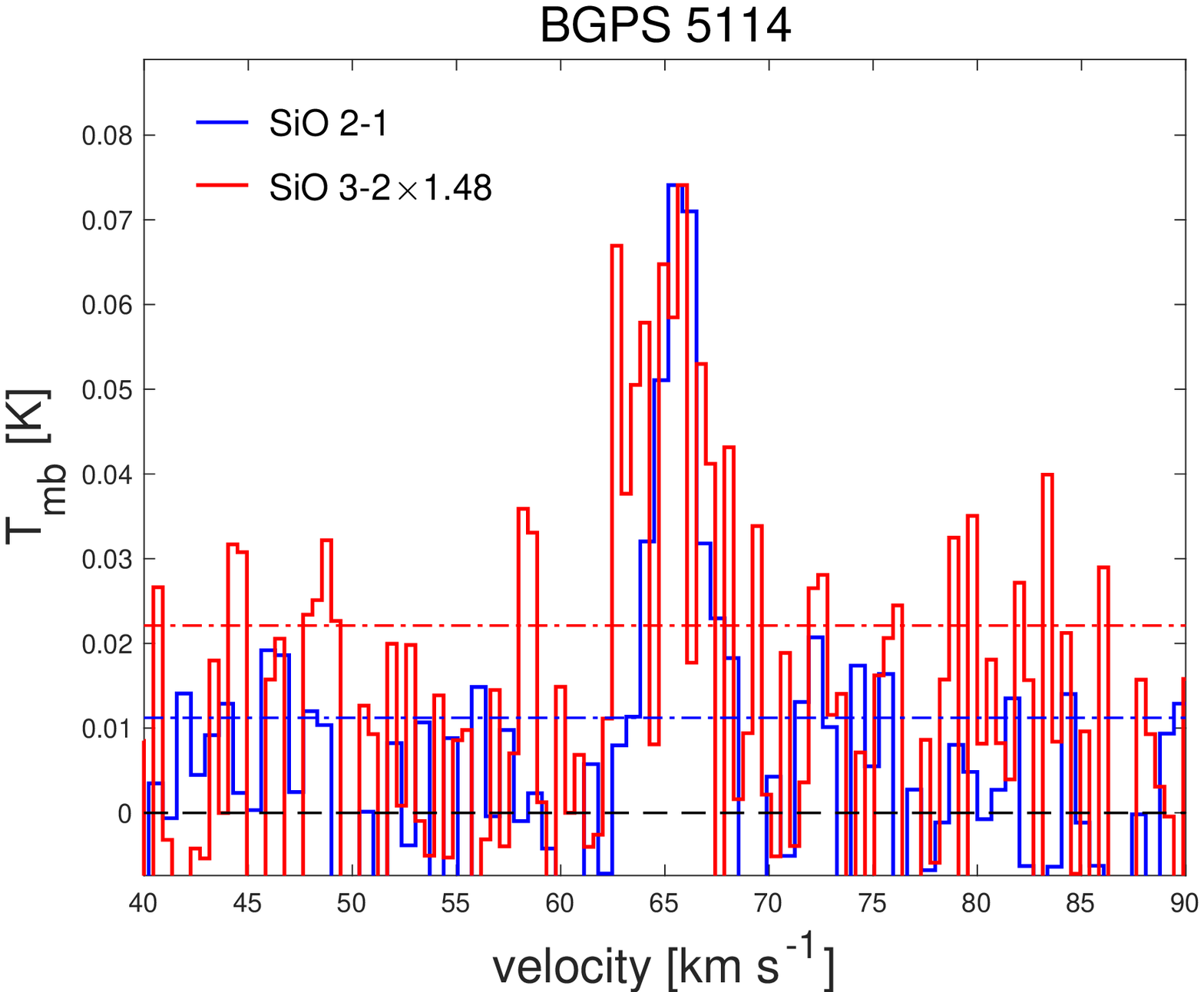}
  \includegraphics[scale=0.35]{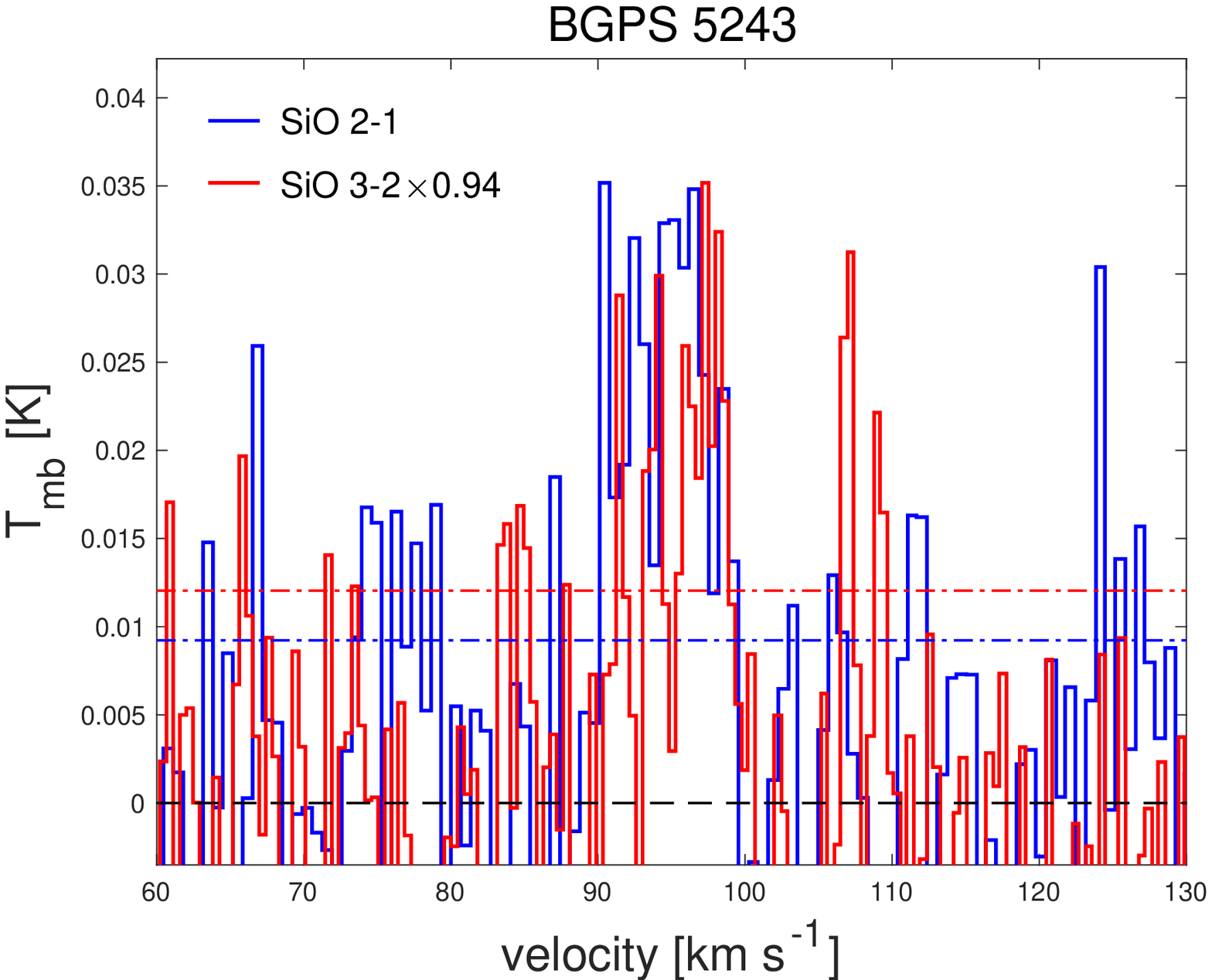}
  \caption{The SiO 2-1 and 3-2 spectra toward the SCCs with H41$\alpha$ detections. The blue and red dash-dotted horizontal lines are the rms noise levels at the velocity resolution for the SiO 2-1 and 3-2 lines, respectively.}\label{fig:SiO_spectrum2}
\end{figure*}

The distributions and spectra of SiO 2-1 and 3-2 lines for the sources with and without H41$\alpha$ detections are shown in Figure \ref{fig:SiO_map1}, \ref{fig:SiO_spectrum1}, \ref{fig:SiO_map2} and \ref{fig:SiO_spectrum2}. The spectra of SiO 2-1 and 3-2 lines correspond to the results given in Table \ref{table_property1}.

%The distributions of SiO 2-1 and 3-2 lines for the sources with and without H41$\alpha$ detections are shown in Figure \ref{fig:SiO_map1} and \ref{fig:SiO_map2}. In all of the 9 sources, the distribution areas of SiO 3-2 line are significantly smaller than those of SiO 2-1 lines because of the higher rms noise level. Hence, only the distribution of SiO 2-1 line is used to trace the distribution of shocked gas in the text. In addition, the SiO 3-2 spectra are vague when they are measured within the spreading areas of SiO 2-1 line. So different regions are chosen for the SiO 2-1 and 3-2 lines to measure the line properties in Table \ref{table_property1}. The SiO 2-1 and 3-2 spectra corresponding to the results in Table \ref{table_property1} are plotted in Figure \ref{fig:SiO_spectrum1} and \ref{fig:SiO_spectrum2}.

\section{Comparison with the KVN 21-m observations}

In Table \ref{table_propertykvn}, the observational results of the SiO 2-1 and 3-2 lines in the current observations are compared with those in previous KVN 21-m observations \citep{zhu20}. The results of the current observations are convolved with the spatial resolution of the KVN 21-m observations. The differences between the central velocities of the SiO lines observed by the two telescopes are typically smaller than 2 km s$^{-1}$. In the comparison of the FWHMs of SiO 2-1 and 3-2 lines, the corresponding differences are $\sim 30\%$ and $40\%$, respectively. And for the velocity-integrated intensities of the SiO 2-1 and 3-2 lines, the differences are $\sim 30\%$ and $25\%$, respectively. Comparing the accuracies of the velocity-integrated intensities and FWHMs in the two sets of observations, no significant  difference between the two observations is found. Therefore, it suggests that the results of the previous KVN 21-m observations toward the sample of 100 SCCs are reliable.

\begin{table*} \tiny %\footnotesize
\centering
\caption{The properties of the SiO lines from IRAM-30m and KVN-21m observations}\label{table_propertykvn}
\begin{threeparttable}
\begin{tabular}{|c|ccc|ccc|}
\hline
   &  \multicolumn{3}{c}{SiO 2-1} & \multicolumn{3}{c}{SiO 2-1(KVN)} \\
\hline
\multirow{2}*{Source} & V$_{lsr}$ & FWHM & $\int T_{mb}dv$ & V$_{lsr}$ & FWHM & $\int T_{mb}dv$ \\
 & [km s$^{-1}$] & [km s$^{-1}$] & [K km s$^{-1}$] & [km s$^{-1}$] & [km s$^{-1}$] & [K km s$^{-1}$] \\
\hline
BGPS 3110 & $17.49\pm0.12$ & $3.66\pm0.29$ & $0.51\pm0.05$ & $17.49\pm0.20$ & $4.43\pm0.57$ & $0.63\pm0.06$ \\
BGPS 3114 & $22.85\pm0.20$ & $2.87\pm0.46$ & $0.26\pm0.05$ & $22.68\pm0.32$ & $4.69\pm0.71$ & $0.45\pm0.06$ \\
BGPS 3118 & $17.08\pm0.58$ & $4.04\pm1.12$ & $0.29\pm0.07$ & $18.10\pm0.76$ & $8.98\pm1.77$ & $0.50\pm0.09$ \\
%BGPS 3686 & $78.51\pm0.95$ & $3.95\pm2.23$ & $0.11\pm0.05$ & $80.55\pm1.67$ & 20.5 & $0.45\pm0.08$ \\
BGPS 3686 & $78.46\pm1.14$ & $5.14\pm2.69$ & $0.15\pm0.08$ & $78.01\pm1.34$ & $7.47\pm4.67$ & $0.25\pm0.09$ \\
BGPS 4029 & $81.56\pm0.75$ & $12.73\pm1.77$ & $0.72\pm0.09$ & $82.85\pm0.68$ & $11.83\pm1.68$ & $0.68\pm0.08$ \\
BGPS 4472 & $49.06\pm0.78$ & $15.83\pm1.84$ & $1.10\pm0.10$ & $49.26\pm0.99$ & $14.36\pm2.63$ & $0.99\pm0.14$ \\
BGPS 5064 & $100.85\pm0.53$ & $7.46\pm1.24$ & $0.40\pm0.06$ & $100.13\pm0.79$ & $10.90\pm2.65$ & $0.50\pm0.08$ \\
BGPS 5114 & $65.34\pm0.26$ & $3.31\pm0.61$ & $0.27\pm0.05$ & $65.17\pm0.25$ & $2.91\pm0.51$ & $0.31\pm0.05$ \\
BGPS 5243 & $95.68\pm0.44$ & $6.50\pm1.03$ & $0.37\pm0.05$ & $96.91\pm0.35$ & $3.67\pm0.64$ & $0.19\pm0.03$ \\
\hline
   &  \multicolumn{3}{c}{SiO 3-2} & \multicolumn{3}{c}{SiO 3-2(KVN)} \\
\hline
\multirow{2}*{Source} & V$_{lsr}$ & FWHM & $\int T_{mb}dv$ & V$_{lsr}$ & FWHM & $\int T_{mb}dv$ \\
 & [km s$^{-1}$] & [km s$^{-1}$] & [K km s$^{-1}$] & [km s$^{-1}$] & [km s$^{-1}$] & [K km s$^{-1}$] \\
\hline
BGPS 3110 & 18.01$\pm$0.20 & $3.79\pm0.46$ & $0.54\pm0.06$ & $17.94\pm0.13$ & $3.38\pm0.32$ & $0.48\pm0.04$\\
BGPS 3114 & $22.98\pm0.21$ & $3.21\pm0.50$ & $0.35\pm0.06$ & $22.21\pm0.29$ & $2.40\pm0.70$ & $0.35\pm0.08$ \\
BGPS 3118 & $17.35\pm0.34$ & $2.19\pm0.81$ & $0.17\pm0.06$ & $17.56\pm0.55$ & $3.42\pm1.01$ & $0.38\pm0.11$ \\
%BGPS 3686 & $78.96\pm2.08$ & $13.72\pm5.61$ & $0.41\pm0.14$ & $80.79\pm1.18$ & 13.8 & $0.45\pm0.09$ \\
BGPS 3686 & $80.99\pm3.35$ & $17.04\pm7.88$ & $0.30\pm0.09$ & $80.79\pm1.28$ & $10.30\pm2.46$ & $0.40\pm0.09$ \\
BGPS 4029 & $82.15\pm1.91$ & $14.66\pm4.49$ & $0.53\pm0.14$ & ... & ... & ... \\
BGPS 4472 & $50.74\pm0.94$ & $20.21\pm2.22$ & $1.65\pm0.16$ & $51.09\pm0.73$ & $11.95\pm1.81$ & $1.24\pm0.15$ \\
BGPS 5064 & $98.67\pm0.96$ & $4.31\pm1.62$ & $0.17\pm0.06$ & $101.24\pm1.24$ & $7.24\pm3.02$ & $0.19\pm0.06$ \\
BGPS 5114 & $65.10\pm0.49$ & $4.38\pm1.16$ & $0.24\pm0.07$ & $65.37\pm0.18$ & $1.66\pm0.42$ & $0.14\pm0.03$ \\
BGPS 5243 & $97.46\pm0.40$ & $2.88\pm0.95$ & $0.21\pm0.06$ & $97.30\pm0.49$ & $3.45\pm0.86$ & $0.17\pm0.04$ \\
\hline
\end{tabular}
\begin{tablenotes}
\item{\textbf{Notes.}} The properties are estimated from line profiles fitted by a single Gaussian distribution. This method is different from that in \citet{zhu20}.
\end{tablenotes}
\end{threeparttable}
\end{table*}

\section{Three components in the HCO$^+$ spectrum toward BGPS 5064}

\begin{figure}
  \centering
  \includegraphics[scale=0.40]{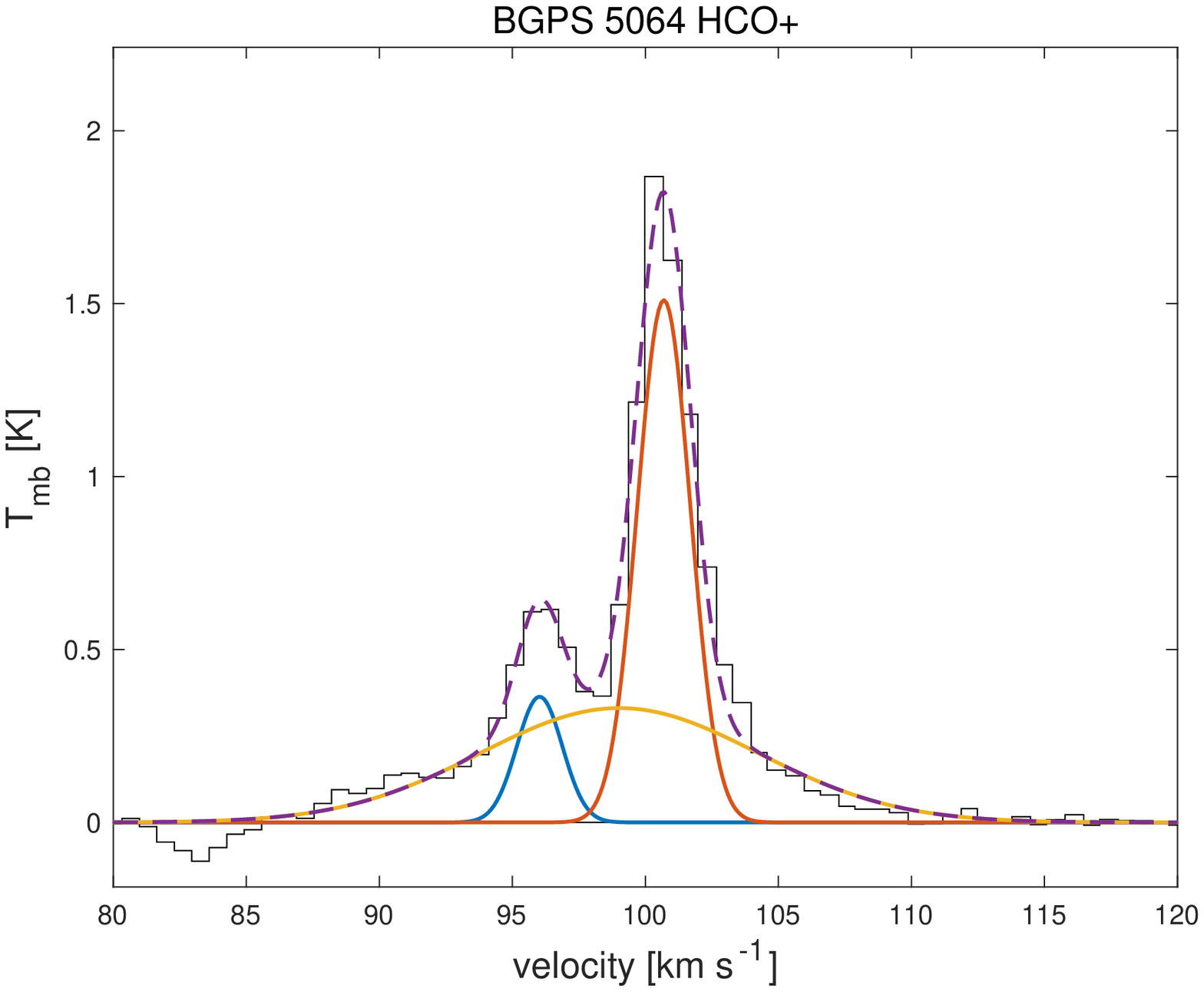}
  \caption{The HCO$^+$ 1-0 line profile toward BGPS 5064 is plotted. The coverage is the same as that of the HCO$^+$ spectrum shown in the top panel of Figure \ref{fig:BGPS5064_HCO+_blrd}. The broad velocity component shown in the yellow curve indicates the presence of protostellar outflows.}\label{fig:BGPS5064_HCO+_spec3}
\end{figure}

The HCO$^+$ line profile toward BGPS 5064 can be divided into three velocity components. The broad velocity component shown in Figure \ref{fig:BGPS5064_HCO+_spec3} should result from protostellar outflows. Then the velocity ranges of the blue- and red-shifted high-velocity components can be appropriately chosen.

%\appendix
%\section{The noise levels and the SiO spectra for individual sources}

%In the current work,

%%%%%%%%%%%%%%%%%%%%%%%%%%%%%%%%%%%%%%%%%%%%%%%%%%

% Don't change these lines
\bsp	% typesetting comment
\label{lastpage}
\end{document}